\documentclass[10pt]{paper}

\usepackage{setspace} 

\usepackage{amssymb}
\usepackage{amsfonts}
\usepackage{graphicx}
\usepackage{subfigure}
\usepackage{enumitem}
\usepackage{amsmath}
\usepackage{setspace}
\usepackage{fancybox}
\usepackage[bottom]{footmisc}
\usepackage[left=3.5cm,top=2.5cm,right=3cm,bottom=2cm]{geometry}
\usepackage[normal]{caption}
\usepackage{rotating}
\usepackage{color}
\usepackage{url}
\usepackage[english]{babel}
\usepackage{booktabs}
\usepackage{siunitx}
\usepackage{footnote}
\usepackage{footmisc}
\usepackage{multirow}
\usepackage{fancyhdr}
\usepackage{verbatim}
\usepackage[normalem]{ulem}

\makeatletter
\renewcommand{\@biblabel}[1]{\quad#1.}
\makeatother

\date{}
\begin{document}


\begin{flushleft}
{\Large
\textbf{How to estimate epidemic risk from incomplete contact diaries data?}
}
\\
Rossana Mastrandrea$^{1,2}$, 
Alain Barrat$^{1,3,*}$
\\
\bf{1} Aix Marseille Univ, Univ Toulon, CNRS, CPT, Marseille, France
\\
\bf{2} IMT Institute of Advanced Studies, Lucca, piazza S. Ponziano 6, 55100 Lucca, Italy
\\
\bf{3} Data Science Laboratory, ISI Foundation, Torino, Italy

$*$ E-mail: alain.barrat@cpt.univ-mrs.fr
\\
\end{flushleft}

\section*{Abstract}
Social interactions shape the patterns of spreading processes in a population. Techniques such 
as diaries or proximity sensors allow to collect data about encounters and to build networks 
of contacts between individuals. The contact networks obtained from these different techniques are however 
quantitatively different. Here, we first show how these discrepancies affect the prediction of 
the epidemic risk when these data are fed to numerical models of epidemic spread: low participation rate, 
under-reporting of contacts and overestimation of contact durations in contact diaries with respect 
to sensor data determine indeed important differences in the outcomes of 
the corresponding simulations {with for instance an enhanced sensitivity to initial conditions}. 
Most importantly, we investigate if and how information gathered from contact diaries 
can be used in such simulations in order to yield an accurate description of the epidemic risk, 
assuming that data from sensors represent the ground truth. The contact networks built from contact sensors and 
diaries present indeed several structural similarities: this suggests the possibility to construct, using only 
the contact diary network information, a surrogate contact network such that simulations using this 
surrogate network give the same estimation of the epidemic risk as simulations using the contact sensor network.
 We present and {compare} several methods to build such surrogate data, 
 and show that it is indeed possible to 
obtain a good agreement between the outcomes of simulations using surrogate and sensor data, as long as the 
contact diary information is complemented by publicly available data describing the heterogeneity of the durations of human contacts.

\section*{Author Summary} 

Schools, offices, hospitals play an important role in the spreading of epidemics. Information about interactions 
between individuals in such contexts can help understand the patterns of transmission and design ad hoc immunization strategies.
Data about contacts can be collected through various techniques such as diaries or proximity sensors. 
Here, we first ask if the corresponding datasets give similar predictions of the epidemic risk when 
they are used to build a network of contacts among individuals. Not surprisingly, the answer is negative:
indeed, if we consider data from sensors as the ground truth, diaries are affected by low participation rate, under-reporting and 
overestimation of durations. Is it however possible, despite these biases, to use data from contact diaries
to obtain sensible epidemic risk predictions? We show here that, thanks to the structural 
similarities between the two networks, it is possible to use the contact diaries to build surrogate versions
of the contact network obtained from the sensor data, such that both yield the same epidemic risk estimation.
We show that the construction of such surrogate networks can be performed using solely the information contained
in the contact diaries, complemented by publicly available data on the heterogeneity of cumulative contact durations between individuals.

\section*{Introduction}

Knowledge of the structure of human interactions is crucial for the study of infectious diseases spread and the design and evaluation of adequate
containment strategies. The structure of contact networks, \cite{BBVbook}, 
the presence of communities \cite{Fortunato:2010}, bridges or \lq\lq linkers\rq\rq{} between communities
\cite{Onnela:2007,Salathe:2010b,Genois2:2015}, \lq\lq super-spreaders\rq\rq{}
\cite{Pastor-Satorras:2001,Galvani:2005,Temime:2009},
the heterogeneity of contact durations \cite{BarCat:2015},
are all important characteristics that determine potential transmission patterns. 
The study of human contacts is particularly relevant in contexts such as schools, working places, hospitals where individuals
might spend several hours in close proximity
\cite{Genois2:2015, Mikolajczyk:2008,Read:2012,Bernard:2009,Salathe:2010,Isella:2011,Hornbeck:2012,Fournet:2014,Sekara:2014,Voirin:2015,Obadia:2015,Eames:2015,Mastrandrea:2015,Mastrandrea2:2015}.  

Interactions and contacts between individuals are conveniently seen within the framework of networks  
in which nodes represent individuals
and (weighted) links correspond to the occurrence of contacts (the weight giving the duration of the contacts). 
Measuring directly such networks
represents an important challenge \cite{Eames:2015}. Many studies have relied on contact diaries or surveys 
\cite{Mikolajczyk:2008,Bernard:2009,Edmunds:1997,Read:2008,Mossong:2008,Goeyvaerts:2010, Melegaro:2011,Conlan:2011,Smieszek:2012,Danon:2012,Potter:2012,Danon:2013},
while technological advances have led to a strong increase in the use of wearable sensors in the recent years 
\cite{BarCat:2015,Salathe:2010,Isella:2011,Hornbeck:2012,Sekara:2014,Voirin:2015,Obadia:2015,Eames:2015,Pentland:2008,Cattuto:2010,Toth:2015}.
Quantitative comparisons between datasets obtained from sensors and self-reported diaries, in terms of 
the numbers and durations of contacts between individuals and of the contact network statistics, are however scarce, mainly 
because very few studies have combined these two data collection means \cite{Mastrandrea:2015,Smieszek:2014}.
These investigations have shown that diaries suffer from small participation rates, under-reporting of contacts, and over-estimation of the contact durations. 
{Underreporting is particularly strong for short contacts, while long ones are better reported, and some studies have
put forward methods to estimate its magnitude and to correct for it \cite{Smieszek:2012,Potter:2015}.}
Interestingly, and
despite the much lower number of nodes and links in contact networks inferred from contact diaries, 
the overall structure of these networks is very similar to the one obtained from wearable sensors. 
Moreover, the links with largest
weights (as measured by sensors), which might play a major role in propagation processes, are reported with high probability in the contact diaries.

In this paper, we go beyond the comparison of the contact networks obtained by these methodologies and 
explore the impact of their differences on the evaluation of the epidemic risk when such datasets are used in 
numerical simulations of infectious disease propagation. 
Our goal is to understand to what extent and how the information gathered from contact diaries can be used in such simulations in order to 
yield an accurate description of the epidemic risk, despite the biases mentioned above. We first compare the outcomes of spreading simulations
performed using data coming from wearable sensors and from contact diaries that describe the contacts between students in the same 
context (a high school) 
and on the same day. Although the two networks are supposed to describe the same reality, we observe important differences 
in the simulations, due to the low participation rate in the diaries and to a stronger community structure in
the contact diaries network than in the contact sensors network. We then design and evaluate
a set of methods to use the information contained in the
contact diaries to build surrogate versions of the contacts that yield, when used in the simulations, a better estimation
of the real epidemic risk as quantified by the distribution of epidemic sizes (considering as ground truth the dataset from the sensors). 
We show that good results are obtained when the contact diary information
is complemented by known stylized facts characterizing human interactions, 
in particular the heterogeneity of contact durations.

\section*{Results}

\subsection*{Data description}

We use two datasets collected in a French high-school in 2013 and made publicly available in \cite{Mastrandrea:2015}. 
The data describe face-to-face contacts between students of $9$ classes as collected by (i) the SocioPatterns infrastructure \cite{SocioPatterns}
based on wearable sensors, during one week and (ii) self-reported contact diaries filled on a specific day of the same week (Dec. $5^{th}$, 2013).
{In the diaries, contact was explicitly defined as close (less than 2 m) face-to-face proximity, in order to match as much as possible
this definition to the contacts detected by sensors.}
Using these data, we build two distinct contact networks for the day in which the diaries were collected: 
the Contact Sensors Network ($\text{CSN}$) and the Contact Diaries Network ($\text{CDN}$). In each network, nodes represent students and a link is drawn between a pair of nodes
$(i,j)$ if at least one contact between students $i$ and $j$ is present in the corresponding dataset during the considered day.  
{We present and compare in the Supplementary Information the main networks' characteristics. 
Note that, in the diaries, some participants reported contacts with non-participants. One could a priori use this information
and build a contact network including both participants and non-participants. However, since by definition the contacts of non-participants are unknown,
this would introduce a potentially strong and most importantly completely 
uncontrolled bias in the measures of the network's structural properties such
as, e.g., its clustering or the node degrees.}

A weight can {moreover} be assigned to each link $(i,j)$: for the $\text{CSN}$, the weight $w_{ij}$ is given by the cumulative duration 
of the contacts registered by the sensors on that day between $i$ and $j$; 
for the $\text{CDN}$ we can use the duration reported by the students in the diaries, building the network $\text{CDN}_\text{D}$,
or use for each link a duration taken at random from the list of durations registered by the sensors, obtaining the network $\text{CDN}_\text{S}$
(see Methods for details). The rationale behind building $\text{CDN}_\text{S}$ comes from the results of 
\cite{Mastrandrea:2015,Smieszek:2014}
that show that durations reported by students tend to be strongly overestimated.
{Since, on average, contacts reported in the diaries as long tend also to be long according to the sensor data, we will also consider
a different assignment of links to the $\text{CDN}$, in which we still take durations
at random from the list of durations registered by the sensors, but assign the longer durations to the
links of $\text{CDN}$ with longer reported durations: we denote the resulting network by
$\text{CDN}_\text{S'}$.}

The contact sensor network counts $295$ nodes (participation rate $77.8 \%$) and $2162$ links, 
while the contact diaries network has $120$ nodes (participation rate $31.6\%$) and $348$ links. 
{Incomplete participation, even in the case of the sensor data, leads to biases in the simulations
using the $\text{CSN}$ with respect to what would be obtained if the whole population had participated, 
due to the fact that contacts with and among non-participants are not detected. This point has been discussed
in \cite{Genois:2015}, together with methods to build surrogate data and obtain estimate of the 
epidemic risk in the case of such population sampling. In order not to confuse the issues of population sampling and comparison between diaries and sensors,
we consider here as ground truth the $\text{CSN}$, collected 
by wearable sensors for which the definition of contact does not depend on a possible
interpretation of the diary question by the students nor on the fact that they might not recall contact events.}

\subsection*{Numerical simulations of epidemic spread}

In the following, we perform simulations of the spread of infectious diseases in the considered population, using as substrate for propagation
events the contact networks described above. 
{It is important to note here that we consider propagation processes on static networks. Indeed, the $\text{CDN}$ 
does not contain information on the timing of the contacts, so that it is natural to compare the outcome of simulations performed on such a static network
with simulations performed on a static version of the $\text{CSN}$. Moreover, when modeling the propagation of infectious diseases with realistic timescales of several days,
it has been shown in \cite{Stehle2:2011} that a daily weighted contact network contains enough information
to obtain a good estimate of the process outcome. When dealing with faster processes, the temporal evolution of the network
would become relevant; in that case, it would be possible to use the techniques put forward in 
 \cite{Genois:2015} to build realistic surrogate timelines of contacts on weighted
networks, using the robustness of the distributions of the durations of single contact events and of the intervals
between successive contacts measured in different contexts. Note also that, even if the networks
do not take into account the timing of the contact events, they still include information on the aggregate contact 
durations through the weights, which are known to play
a crucial role in the outcome of spreading processes \cite{Stehle2:2011,Smieszek:2009,Eames:2009,Kamp:2012}.
}

For simplicity, we consider the paradigmatic Susceptible-Infected-Recovered (SIR) model of epidemic propagation. In this model, each
Susceptible node $i$ can be infected by an Infected neighbour $j$ with probability $\beta *w_{ij}* dt$ for each small time step $dt$.
Infected people recover with rate $\mu$ and enter in the Recovered category. Recovered individuals cannot be infected again. 
The process starts with a single Infected individual chosen at random (the seed) and ends when there are no more Infected nodes.
The epidemic risk in the population, which depends on the interplay of the ratio $\beta/\mu$ and the network's structure and weights,
is measured by the distribution of the final size of epidemics (i.e., of the fraction of individuals in the Recovered category at the end of the process), 
obtained by repeating the simulations with randomly chosen seeds. 
Note that, since we consider static networks, only the ratio $\beta/\mu$ is relevant, 
and multiplying both by a certain factor only changes the timescale on which the epidemic unfolds.
The shape of the distribution of epidemic sizes depends on the features of the underlying network structure in terms of
possible patterns of contagion. The comparison of these distributions
gives hints about similarities and discrepancies of various datasets for the evaluation of the epidemic risk.

\begin{figure}[!ht]
\centering
{\includegraphics[width=1\textwidth]{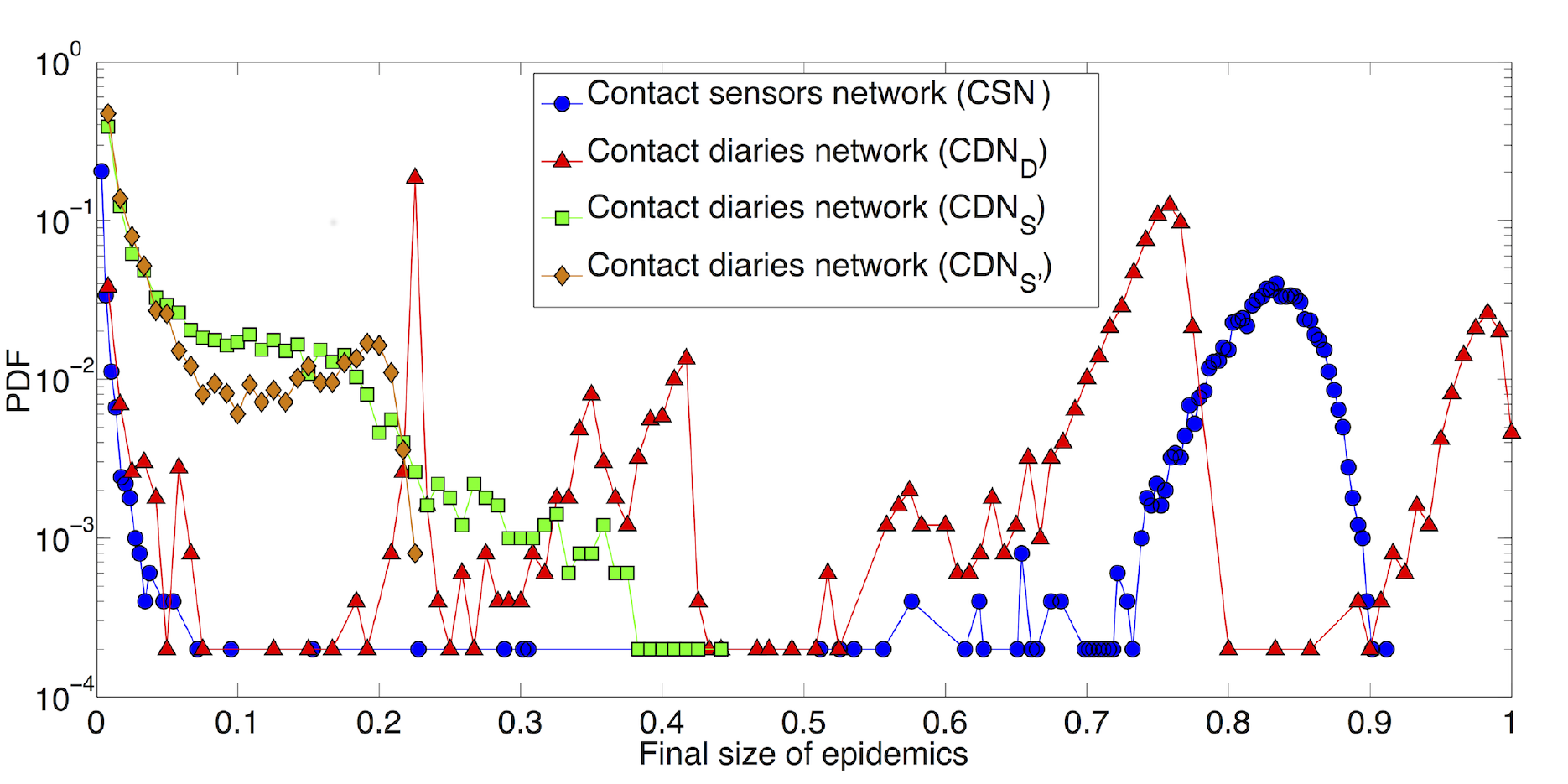}}
\caption{{{\bf Distribution of final size of epidemics.}  1000 SIR simulations performed on the original contact 
sensors network (CSN) and the original contact diaries network with durations respectively reported by students ($\text{CDN}_{\text{D}}$) 
and registered by sensors ($\text{CDN}_{\text{S}}$ 
{and $\text{CDN}_{\text{S'}}$}). Each process starts with one random infected seed. $\beta/\mu=30$.}
\label{pre}}
\end{figure}

We first compare in Fig \ref{pre} the outcome of simulations of the SIR model performed on the $\text{CSN}$ and on the two versions of the 
$\text{CDN}$ described above ($\text{CDN}_\text{D}$ with weights reported by students and $\text{CDN}_\text{S}$ with weights registered by sensors
assigned randomly to the links), for one specific value of $\beta/\mu=30$. The three distributions of epidemic sizes are very different from each other.
The outcome of simulations performed using $\text{CSN}$ is quite standard, with a fraction of small outbreaks that reach only a small fraction of the population
and another peak corresponding to large outbreaks. {As shown in the SI, the 
outcome does not depend on the class of the initial seed.} The shape of the
distribution obtained when using the $\text{CDN}_\text{D}$ is more peculiar, with a series of peaks, including one at very large epidemic sizes.
Such structure is typical of spreading processes on networks with a strong community structure \cite{Salathe:2010b}, which corresponds
to the results of \cite{Mastrandrea:2015}:
(i) due to the low participation rate and the under-reporting, the community structure of the CDN is stronger than the one of the $\text{CSN}$,
with few links between classes; depending on the seed, the simulated disease can thus remain confined in one class or in a group of few classes, leading to the peaks 
at intermediate values of the epidemic size; {we moreover show in the SI
that the outcome depends on the class of the initial seed for the $\text{CDN}$ but not for the $\text{CSN}$};
(ii) on the other hand, as contact durations are overestimated, the propagation probability on each link is also overestimated and, if the 
disease manages to spread between classes, almost all individuals are affected, leading to the peak at large epidemic sizes.
The $\text{CDN}_{\text{S}}$ case shows a different result: no more than half of the whole population is affected by the
spread. As the weights have in this case the same statistics as the $\text{CSN}$, this is simply due to the low participation
rate \cite{Genois:2015} and the much smaller average degree in the $\text{CDN}$ with respect to the $\text{CSN}$. We also note that,
since the weights are assigned randomly to the links between students, the structure of the contact matrix giving the average durations
of contacts between students of different classes can strongly differ between the $\text{CDN}_{\text{S}}$ and both the $\text{CSN}$ and
the $\text{CDN}_{\text{D}}$, leading to different patterns of propagation between classes (see Supporting
Information). {We finally note that the simulations on the $\text{CDN}_{\text{S'}}$, which keeps the 
distribution of the weights from $\text{CSN}$ and in which larger weights are assigned to links with longer reported durations,
yield even smaller outbreaks. This is probably due to the fact that the large weights reported in the diaries tend to be within
classes, so that the links bridging classes and favoring the spread tend to have smaller weights in the $\text{CDN}_{\text{S'}}$ than
in the $\text{CDN}_{\text{S}}$.}
{We also show in the SI the temporal evolution of the 
density of infectious individuals for the various cases considered here.}

\subsubsection*{Matching networks}

In order to discard the differences due simply to the population sampling, and to focus on the impact of under-reporting (i.e., of unreported
links) and overestimation of durations on the estimation of the epidemic risk, we consider ``matched'' versions
of the networks, in which we keep only
the nodes present in both $\text{CSN}$ and $\text{CDN}$. We obtain the \textit{matched} networks:
$\text{CSN}^{m}$, $\text{CDN}^{m}_{\text{D}}$ and $\text{CDN}^{m}_{\text{S}}$. Among the $120$ students who filled in the diaries, 
$11$ in fact did not wear sensors on the day of interest: we obtain thus $109$ nodes, distributed in $7$ of the $9$ classes of the 
$\text{CSN}$. We moreover discard one of the classes in which only one student filled in the diary. We end up with matched networks of $108$
nodes in $6$ classes.

\begin{figure}[!ht]
\centering
{\includegraphics[width=1\textwidth]{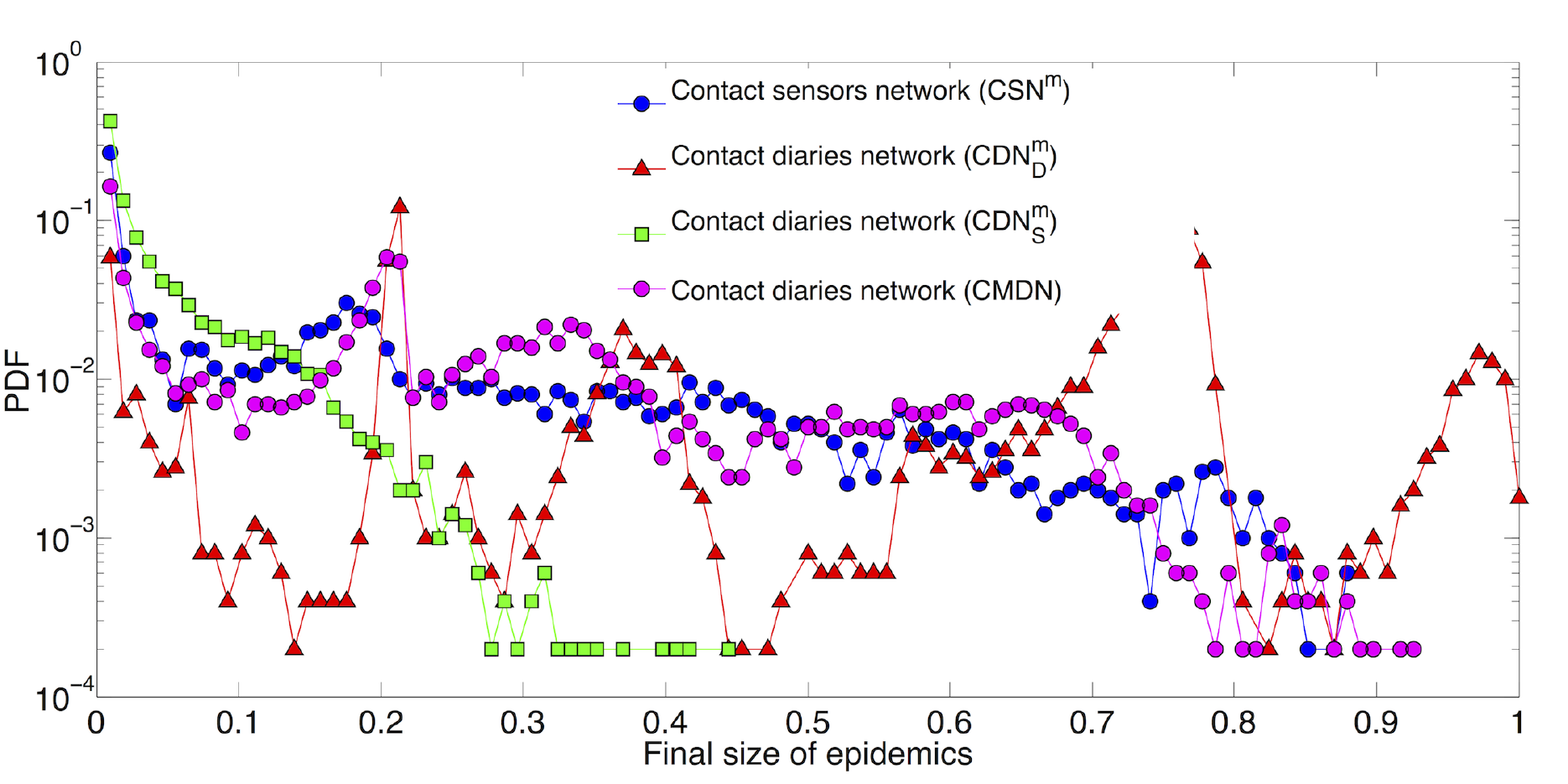}}
\caption{{{\bf Distribution of final size of epidemics.} 1000 SIR
    simulations performed on the matched contact sensors network
    ($\text{CSN}^m$), the matched contact diaries network with
    durations respectively reported by students
    ($\text{CDN}^m_{\text{D}}$) and registered by sensors
    $\text{CDN}^m_{\text{S}}$) and the contact diaries network with
    weights obtained by a negative binomial fit of contact durations
    reported by students between and within classes (CMDN). Each
    process starts with one random infected seed.  $\beta/\mu=30$.}
\label{sim}}
\end{figure}

Fig \ref{sim} displays the outcome of SIR simulations performed on the three matched networks. 
Comparison with Fig \ref{pre} shows that the outcomes for
$\text{CDN}^{m}_{\text{D}}$ and $\text{CDN}^{m}_{\text{S}}$ are similar to the cases of 
$\text{CDN}_{\text{D}}$ and $\text{CDN}_{\text{S}}$, which is expected as these networks do not differ strongly (only $12$ nodes and $62$ links have been removed
in the matching procedure). On the other hand, the epidemic risk is strongly underestimated in the $\text{CSN}^m$ with respect to the $\text{CSN}$:
this is due to the strong reduction in the number of nodes and links \cite{Genois:2015} and hence in the number of potential transmission routes
between students and classes. However, the distribution does not exhibit peaks as for the $\text{CDN}^m_\text{D}$ case: the community structure
remains indeed weaker in the $\text{CSN}^m$ with respect to the $\text{CDN}^m$, with higher densities of links between different classes. 
We also note that the underestimation obtained by using $\text{CSN}^m$ is less strong than in the case of a random removal of the same number of nodes
\cite{Genois:2015} (not shown). This is due to the fact that the students who filled in the diaries tend to be more connected than the others in the $\text{CSN}$:
as a result, the  $\text{CSN}^m$ has $970$ links while a random removal of the same number of nodes from the $\text{CSN}$ leads on average
to a network with $\approx 560$ links. 
Finally, although both  $\text{CSN}^m$ and $\text{CDN}^m_{\text{S}}$ have the same distributions of weights and lead both
to strong underestimations of the epidemic risk, the resulting distributions
do not coincide, in particular because the $\text{CDN}^m_{\text{S}}$ has a much smaller number of links.

We also show in Fig \ref{sim} the outcome of simulations performed using another representation of the contact diaries network,
namely the Contact Matrix Distribution Network (CMDN) introduced in \cite{Machens:2013} and built as follows: as explained in Methods,
we perform a fit of the distributions of contact durations reported by students by a negative binomial functional form, distinguishing between
contacts between students of the same class or of different classes. We then use these fitted distributions to randomly assign weights to each pair 
of students. Note that these weights can be equal to $0$, in which case no link is drawn between the students. This procedure yields a network
with global link density close to the $\text{CDN}^m$ and such that the contact matrices of link densities and of average contact durations
between classes are also similar to the ones obtained from the $\text{CDN}^m$ (see Supporting Information). The overall result is a
distribution of epidemic sizes more similar to the case of the $\text{CSN}^m$ (results for other values of $\beta/\mu$ are shown in the Supporting Information).

\subsubsection*{Building surrogate networks to estimate the epidemic risk}

We now address the issue of how the data coming from contact diaries could be used to provide an accurate estimation of the epidemic risk, despite
the discrepancies obtained when these data are used directly in simulations. To this aim, we propose several procedures
to build surrogate contact diaries networks that overcome the issues of low participation rate and overestimation of contact durations, 
which bear a strong impact on the simulation outcome, as shown above. 
In the same spirit as \cite{Genois:2015}, we start from the available dataset and extend it 
by adding the missing nodes to the contact network, as well as surrogate links.
We build these surrogate networks using only information known in the $\text{CDN}^m$. We note that we do not try to infer the true missing links but
to build a \lq\lq plausible\rq\rq{} version of these links, such that the simulations of epidemic spread on the resulting network yield an accurate estimation of the epidemic risk.

\begin{figure}[!ht]
\centering
{\includegraphics[width=\textwidth]{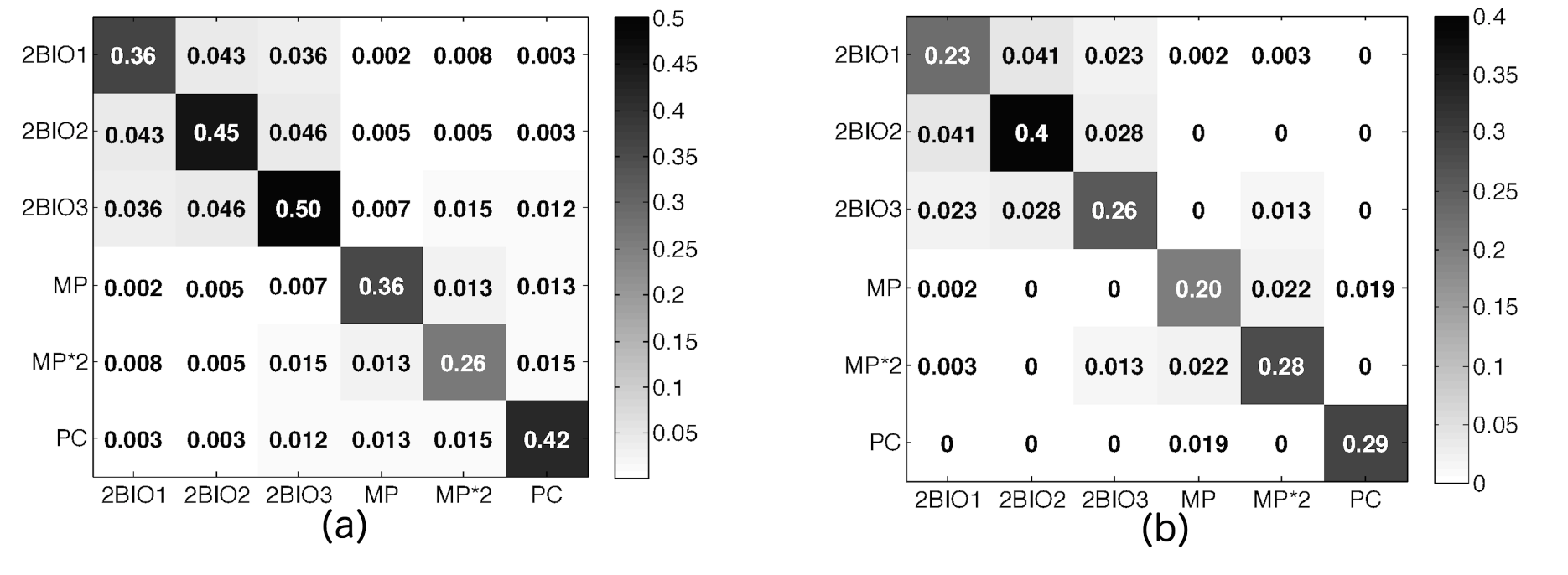}}
\caption{{\bf Contact matrices of edge densities.}  The entry at row $X$ and column $Y$ of the matrix is given 
by the total number of links between students in class $X$ and students in class $Y$, 
  normalized by the maximum number of observable links ($n_Xn_Y$ or $n_X(n_X-1)/2$ if $X=Y$, 
  with $n_X$ the cardinality of the class $X$) giving the edge link densities for 
  (a) the contact sensors network and (b) the matched contact diaries network, both limited to the $6$ classes considered. 
The cosine similarity between the two matrices is equal to 97\%.
\label{cont}}
\end{figure}

The rationale behind the procedures we propose comes from (i) the observed similarity between the overall structure of the contact networks measured
by sensors and by diaries, as quantified by the contact matrices of the densities of links between classes \cite{Mastrandrea:2015} 
shown in Fig \ref{cont}, and (ii) the results of \cite{Machens:2013} that show how such contact matrices, together with information
on the heterogeneity of contact durations, play a crucial role in determining propagation patterns in a structured population. Note that,
as $3$ of the $9$ classes
of the $\text{CSN}$ are not present in the $\text{CDN}$, we consider here a version of the  $\text{CSN}$ limited to the remaining $6$ classes (the resulting
$\text{CSN}$  has $204$ nodes and $1600$ links). 

We propose a three-steps procedure to build surrogate contact networks for the $6$ classes considered, 
 starting from the $\text{CDN}^m$, which contains only a fraction of the 
nodes of these classes (see details in Methods): 
(i) we first add the missing nodes in each class; 
(ii) we randomly add links in each class and between classes in order to maintain the contact matrix of edge densities fixed to its measured value in the 
$\text{CDN}^m$, shown in Fig \ref{cont} (b);
(iii) we associate weights to the links of the resulting surrogate network $\text{CDN}^{s}$.

Both steps (ii) and (iii) can be performed in different ways. 
With respect to step (ii), we notice that the empirical contact matrix (Fig \ref{cont}(b)) contains some elements equal to zero, corresponding to a total absence of links
between classes. This corresponds to an unrealistically strong community structure and is due to the low sampling rate and to the under-reporting of contacts. We
thus consider two cases: (a) we strictly keep the contact matrix with its zero elements; (b) we replace the zeros with random values drawn from a uniform distribution between
the minimum and maximum values of the non-zero off-diagonal elements of the matrix (see Methods).
In what follows, we will refer to these cases respectively with the subscripts  \textit{z} (\lq zero\rq{}) and \textit{nz} (\lq no zero\rq{}), 
obtaining $\text{CDN}^{s}_{\text{z}}$  and $\text{CDN}^{s}_{\text{nz}}$.

\begin{figure}[!ht]
\centering
{\includegraphics[width=1\textwidth]{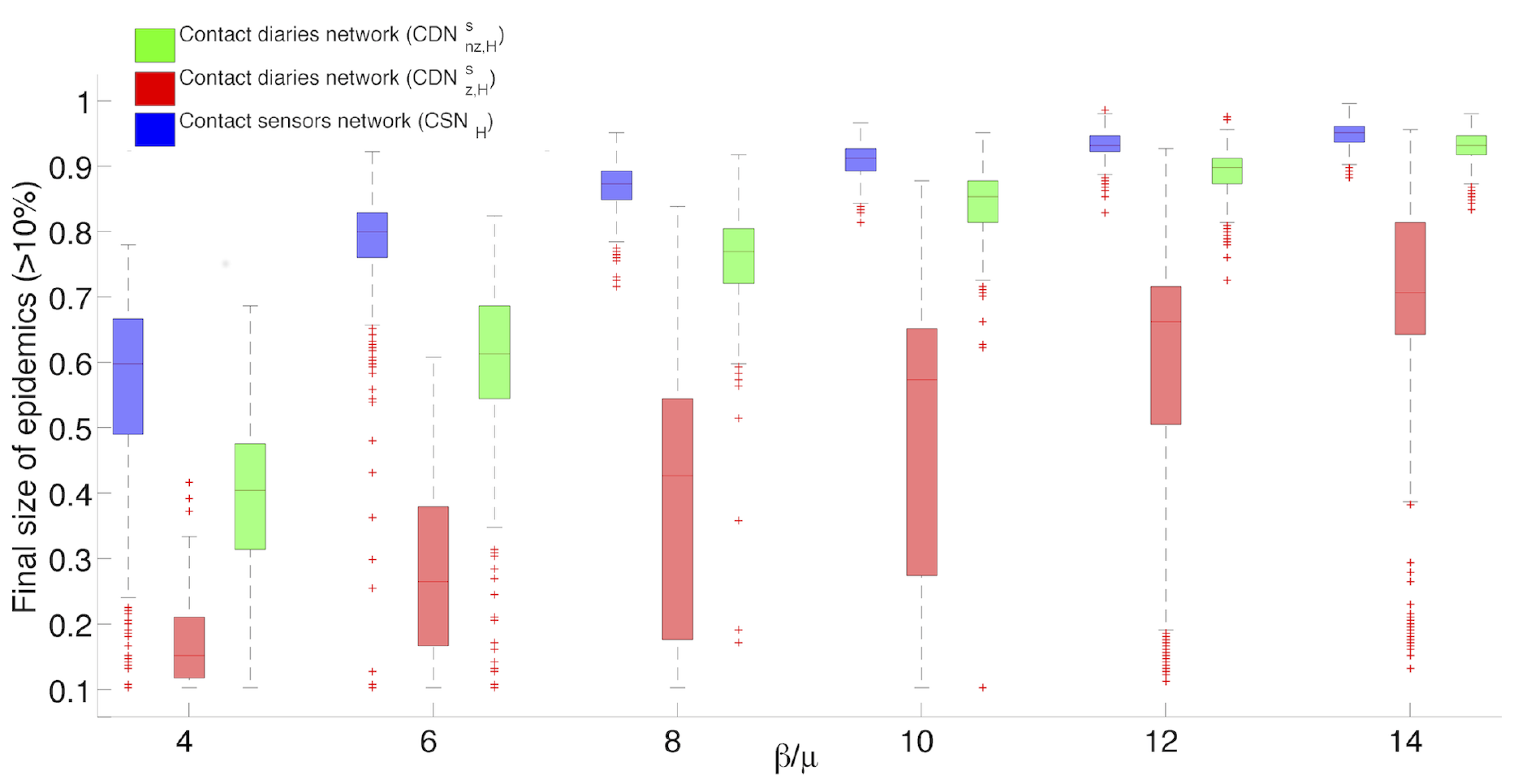}}
\caption{{\bf Box-plots of epidemic size distributions.} Comparison of the distributions of epidemic sizes for epidemics
reaching more than $10\%$ of the population, resulting from 
SIR simulations performed on the contact sensors network with homogeneous weights ($\text{CSN}_{\text{H}}$) and
two surrogate contact diaries networks with homogeneous durations ($\text{CDN}^s_{\text{z,H}}$, $\text{CDN}^s_{\text{nz,H}}$). 
For each boxplot, the central mark stands for the median, its edges represent the $25^{th}$ and $75^{th}$ percentiles.  
The whiskers extend to the most extreme data points not considered outliers, while outliers are plotted individually. 
Points are drawn as outliers if they are larger than $a + h(b - a)$ or smaller than $a - h(b - 
a)$, where $a$ and $b$ are the 25th and 75th percentiles, respectively and $h$ is the maximum 
whisker length set by default to $h=1.5$.
(1000 simulations for each value of the ratio $\beta/\mu \in \{4, 6,8,10,12,14\}$).
\label{hom}}
\end{figure}

We first focus on the structure obtained through this procedure.
We show in the SI some statistical characteristics of the surrogate networks, compared to the empirical
$\text{CDN}$  and $\text{CSN}$: in particular, the structural properties of $\text{CDN}^{s}_{\text{nz}}$ are much closer
to the $\text{CSN}$ than the empirical $\text{CDN}$.
Moreover, we start
by simply assigning homogeneous weights in step (iii) and compare the outcome of simulations
of the SIR model with simulations performed on a version of the  $\text{CSN}$ with as well homogeneous weights, denoted $\text{CSN}_\text{H}$. This amounts to the assumption 
that each student spends the same amount of time with all his/her contacts, a minimal assumption corresponding to an absence of information about contact durations.
We report in Fig \ref{hom}  boxplots for the distributions of epidemic sizes larger than $10\%$, 
obtained from SIR simulations at various values of $\beta/\mu$ on the resulting homogeneous networks
$\text{CDN}^{s}_{\text{z,H}}$ (homogeneous weights and contact matrix zeros kept) and $\text{CDN}^{s}_{\text{nz,H}}$ 
(homogeneous weights and contact matrix zeros replaced). We also report in the Supporting Information the fraction of epidemics reaching more
than $10\%$ of the population, as a function of $\beta/\mu$.

Fig  \ref{hom}  and Fig S7 of the Supporting Information clearly show that, despite the high similarity between 
the contact matrices of the surrogate network and of the $\text{CSN}$, an important underestimation of the epidemic size is obtained. 
Replacing the zeros in the contact matrix gives better results but still yields a clear underestimation of the risk with respect to the
$\text{CSN}$ reference.

Let us now turn to the more realistic hypothesis of heterogenous cumulative durations 
of contacts among students. It is indeed known that these durations are very heterogeneous: most are short, 
but durations orders of magnitude longer than the average are not uncommon \cite{BarCat:2015}. Within the usual hypothesis of 
a transmission probability proportional to the contact duration, this implies that 
different contacts can correspond to very different transmission probabilities, and hence that they should not be treated as equivalent. The
importance of the diversity of contact durations has indeed been 
assessed for instance in \cite{Stehle2:2011,Smieszek:2009,Eames:2009,Kamp:2012}.
In this case, step (iii) of the surrogate network building procedure,
which regards the assignment of weights to links, needs to be precised. 
We consider here two possibilities: we use either the list of weights (daily cumulative durations)
reported in the diaries, or the list of weights registered by the sensors. 
In both cases, weights are randomly drawn from the empirical list and assigned at random to the links in the surrogate network (see Methods). Taking into account the 
two possibilities of keeping or replacing the zeros in the contact matrix of link densities, we 
end up with four surrogate contact networks: $\text{CDN}^{s}_{\text{z,D}}$ and $\text{CDN}^{s}_{\text{nz,D}}$, both with weights randomly drawn from the 
list of durations reported by students (note that we do not find different results 
between keeping fixed the original weighted structure of the $\text{CDN}$ and 
assigning random weights also to the corresponding links); 
$\text{CDN}^{s}_{\text{z,S}}$ and $\text{CDN}^{s}_{\text{nz,S}}$, both with weights randomly picked from the cumulative durations registered by sensors.

\begin{figure}[!ht]
\centering
{\includegraphics[width=\textwidth]{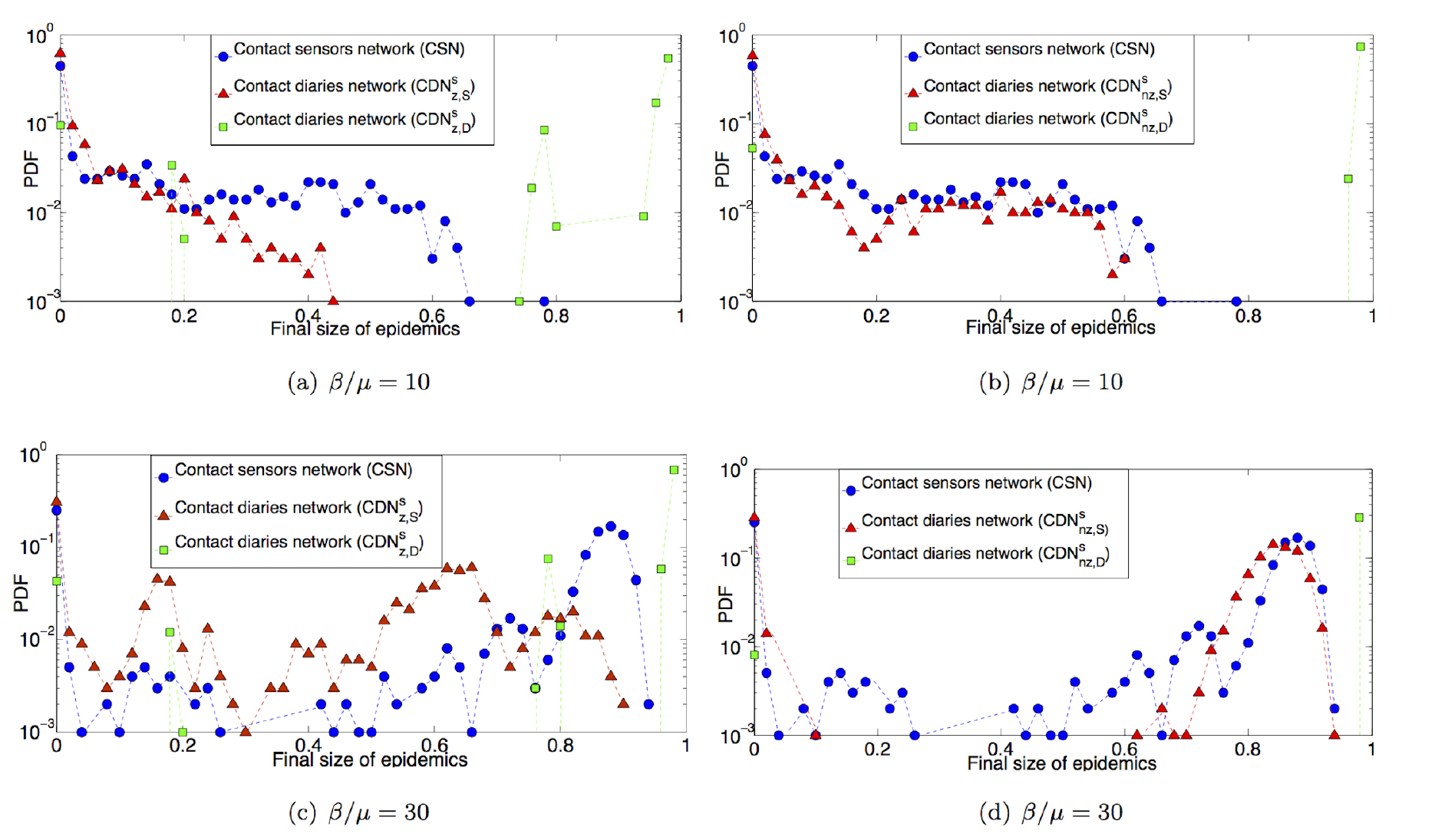}}
\caption{{{\bf\bf Distribution of final size of epidemics.}  $1000$ SIR simulations performed on the contact sensors network (CSN) and 
the four surrogate contact diaries network with and without zeros in the contact matrix of link densities,
and with durations extracted at random from the lists of values respectively reported by students and registered by sensors:  
(a), (c) $\text{CDN}^s_{\text{z,D}}$ and $\text{CDN}^s_{\text{z,S}}$; (b), (d)  $\text{CDN}^s_{\text{nz,D}}$ and $\text{CDN}^s_{\text{nz,S}}$. 
Each process starts with one random infected seed. $\beta/\mu \in \{10,30\}$.}
\label{comp1}}
\end{figure}

Fig \ref{comp1} presents the outcome of SIR simulations on these surrogate networks, compared to the distributions of epidemic sizes obtained with the
$\text{CSN}$, for two values of $\beta/\mu$. First, the overestimation of the contact durations in the diaries, combined with the replacement of zeros in the contact matrix,
leads to a very strong overestimation of the epidemic risk when $\text{CDN}^{s}_{\text{nz,D}}$ is used. The $\text{CDN}^{s}_{\text{z,D}}$ in turn yields
a peculiar shape of the distribution with intermediate peaks, due to its unrealistically strong community structure,
in a way similar to the $\text{CDN}_{\text{D}}$ case.
Distributions obtained with $\text{CDN}^{s}_{\text{z,S}}$ are also 
impacted by this structure and lead to an underestimation of the risk together with the intermediate peaks
due to the strong community structure.
Finally, simulations performed using the $\text{CDN}^{s}_{\text{nz,S}}$  give a much better 
prediction of the epidemic risk associated to the $\text{CSN}$ (Fig \ref{comp1} (b), (d)). 

Some differences between the outcomes of simulations using the contact sensor network and the $\text{CDN}^{s}_{\text{nz,S}}$ are 
nonetheless observed at large $\beta/\mu$, in particular
for intermediate epidemic sizes (epidemics involving between $20\%$ and $70\%$ of the population): a non-negligible contribution to the epidemic 
size distribution is observed for the $\text{CSN}$ 
but not for the $\text{CDN}^{s}_{\text{nz,S}}$. We show in the Supporting Information that the distribution of sizes obtained on a version
of the $\text{CSN}$ in which weights are randomly reshuffled looses this contribution of intermediate size epidemics. 
This shows that the observed discrepancies
result  from the random assignment of weights to links in the $\text{CDN}^{s}_{\text{nz,S}}$, which does not preserve 
correlations between structure and weights present in the $\text{CSN}$.

\begin{figure}[!ht]
\centering
{\includegraphics[width=1\textwidth]{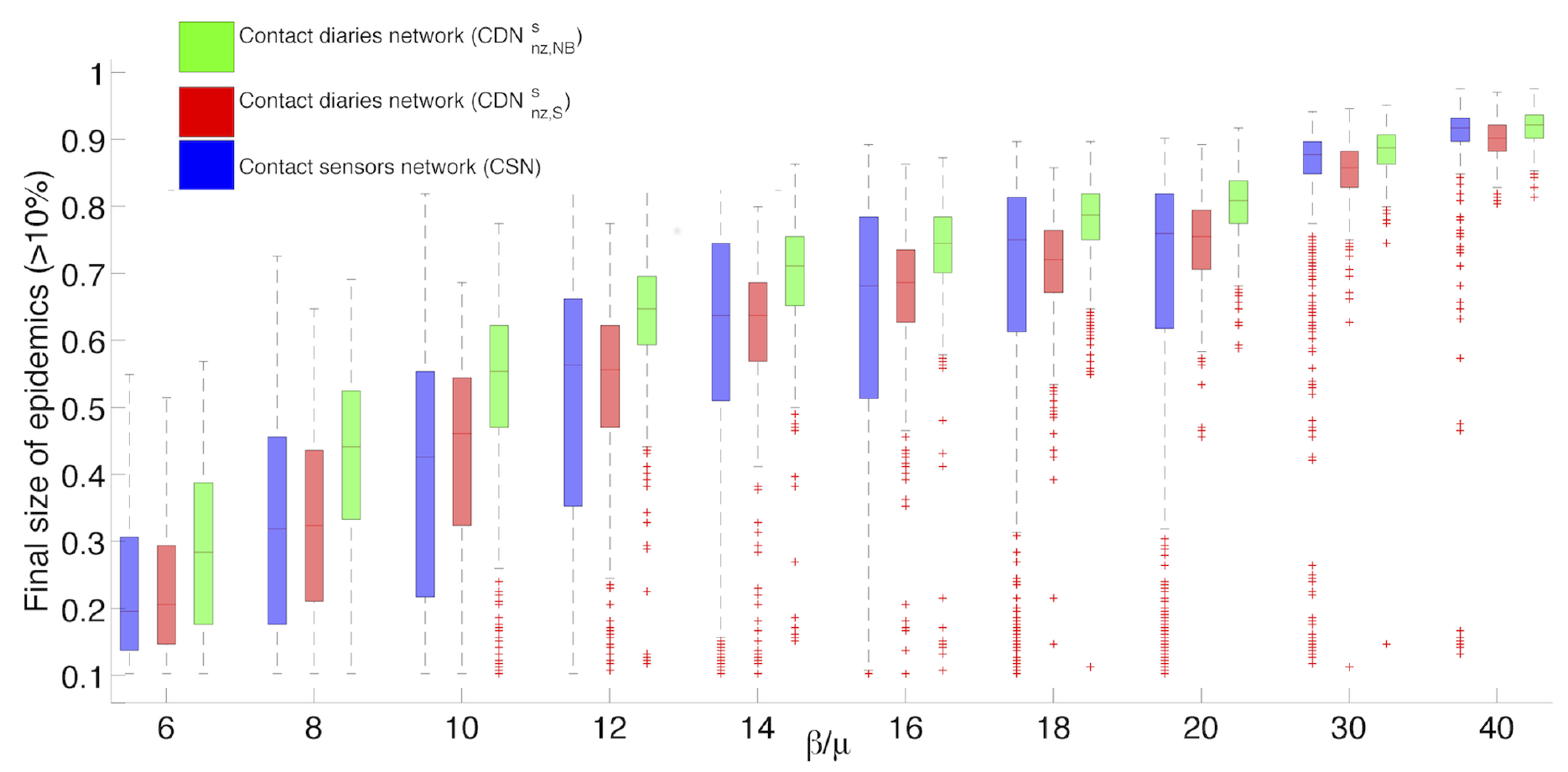}}
\caption{{\bf Box-plot of the distribution of epidemic sizes larger
    than $10\%$.}  Comparison of the distribution of epidemic sizes
  for SIR simulations performed on the contact sensors network
  ($\text{CSN}$), the surrogate contact network without
  zeros in the contact matrix of link densities and with weights randomly drawn from the distribution
  of contact durations registered by sensors
  ($\text{CDN}^s_\text{nz,S}$), and the same surrogate contact
  network but with weights randomly drawn from
  a negative binomial fit of the distribution of contact durations registered by sensors
  in several similar environments ($\text{CDN}^s_{\text{nz,NB}})$. For each
  box, the central mark stands for the median, its edges represent the
  $25^{th}$ and $75^{th}$ percentiles. The whiskers extend to the
  most extreme data points not considered outliers, while outliers are
  plotted individually. Points are drawn as outliers if they are larger than $a + h(b - a)$ or smaller than $a - h(b - 
a)$, where $a$ and $b$ are the 25th and 75th percentiles, respectively and $h$ is the maximum 
whisker length set by default to $h=1.5$.
($1000$ simulations for each value of the ratio
  $\beta/\mu$. $\beta/\mu \in \{6,8,10,12,14,16,18,20,30,40\}$).
\label{fig7}}
\end{figure}

Fig \ref{fig7} shows the robustness of our results {concerning the comparison of outcomes of simulations
on the various networks} when $\beta/\mu$ is varied, by presenting the boxplots of the distribution of epidemic 
sizes for epidemics that involve more than $10\%$ of the population. The fraction of such epidemics as a 
function of $\beta/\mu$ is shown in the Supporting Information. 
Overall, a good agreement is observed for all values of $\beta/\mu$, with however 
a systematic small underestimation of the largest epidemic sizes as well as an underestimation of intermediate sizes, especially 
at large $\beta/\mu$ (see Supporting Information),  and a narrower peak at large sizes when the $\text{CDN}^{s}_{\text{nz,S}}$ is used.

\newpage

In order to build $\text{CDN}^{s}_{\text{nz,S}}$, we have used in step (iii) the distribution of aggregate contact durations measured by the sensors. We however need to consider
the possibility that only diaries have been collected in a given setting, so that such a distribution is not available. To this aim, we take advantage
of the robustness of such distributions, as discussed for instance in \cite{BarCat:2015}. We investigate this issue in some more details here, to understand if distributions of
contact durations are similar enough in different contexts: our aim is to use publicly available data on contact duration distributions in a context-independent way in the step (iii)
of our procedure. We consider five publicly available datasets, corresponding to contacts measured by wearable sensors in: a French and an American
primary school \cite{Toth:2015,Stehle:2011}, an office building \cite{Genois2:2015} a hospital \cite{Mastrandrea2:2015} and a conference \cite{Stehle2:2011}.
All these data have been collected by the SocioPatterns collaboration \cite{SocioPatterns}, except for the case of the 
American primary school, in which a different infrastructure was used \cite{Toth:2015}. 

\begin{figure}[!ht]
\centering
{\includegraphics[width=\textwidth]{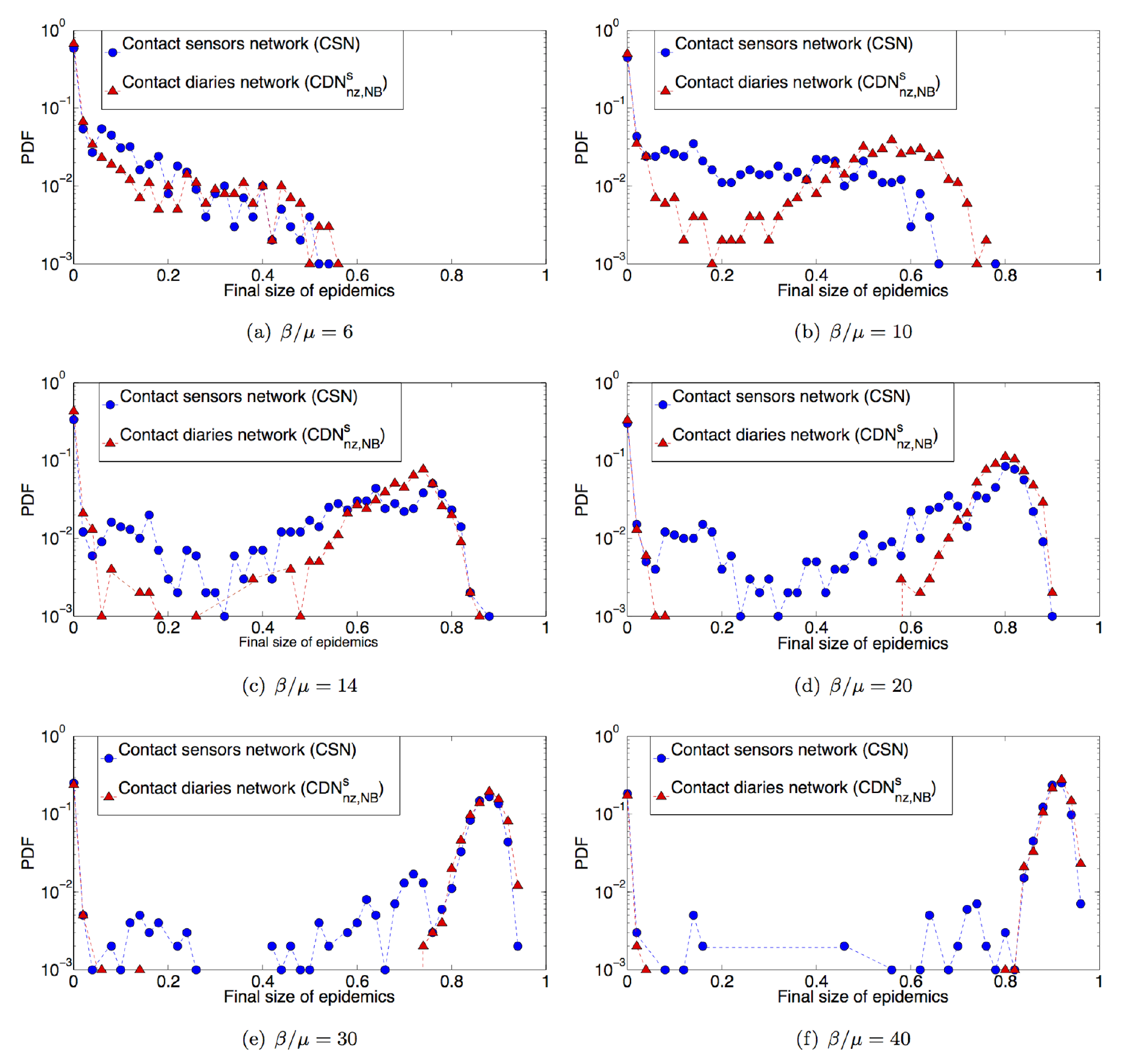}}
\caption{{\bf Outcome of the spreading processes.} 
Comparison of the distributions of epidemic sizes obtained for $1000$ SIR simulations performed on the contact sensors network and on 
the surrogate contact network with weights randomly drawn from a negative binomial fit of the 
distribution of contact durations registered by sensors in several environments ($\text{CDN}^s_\text{nz,NB}$).
Each process starts with one random infected seed. (a) $\beta/\mu = 6$,
(b) $\beta/\mu = 10$,
(c) $\beta/\mu = 14$,
(d) $\beta/\mu = 20$,
(e) $\beta/\mu = 30$,
(f)  $\beta/\mu = 40$.
\label{pool}}
\end{figure}

For all these datasets, the  distributions of cumulated contact durations are broad and, 
as also discussed in \cite{Machens:2013}, can be modeled by negative binomial functional forms. We show in the Supporting Information that similar parameters
of the negative binomial fit are obtained for each dataset and for the combined one.
Therefore, to further generalize the procedure and avoid relying on a single dataset, 
we consider in the following the fit of the five combined datasets. We then assign
to the links of the $\text{CDN}^{s}_{\text{nz}}$ weights drawn at random from this fitted distribution, obtaining $\text{CDN}^{s}_{\text{nz,NB}}$.
Fig \ref{pool} and Fig \ref{fig7} compare the distributions of epidemic sizes obtained when the 
SIR model is simulated on the resulting surrogate network
{and on the $\text{CSN}$  (see also the Supporting Information,
in which we moreover show the outcome of simulations for different initial seeds, as well as the temporal
evolution of the density of infectious individuals in the population when using the  $\text{CDN}^{s}_{\text{nz,NB}}$.)}

Despite using less information on the specific context than $\text{CDN}^{s}_{\text{nz,S}}$, since we do not rely on the specific distribution of weights measured there, 
the surrogate contact network $\text{CDN}^{s}_{\text{nz,NB}}$ leads to a good prediction of the epidemic risk. 
{In particular, the emergence, location and amplitude of the peak of the distribution at large epidemic values is correctly predicted.}
{However}, the distribution of final epidemic sizes is systematically shifted towards higher shares of population 
with respect to the $\text{CDN}^s_\text{nz,S}$ case. 
Thus, for small values of $\beta/\mu$ the outcome of simulations using $\text{CDN}^s_\text{nz,S}$ are in slightly better 
agreement with the $\text{CSN}$ case. However,
for higher $\beta/\mu$ and for the largest epidemic size reached, the $\text{CDN}^{s}_{\text{nz,NB}}$ performs better.
Overall,  both surrogate networks yield satisfactory predictions of the epidemic risk associated to a propagation on the 
$\text{CSN}$.

\section*{Discussion}

Data on the contact patterns of individuals collected by different methods lead to different contact network structures, and some studies have started to investigate
this issue through detailed quantitative  comparisons \cite{Mastrandrea:2015,Smieszek:2014}. In the present paper, we have gone 
further by comparing the outcome of simulations of spreading processes performed on contact networks gathered either through 
wearable sensors or through contact diaries. 
Not surprisingly, we have shown that the results differ strongly, due to the low participation rate to the diaries,
the under-reporting of contacts and the overestimation of contact durations in diaries. In particular, the direct use of the links and durations reported
in the diaries yields a peculiar distribution of epidemic sizes suggesting a very strong community structure that might
lead to the design of inadequate containment strategies. On the other hand, using the links reported in the diaries but more realistic weights
yields a strong underestimation of the epidemic risk.

In a second and more important step, we have asked if, despite this low participation rate and
these biases, the information contained in the contact network built from the contact diaries can be used to build a surrogate contact network
whose properties are close enough to the real contact network (considered here to be the one obtained from the wearable sensors) to yield 
a correct estimation of the epidemic risk when used in simulations of spreading processes. The rationale comes from the structural similarities
found in the contact matrices giving the densities of links between individuals of different classes obtained using both sensors and
diaries \cite{Mastrandrea:2015}. These similarities suggest to build a surrogate contact network starting from the contact diary network, adding nodes
and links in order to maintain this matrix fixed, and assigning weights to the links. We note that two recent works
\cite{Potter:2015,Genois:2015} 
have considered related but different issues. In \cite{Potter:2015}, only diary data is available, and the authors
present a synthetic network model based on data and adjusting for under-reporting. This adjustment for reporting errors leads in this case only
to a small difference in epidemic predictions.
In \cite{Genois:2015} {on the other hand, only sensor data is considered}, and the authors assume to have an incomplete information 
on the contact network registered by sensors due to an uniform population sampling (hence, all contacts between participating individuals
are assumed to be known). Here on the other hand, the available dataset is  
given by diaries, in which population sampling is not uniform (actually, the students who filled in diaries tend to have more contacts than the others) 
and in which under-reporting implies that many links between participating
individuals are also missing. Moreover, we face two additional issues (i) the low sampling rate yields a contact matrix of link densities which contains
zeros, in an unrealistic way, and
(ii) various possibilities can be considered when assigning weights to the links of the surrogate networks as weights reported in the diaries
are strongly overestimated. Despite these issues, the surrogate network we build yields, when used in simulations, a good agreement
with simulations performed on the whole contact sensor network in terms of epidemic risk prediction, under the condition of using the list of weights
(cumulative contact durations) gathered by the wearable sensors. In order to get rid of this condition, we argue that 
this list comes from a distribution that has been shown in previous works to be very robust across contexts \cite{BarCat:2015}. We therefore consider
weights taken at random from a pool of publicly available datasets, and show that using these weights gives also satisfactory results. Overall,
we thus have presented a procedure that uses only the information contained in the contact diaries and in public data, which allows to obtain
a good prediction of the epidemic risk, as measured by the distribution of epidemic sizes,  when used in simulations of a spreading process.

In the Supporting Information, we moreover consider the issue of using, instead of contact diaries, data coming from friendship surveys, in order
to build the surrogate contact network used in the simulations. We show that the epidemic risk prediction obtained through this procedure is not
accurate. This could be expected as daily encounters in the school are not necessarily related to the existence of a relationship between students:
contacts occur between non-friends due to daily activities, while friends do not meet necessarily every day. This outcome highlights the importance
of taking into account the different nature of social ties \cite{Mastrandrea:2015, Coviello:2015}, which can each be relevant for specific processes.

Some limitations of our work are noteworthy. First, our results rely on an assumption made in replacing the zeros observed in the contact matrix
of link densities by random values. In the context under scrutiny, zero values can indeed easily be considered as unrealistic. In other contexts, it might
however happen that different groups in the population really do not mix. In such a case, one might expect that this kind of information could
be gathered from other sources (schedules, location of classrooms or offices, etc) \cite{Smieszek:2013} and thus integrated into the procedure. 
{Second, we have considered as ground truth the contact sensor network. On the one hand, this network in fact suffers from an incomplete participation,
so that the outcome of spreading processes is underestimated with respect to hypothetical data containing information on the whole population. However,
such underestimation can be compensated through the procedure presented in \cite{Genois:2015}. On the other hand, it is important to note that it
is not yet completely clear whether the contacts measured by wearable sensors are the best proxy
for potentially infectious contacts. We work therefore under this hypothesis, which is indeed quite widely used but should be kept in mind.}
{Third, we have considered here static networks. As discussed in \cite{Stehle2:2011}, the outcome of simulations is then close enough
to the one of simulations taking into account the full contact dynamics if we consider slow enough processes. For fast processes, the
burstiness of contacts becomes very relevant; in this case, it would be crucial to supplement our procedure by the construction of surrogate
timelines of contacts and intervals between contacts at high temporal resolution, as done in \cite{Genois:2015}.
}
{Finally}, we cannot at this point investigate the efficiency of our procedure in other contexts, for lack of datasets reporting contacts
measured by both wearable sensors and contact diaries in the same context and at the same date. Hopefully such datasets will become more
available in the future, yielding new testing grounds for our method. {Many populations of interest can indeed be divided into groups or
categories that do not mix homogeneously, often with more contacts within groups than between groups, 
and for which the contact matrix formalism and the procedures we present to construct surrogate
networks are therefore relevant \cite{Genois:2015}.}
We conclude by mentioning that future work could also investigate other dynamical processes 
on networks, such as information spreading or opinion formation processes.

\section*{Methods}

\subsection*{Data description}

The datasets we use have been presented and made publicly available in \cite{Mastrandrea:2015}. They correspond to 
contacts between students of $9$ classes in a high school in France, collected through wearable sensors on the one hand and contact diaries on the other
hand. The sensors registered contacts with a temporal resolution of $20s$ for $327$ participating students (out of $379$ in the $9$ classes, i.e.,
a $86.3\%$ participation rate) during the week of Dec. 2-6, 2013. Contact diaries contain data reported by students about 
encounters and their cumulative durations for Dec. 5, 2013. In these diaries, 
students were asked to report the cumulative durations of their contacts choosing among four intervals: at most 5 minutes, 
between 5 and 15 minutes, between 15 minutes and 1 hour, more than one hour. 
The students belong to $9$ classes with different specializations: ``MP'' classes focus more on mathematics and physics, ``PC'' classes on physics and
chemistry, ``PSI'' classes on engineering studies and ``BIO'' classes on biology.  We collected data among students of nine classes
corresponding to the second year of such studies: 3 classes of type ``MP'' (MP, MP*1, MP*2),
two of type ``PC'' (PC and PC*), one of type ``PSI'' (PSI*) and 3 of type ``BIO'' (2BIO1, 2BIO2, 2BIO3).

Using these datasets, we build two networks of contacts among students for the same day (Dec 5, 2013): 
the Contact Sensors Network (CSN) and the Contact Diaries Network (CDN). 
In each network, nodes represent students, and a links is drawn between two students if:
\begin{enumerate}
\item sensors register at least one contact during the relevant day (CSN);
\item at least one of the two students reported an encounter (CDN).
\end{enumerate}
The resulting networks have $295$ nodes for the CSN (other students were absent or did not wear the sensors on that day) and
$120$ nodes for the CDN. In particular, no student from classes PC* and PSI* filled a diary, and only one from MP*1. We thus discarded these
classes in most of the analysis and in particular in the contact matrices, remaining with 6 classes. 

Each link carries a weight. In the CSN it represents the cumulative duration of contacts registered by sensors during the day.
For the CDN we consider several possibilities. In the $\text{CDN}_{\text{D}}$ we use weights reported in the diaries:
we associate to each time-interval its maximum possible value (5, 15, 60 minutes respectively for the first three intervals  
and 4 hours for the last one. This choice takes into account data reported and registered and the school schedule) and, 
if two students reported different durations for their
encounter, we use the average of the reported values. In the $\text{CDN}_{\text{S}}$ on the other hand, we consider weights
randomly drawn from the distribution of contact durations registered by sensors. Results are averaged over $1000$ such 
weight assignments. {For the $\text{CDN}_{\text{S'}}$ finally, we
start from $\text{CDN}$ and rank the $E$ links in decreasing order of their reported weights 
(assigned as in $\text{CDN}_{\text{D}}$, and with random order for equal weights). We extract $E$ 
weights from the distribution of contact durations registered by sensors, rank them as well in decreasing order, and
assign the weights to the links of $\text{CDN}$ in such a way to match the two orderings (i.e., assigning the largest weights
to the links with largest reported weights).}

\subsection*{Matching networks} 

The CSN, the $\text{CDN}_{\text{D}}$ and the $\text{CDN}_{\text{S}}$ are matched to retain only nodes appearing in both CSN and CDN (see table \ref{table:matched} for details about classes size before and after matching).
We refer to them as the \textit{matched} networks: $\text{CSN}^m$, $\text{CDN}^m_{\text{D}}$ and $\text{CDN}^m_{\text{S}}$.

\begin{savenotes}
\begin{table}[!ht]
\centering
\medskip
\begin{tabular}{c|cc|c|cc}
\hline
\toprule
    &{\bf CSN} & {\bf CDN} & & {\bf $\text{CSN}^m$} & {\bf $\text{CDN}^m$}\\
\hline
\midrule
 {\bf 2BIO1} &      $35$  &  $22$   & & $20$ & $20$ \\
{\bf 2BIO2} &      $30$  &  $13$   & & $11$ & $11$ \\
{\bf 2BIO3} &      $33$  &  $15$   & & $13$ & $13$ \\
{\bf MP} &      $32$  &  $23$   & & $23$ & $23$ \\
{\bf MP*1} &      $28$  &  $1$   & & $0$ & $0$ \\
{\bf MP*2} &      $34$  &  $19$   & & $18$ & $18$ \\
{\bf PC} &      $40$  &  $27$   & & $23$ & $23$ \\
{\bf PC*} &      $35$  &  $0$   & & $0$ & $0$ \\
{\bf PSI*} &      $28$  &  $0$   & & $0$ & $0$ \\
\hline
\hline
{\bf Tot} & $295$ & $120$ && $108$ & $108$ \\
\bottomrule
\end{tabular}
\caption{{\bf Comparison of network properties.} Number of nodes in each class in the 
contact sensors network and in the
contact diaries network in the original (respectively CSN, CDN) and the matched forms
(respectively $\text{CSN}^m$, $\text{CDN}^m$).}
\label{table:matched}
\end{table}
\end{savenotes}
The CMDN is built by using a Contact Matrix Distribution (CMD). Following \cite{Machens:2013}, we consider  a CMD where each entry, $(X,Y)$, is the empirical distribution of durations reported by diaries for contacts between all students in class $X$ and class $Y$, including zero durations 
(corresponding to an absence of link between two students). We fit each such distribution by a negative binomial functional form.
Then, for each pair of nodes, we draw at random a weight using the corresponding negative binomial fit. 
Note that in this way we do not maintain fixed the link structure of the CDN. We however keep on average the same density of links
between different classes. 

\subsection*{Construction of surrogate networks}

The basic steps for building a binary surrogate contact network $\text{CDN}^s$ for the six considered classes, 
starting from the matched contact diaries network, are:
\begin{enumerate}

\item we add all missing nodes in each class (we know the number of students in each class in the CSN): the number of nodes grows from $108$
in the $\text{CDN}^m$ to $204$ in the $\text{CDN}^s$;

\item we add new links within and between classes in order to keep fixed the observed contact matrix of edge densities 
for the contact diaries network (given in Fig \ref{cont} (b)).  To this aim, 
we randomly pick up pairs of nodes, $i$ belonging to class $X$ and $j$ to $Y$. If $i$ and $j$ are not yet linked and if the current
density of links between classes $X$ and $Y$ is smaller than the corresponding entry of the empirical matrix (Fig \ref{cont} (b)), we add a link
between $i$ and $j$.

\item the previous step is repeated until we obtain link densities within and between classes equal to the ones of the $\text{CDN}^m$ (Fig \ref{cont} (b)).

\end{enumerate}
Results are averaged over 500 realisations of this procedure.

As explained in the main text, we moreover deal in two different ways with the zero values of the link densities between several class-pairs
in the CDN. We either keep these densities or replace them with values drawn at random 
from a uniform distribution of values between the minimum and maximum values (diagonal excluded) of the contact matrix of Fig \ref{cont}(b). 
In this way the contact matrix structure is preserved, with more interactions within than between classes.

To assign weights to the links of the surrogate networks, we consider several possibilities.
We first assume homogeneous contact durations and  assign to each link a weight equal to the average of cumulative 
durations registered by sensors. This yields two versions of the surrogate contact networks:
\begin{itemize}
\item $\text{CDN}^{s}_{\text{z,H}}$: with homogeneous contact durations and keeping zero densities in the contact matrix of edge densities;

\item $\text{CDN}^{s}_{\text{nz,H}}$: with homogeneous contact durations and zero densities replaced in the contact matrix of edge densities.

\end{itemize}
We refer to the contact sensors network under the homogeneous duration hypothesis by $\text{CSN}_\text{H}$.

If instead we assume heterogeneous contact durations, we obtain two possible surrogate contact diaries networks: 
we assign weights at random to the links of  $\text{CDN}^{s}$, drawn at random with replacement from 
the list of durations either reported by students or registered by sensors. We thus obtain four versions of the surrogate contact networks: 
\begin{itemize}

\item $\text{CDN}^{s}_{\text{z,D}}$, with durations drawn from the ones reported by students and keeping zero densities in the contact matrix of edge densities;

\item $\text{CDN}^{s}_{\text{nz,D}}$, with durations drawn from the ones reported by students and zero densities replaced in the contact matrix of edge densities;

\item $\text{CDN}^{s}_{\text{z,S}}$, with durations drawn from the ones registered by sensors and keeping zero densities in the contact matrix of edge densities;

\item $\text{CDN}^{s}_{\text{nz,S}}$, with durations drawn from the ones registered by sensors and zero densities replaced in the contact matrix of edge densities.

\end{itemize}

Finally, the surrogate contact network obtained by assigning weights randomly drawn from the negative binomial fit of the 
distribution of publicly available contact durations registered by sensors is indicated by the acronym $\text{CDN}^s_\text{nz,NB}$.

\section*{Supplementary Information}
Supplementary Information Legend: Supplementary pdf file containing all supplementary figures and
tables.

\section*{Acknowledgments}

We are grateful to the SocioPatterns collaboration \cite{SocioPatterns} for privileged access to 
the datasets on face-to-face interactions collected during a two-days conference and in a French hospital.
This work was supported by the A*MIDEX project (ANR-11-IDEX-0001-02) funded by the ''Investissements d'Avenir'' 
French Government program, managed by the French National Research Agency (ANR), to A.B. and R.M. 
A.B. is also partially supported by the French ANR project HarMS-flu (ANR-12-MONU-0018) and by the EU FET project Multiplex 317532.


\newpage

\setcounter{figure}{0}
\renewcommand{\thefigure}{S\arabic{figure}}

\setcounter{table}{0}
\renewcommand{\thetable}{S\arabic{table}}

\renewcommand{\thesection}{S\arabic{section}}

\begin{center}
{\Large
\textbf{How to estimate epidemic risk from incomplete contact
diaries data? Supplementary Information}
}\\
\end{center}

\section{Structure of the networks and of the contact matrices}

\subsection{Networks' main characteristics}

\begin{figure}[!ht]
\centering
\subfigure[]
{\includegraphics[width=0.458\textwidth]{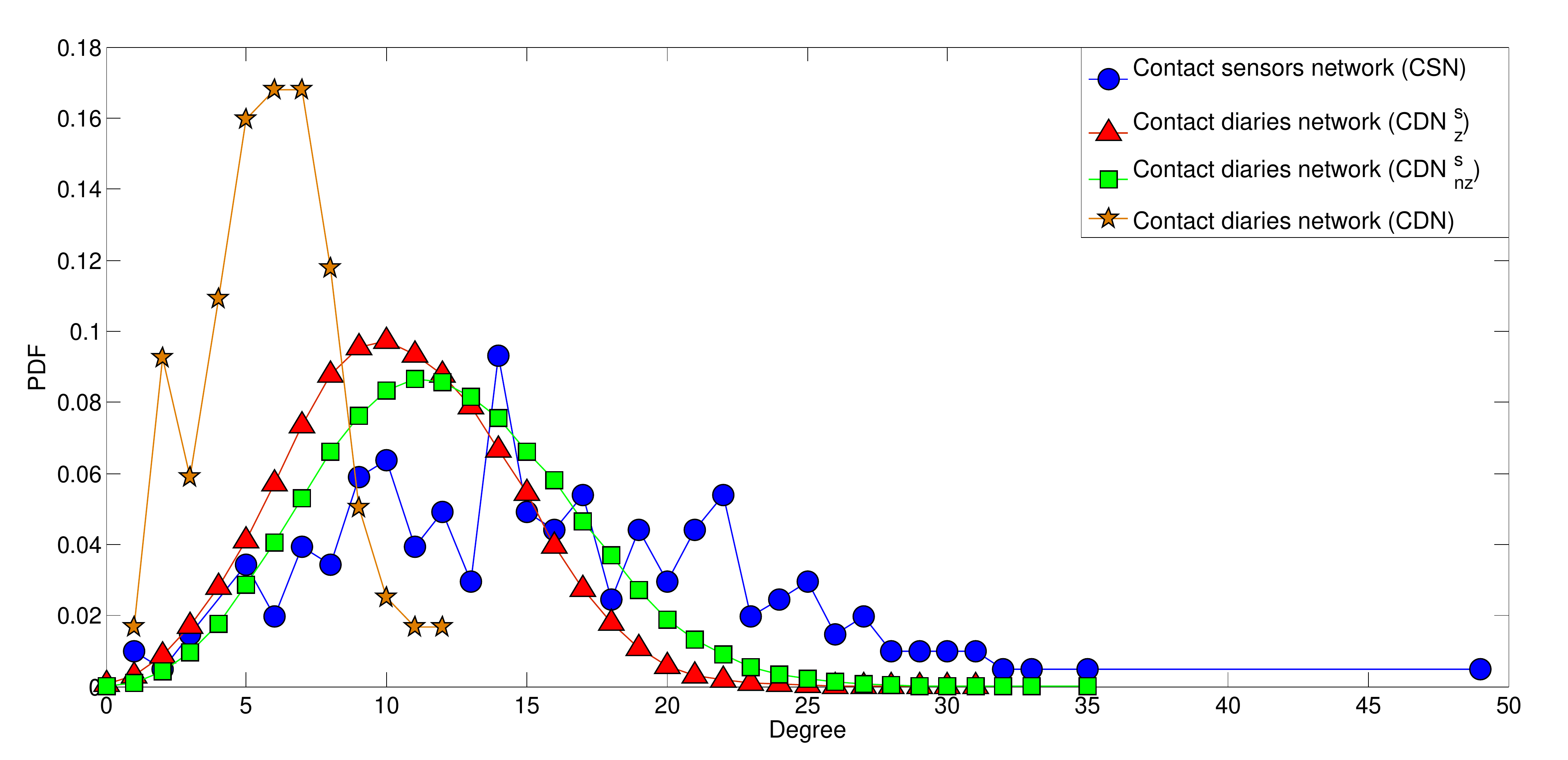}}
\hspace{-3mm}
\subfigure[]
{\includegraphics[width=0.5\textwidth]{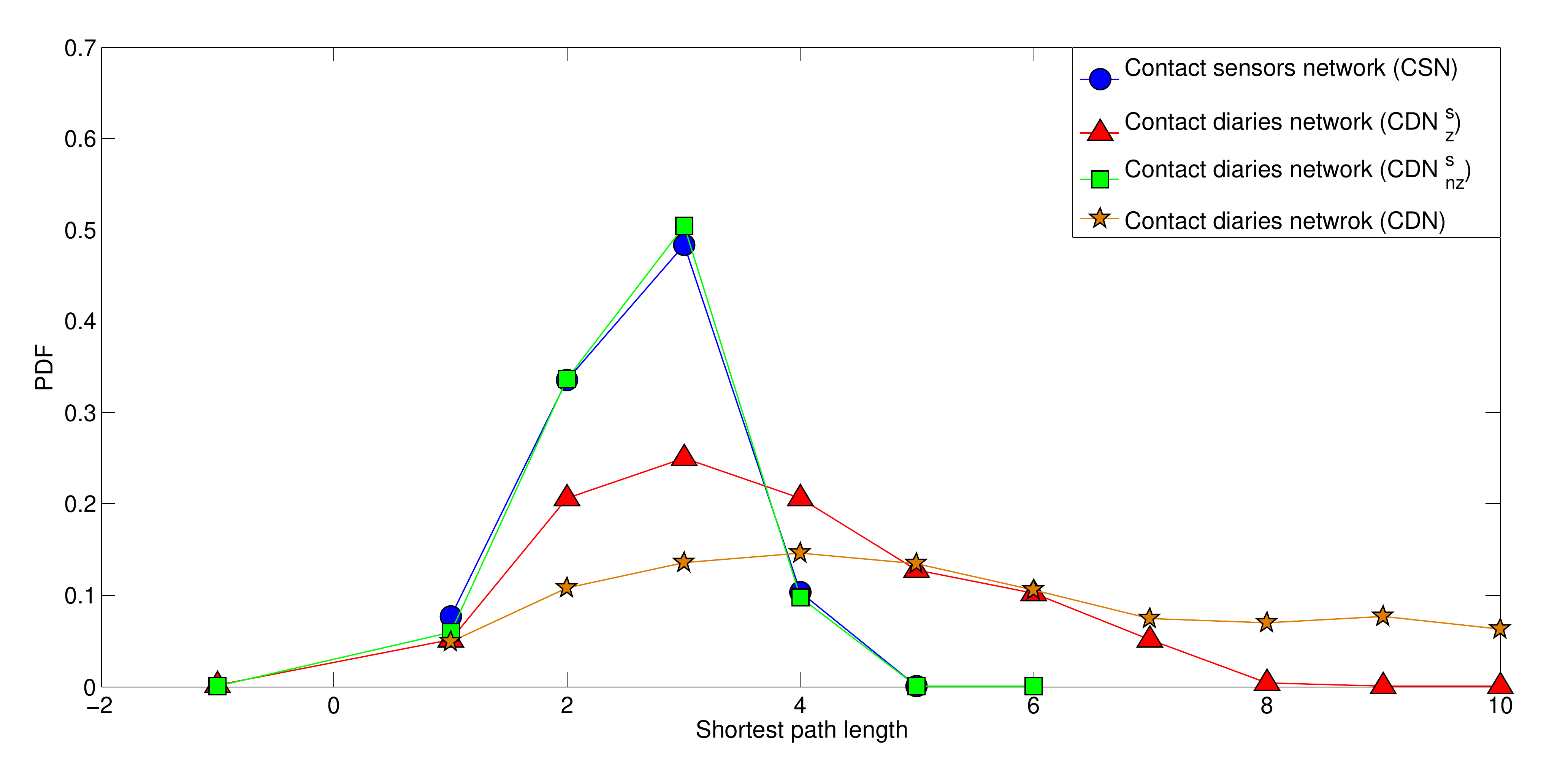}}
\subfigure[]
{\includegraphics[width=0.5\textwidth]{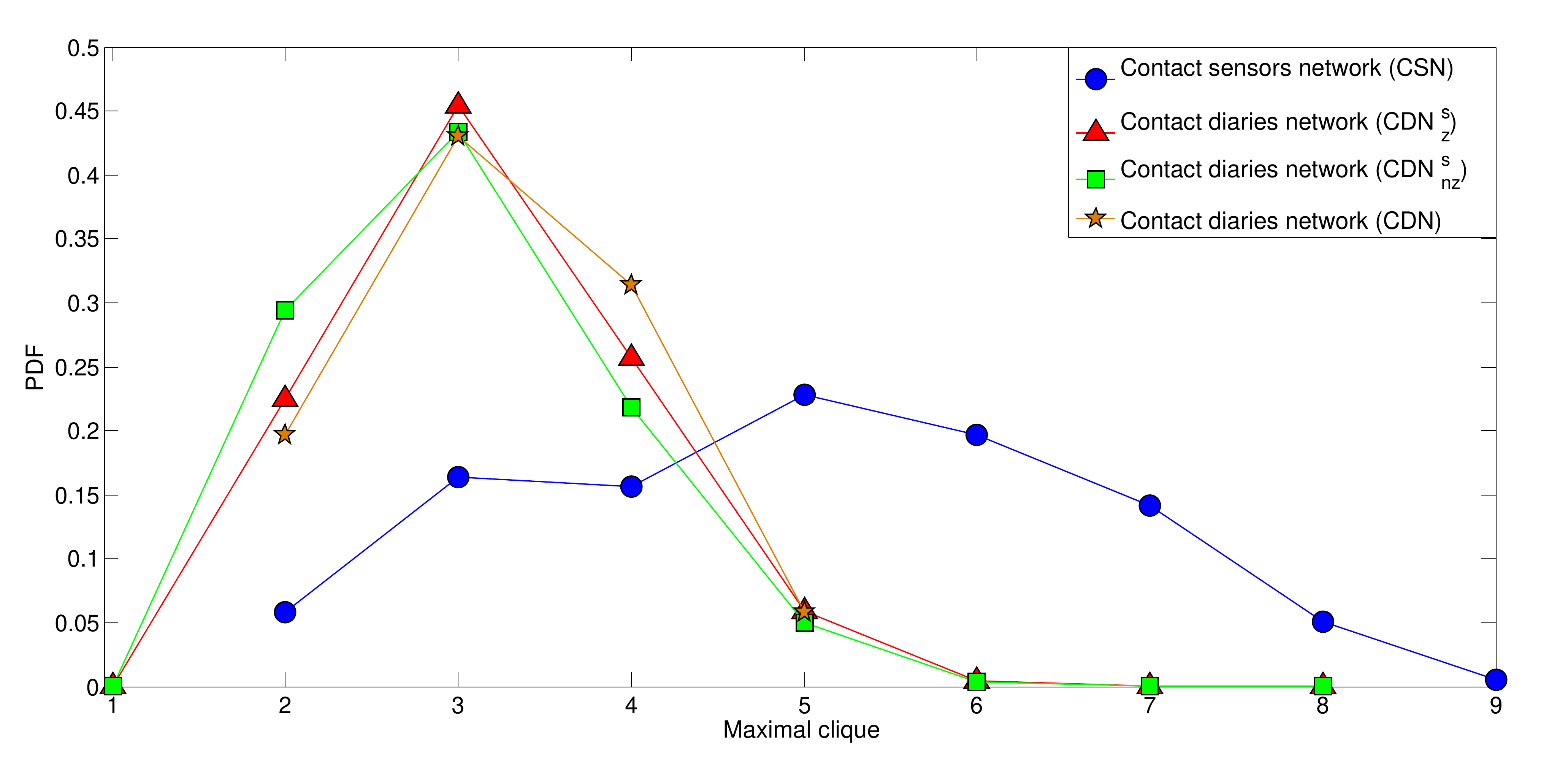}}
\hspace{-3mm}
\subfigure[]
{\includegraphics[width=0.5\textwidth]{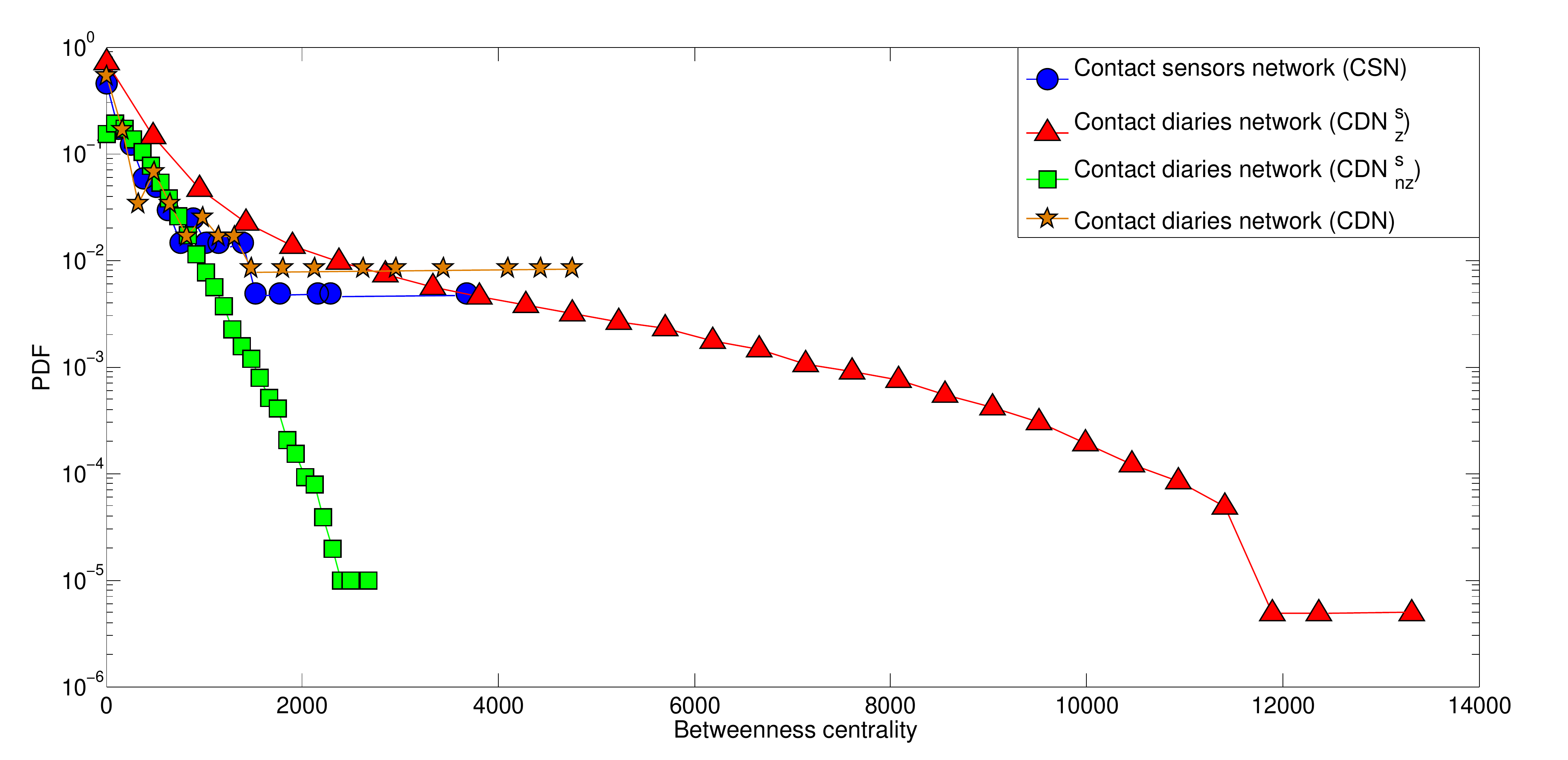}}
\caption{{\bf Networks statistical properties.} Distributions of degrees, shortest path lengths, 
betweenness centralities and maximal clique sizes for the contact sensor network and the contact diaries
network reduced to $6$ classes, as well as for the surrogate networks.
\label{fig:networks}}
\end{figure}

Tables \ref{table:networks}-\ref{table:contact_vs_diaries} 
and Fig. \ref{fig:networks} report some standard statistical characteristics
of the empirical and surrogate networks considered in the main text. Moreover, in
Tab. \ref{table:contact_vs_diaries} 
and Fig. \ref{fig:networks}, we consider the versions
of CSN and CDN restricted to $6$ classes, and we use unweighted measures, therefore we consider
the surrogate networks $\text{CDN}^s_\text{z}$ and $\text{CDN}^s_\text{nz}$ without referring to a weight
assignment.

The contact diaries network has less nodes and links than the contact sensor network. Moreover, the average
degree is much smaller, with the whole degree distribution shifted to smaller values. The maximal cliques
are also smaller, while the shortest paths are longer in the CDN. All this is 
consistent with its more dilute structure with in particular less connections
between classes. This also leads to more extreme values of the betweenness centrality (corresponding to the
bridges between classes). 

The $\text{CDN}^s_\text{z}$ surrogate network becomes closer to the contact sensor network for the degree
distribution and the shortest path lengths. As the zero values in the contact matrix are not replaced in this case,
it however keeps a very strong modular structure, with large values of the betweenness centrality and 
larger shortest path lengths than the CSN. The $\text{CDN}^s_\text{nz}$ on the other hand yields
 shortest path lengths in perfect agreement with the CSN structure and closer distributions of degrees and
betweenness centralities (properties of crucial
importance in terms of propagation processes. As the construction of the  $\text{CDN}^s_\text{nz}$ does not take into account
clustering or cliques, these properties remain however smaller than for the original CSN.

\begin{savenotes}
\begin{table}[!ht]
\medskip
\begin{tabular}{c|cc|}
\hline
\toprule
    &{\bf CSN} & {\bf $\text{CDN}$} \\
\hline
\midrule
 {\bf Nodes} &      $295$  & $120$  \\
{\bf Links} &       $2162$ & $348$ \\
{\bf Density} &    $0.05 $ &  $0.05$ \\
{\bf Avg. Degree} &$15 (8)$ & $6 (2)$ \\
{\bf Avg. Clustering} & $0.38$ & $0.45$   \\
{\bf Avg. Betweenness} & $532 (797)$ & $519 (983)$  \\
{\bf Avg. SPL} &  $2.81 (0.8)$ & $5.36 (2.73)$ \\
{\bf Maximal Clique Size} & $9$ &  $5$  \\
\bottomrule
\end{tabular}
\caption{{\bf Comparison of some properties of the 
initial contact sensor network and surrogate contact diaries network.} 
SPL: shortest path length. In parenthesis, standard deviations.}
\label{table:networks}
\end{table}
\end{savenotes}

\begin{savenotes}
\begin{table}[!ht]
\medskip
\begin{tabular}{c|cccc|}
\hline
\toprule
    &{\bf CSN} (6 classes) & {\bf CDN} (6 classes)& {\bf $\text{CDN}^s_\text{z}$} & {\bf $\text{CDN}^s_\text{nz}$} \\
\hline
\midrule
 {\bf Nodes} &      $204$  & $119$ & $204$ &  $204$   \\
{\bf Links} &       $1600$ & $347$ & $1076$ & $1324$   \\
{\bf Density} &    $0.08 $ & $0.05$ &  $0.05$ & $0.06$  \\
{\bf Avg. Degree} &$16 (7)$ & $6 (2)$ & $10 (4)$ & $12 (4)$   \\
{\bf Avg. Clustering} & $0.44$ & $0.45$ & $0.25$  & $0.2$  \\
{\bf Avg. Betweenness} & $328 (523)$ & $515 (974)$ & $544 (1024)$ & $333 (264)$   \\
{\bf Avg. SPL} &  $2.61 (0.8)$ & $5.37 (2.74)$ & $3.7 (1.6)$ & $2.64 (0.74)$  \\
{\bf Maximal Clique Size} & $9$ & $5$ &  $6$ & $6$  \\
\bottomrule
\end{tabular}
\caption{{\bf Comparison of some properties of the contact sensor network 
and contact diaries network reduced to $6$ classes, and for the surrogate contact diaries networks
obtained either keeping the zero elements in the contact matrix, or replacing them.} 
SPL: shortest path length. In parenthesis, standard deviations.}
\label{table:contact_vs_diaries}
\end{table}
\end{savenotes}

\clearpage
\newpage

\subsection{Original contact sensors and diaries networks}

\begin{figure}[!ht]
\centering
\subfigure[]
{\includegraphics[width=0.458\textwidth]{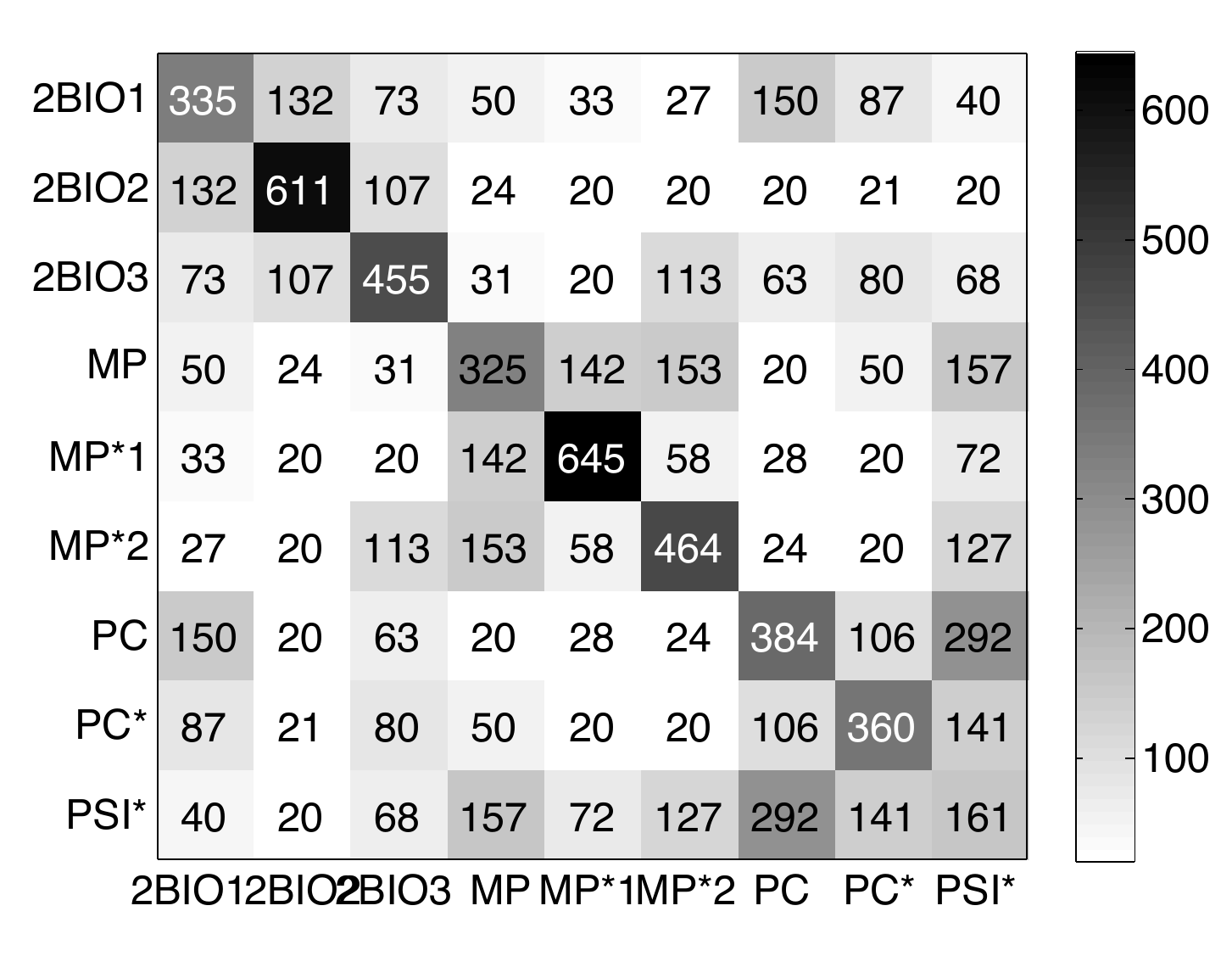}}
\hspace{-3mm}
\subfigure[]
{\includegraphics[width=0.5\textwidth]{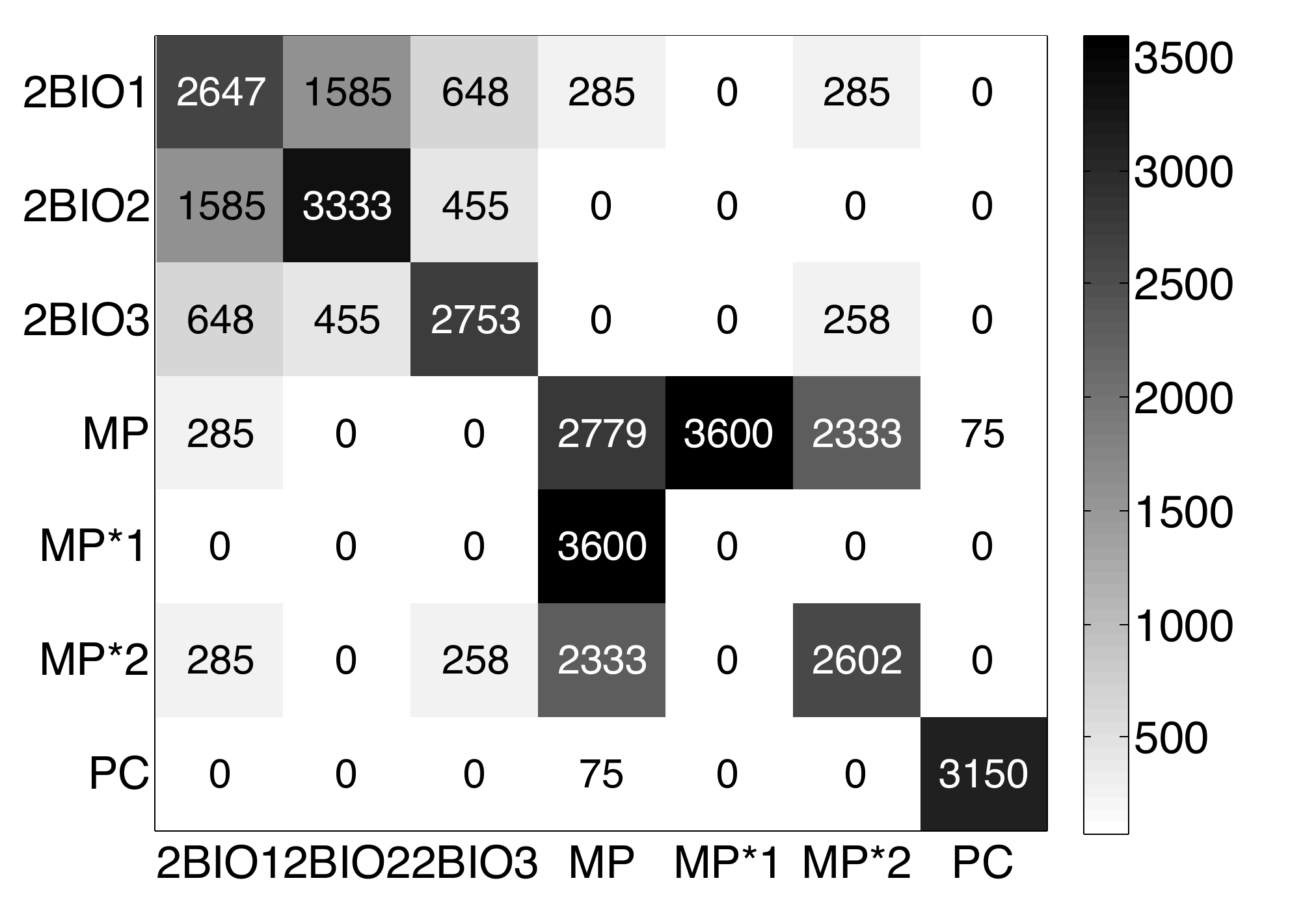}}
\subfigure[]
{\includegraphics[width=0.3\textwidth]{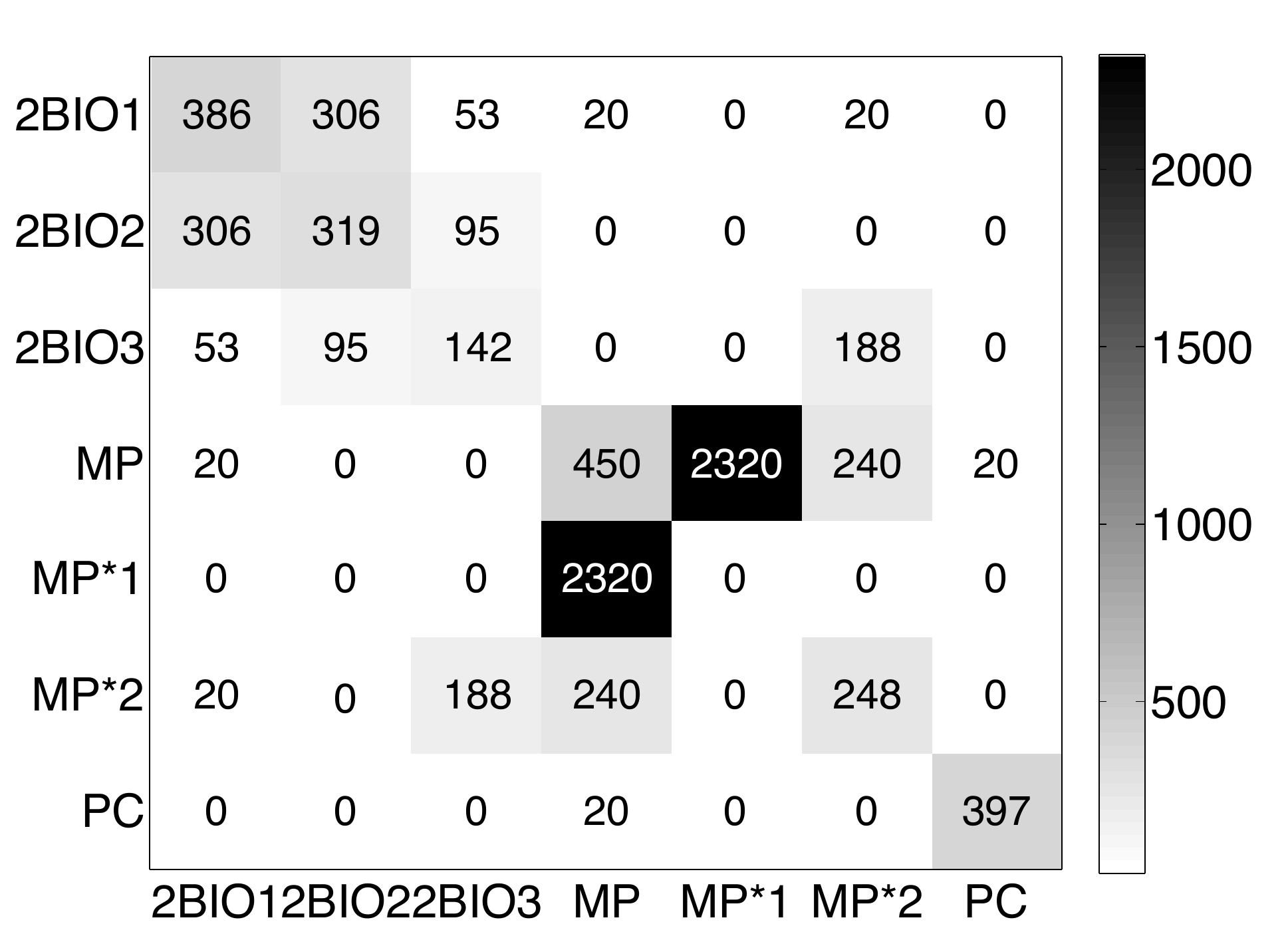}}
\subfigure[]
{\includegraphics[width=0.295\textwidth]{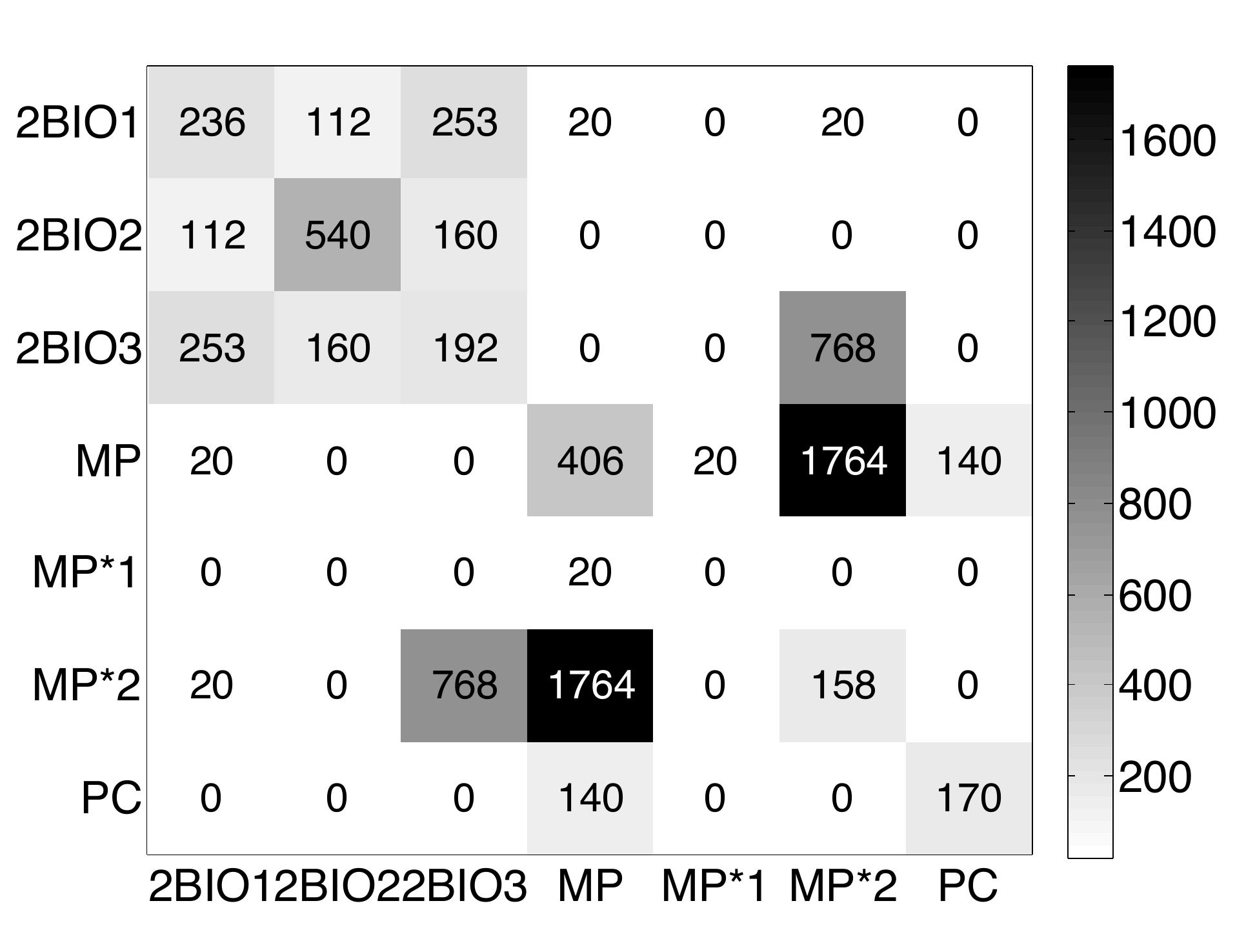}}
\subfigure[]
{\includegraphics[width=0.297\textwidth]{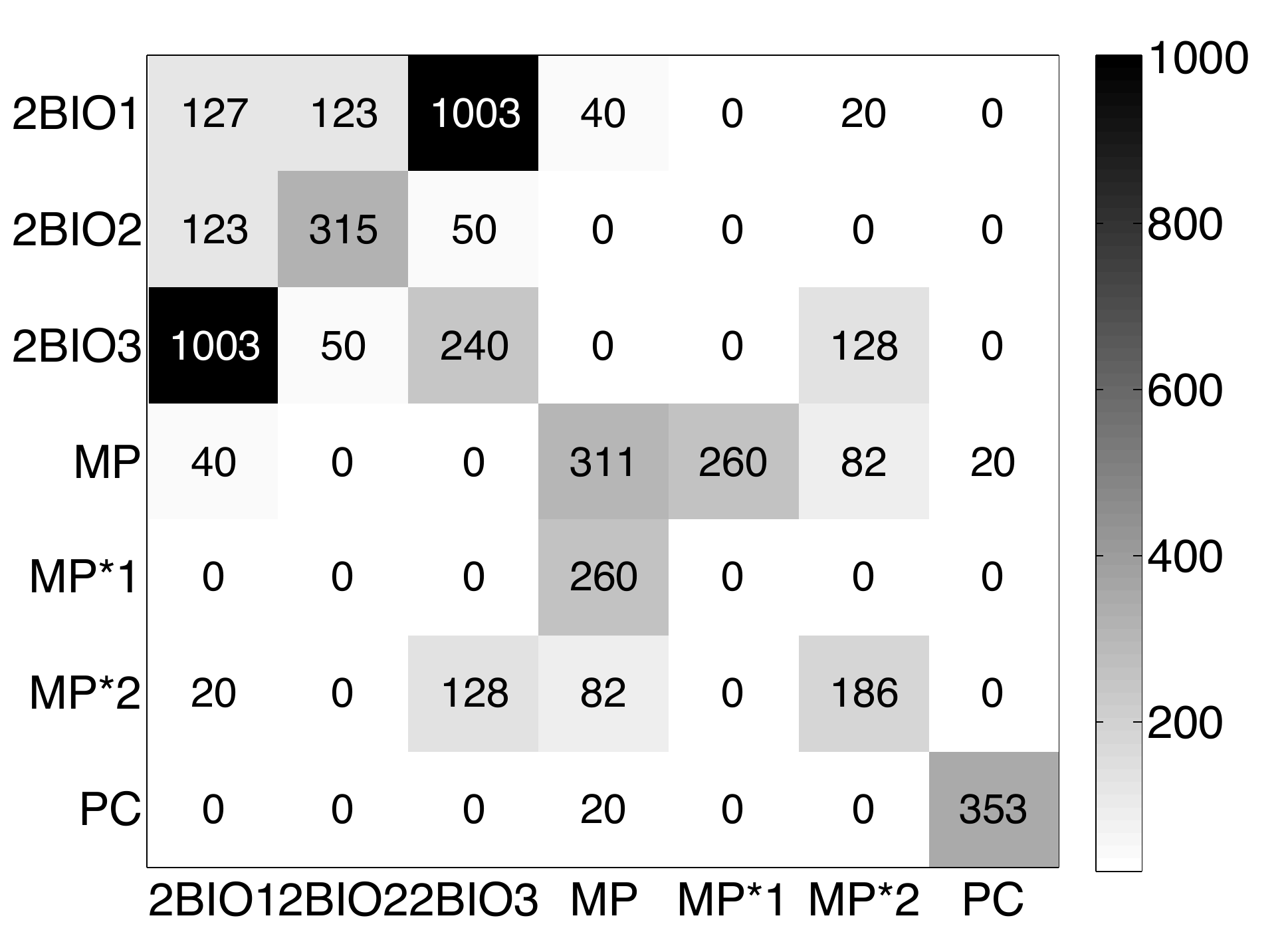}}
\caption{{\bf Contact matrices of average durations.} (a) Contact sensors network (CSN); (b) Contact diaries network with durations reported by students ($\text{CDN}_\text{D}$); (c), (d), and (e) three examples of contact diaries network with weights randomly drawn from the distribution of contact 
durations registered by sensors ($\text{CDN}_\text{S})$. 
\label{AvgDur}}
\end{figure}

In Fig. \ref{AvgDur} we report the contact matrices of average durations of contacts between students of different classes. 
Each entry of the matrix, $C_{XY}$, is given by the average of the total contact durations between students in class $X$ and 
students in class $Y$. 

The contact sensors network and the contact diaries network with weights reported by students have a  diagonal 
structure: people in the same class are in general more likely to stay in contact longer than people belonging to different classes (except for MP*1
for which only one student filled in the diaries). 
In the $\text{CDN}_{\text{S}}$ case, this
structure is not respected due to the random assignment of weights drawn from the distribution of durations registered by sensors (Fig. \ref{AvgDur} (c), (d), (e)). 

\clearpage
\newpage

\subsection{Matched contact sensors and diaries networks}

\begin{figure}[!ht]
\centering
\subfigure[]
{\includegraphics[width=0.5\textwidth]{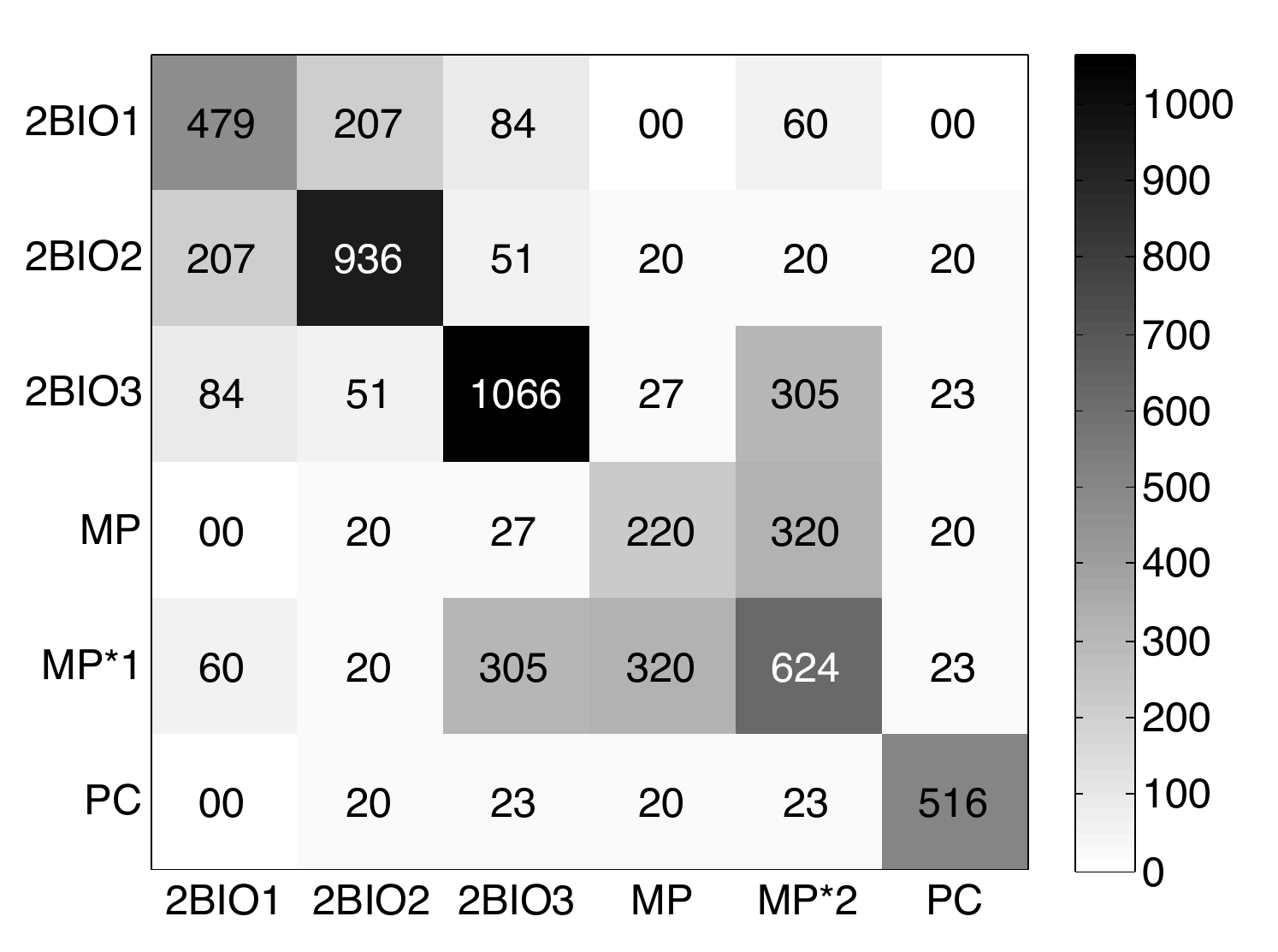}}
\hspace{-3mm}
\subfigure[]
{\includegraphics[width=0.5\textwidth]{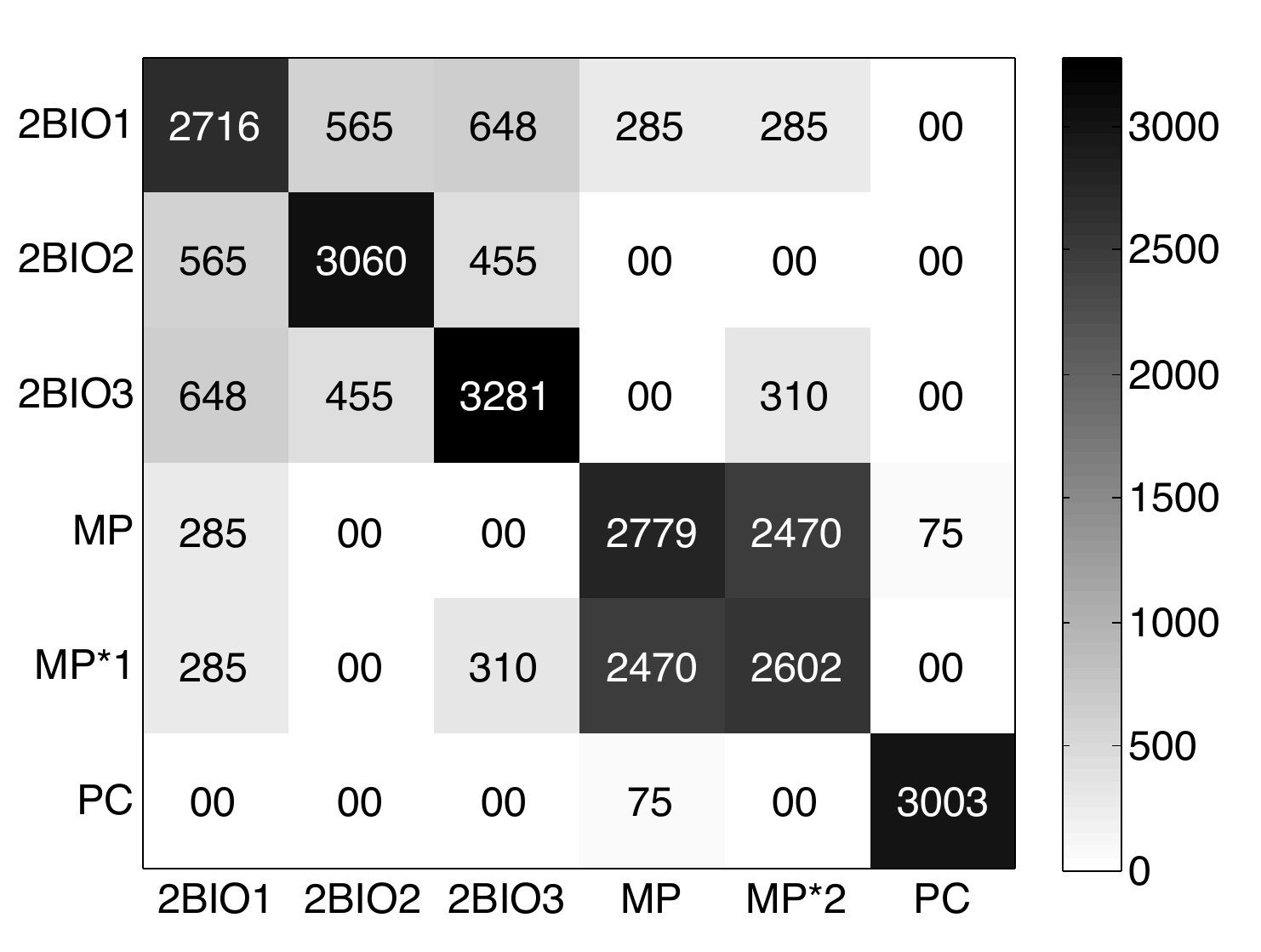}}
\subfigure[]
{\includegraphics[width=0.5\textwidth]{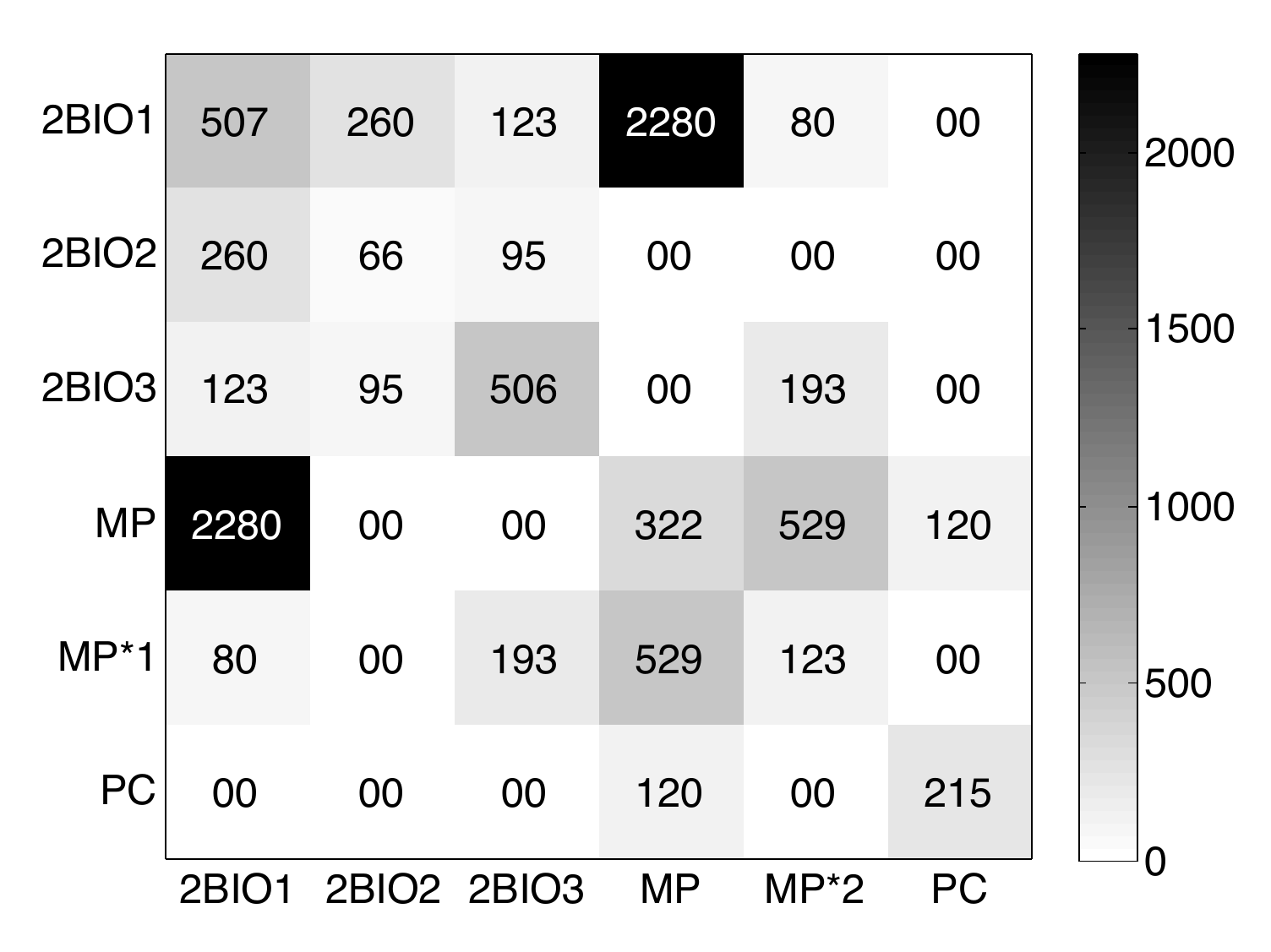}}
\hspace{-3mm}
\subfigure[]
{\includegraphics[width=0.5\textwidth]{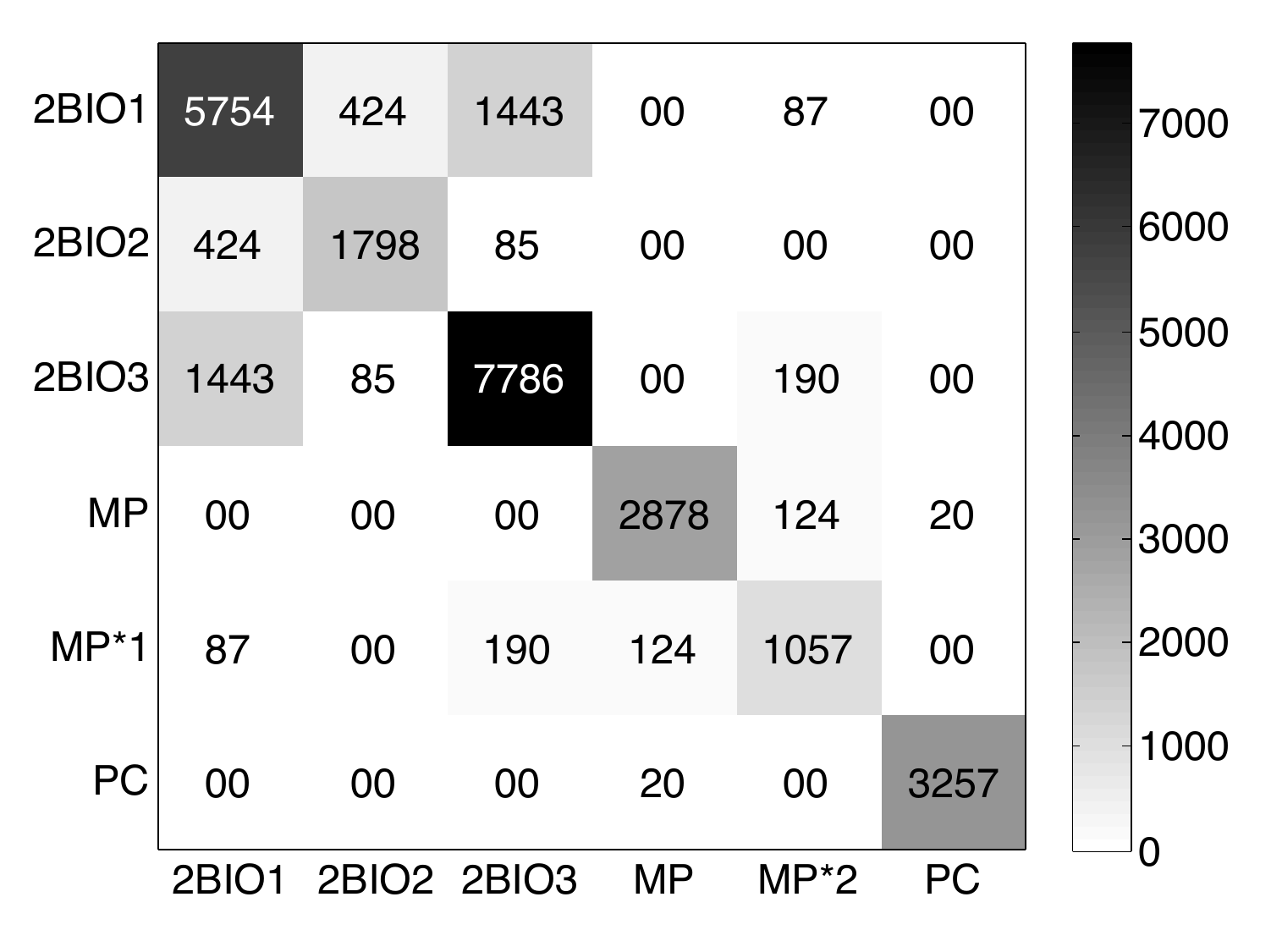}}
\caption{{\bf Contact matrices of average durations for matched networks.} (a) Contact sensors network $\text{CSN}^m$; 
(b) Contact diaries network with durations reported by students ($\text{CDN}^m_\text{D}$); 
(c) one example of contact diaries network with weights randomly drawn from the distribution of contact durations registered by sensors 
($\text{CDN}^m_\text{S}$); 
(d) one example of contact diaries network with weights (including zero values, corresponding to an absence of link),
randomly drawn from the negative binomial fit of the distributions of durations reported by students within and between classes ($\text{CMDN}$). 
\label{AvgDur2}}
\end{figure}

Here, we report the contact matrices of average durations of contacts for the matched versions of the contact sensors and diaries networks. The weighted 
diagonal structure is evident in the $\text{CSN}^m$, $\text{CDN}^m_{\text{D}}$ and $\text{CMDN}$ (respectively Fig.s \ref{AvgDur2} (a), (b) and (d)), 
while it is not respected in the $\text{CDN}^m_{\text{S}}$ case. 
This depends again on the random assignment of durations registered from sensors to the links reported in the diaries.

\clearpage

\section{Impact of the initial seed}

\begin{figure}[!ht]
\centering
{\includegraphics[width=0.8\textwidth]{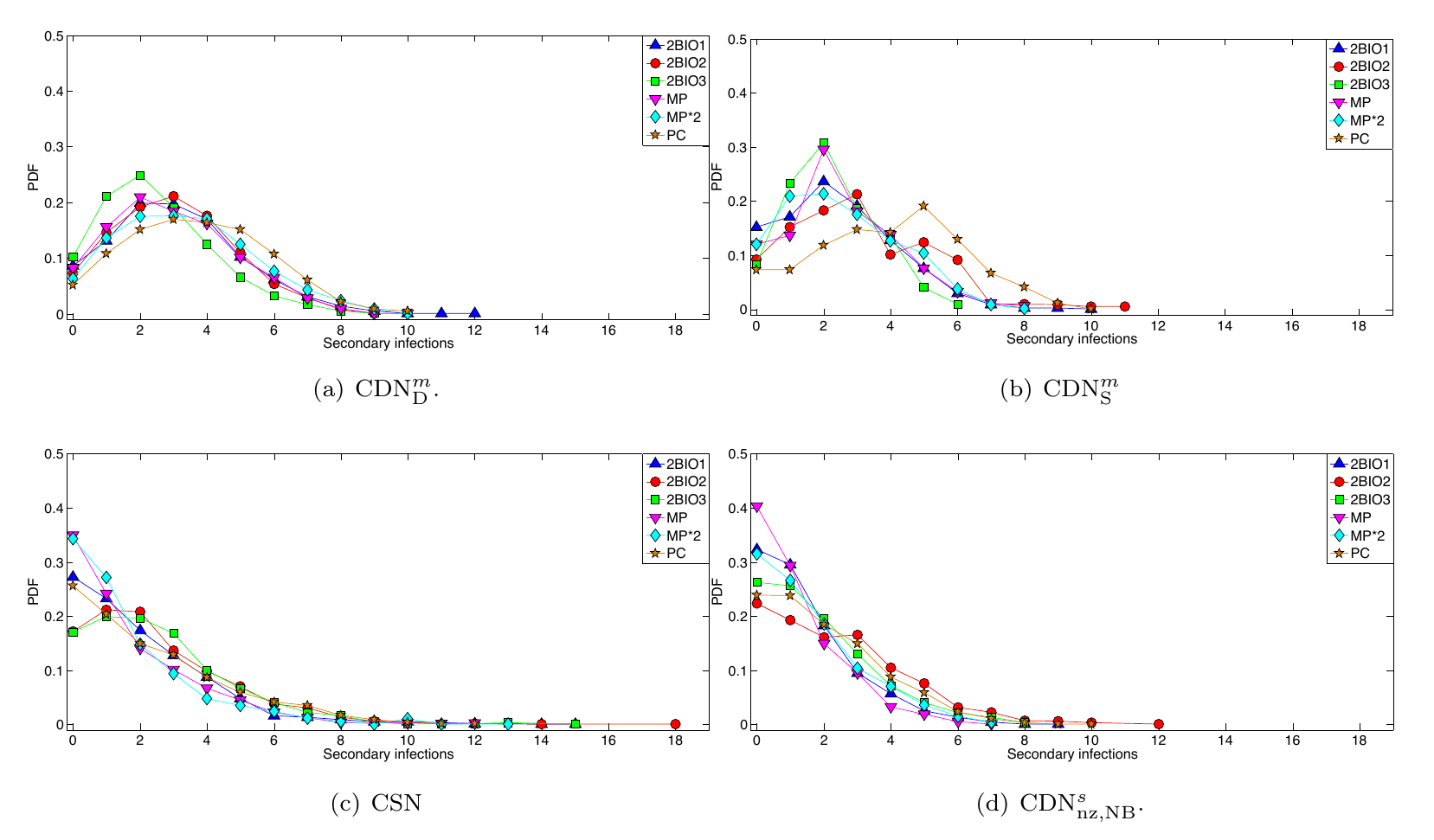}}
\caption{{\bf Distribution of the number of secondary infections by the initial seed.}  
Distributions are averaged
over $1000$ realizations and separated depending on the class of the
initial seed, for the contact diaries networks with weights durations respectively reported by students
    ($\text{CDN}^m_{\text{D}}$) and registered by sensors
    $\text{CDN}^m_{\text{S}}$), the contact sensor network restricted to $6$ classes and
the surrogate network  $\text{CDN}^m_{\text{nz,NB}}$. $\beta/\mu=30$.
\label{R0_seed}}
\end{figure}

\begin{figure}[!ht]
\centering
{\includegraphics[width=0.8\textwidth]{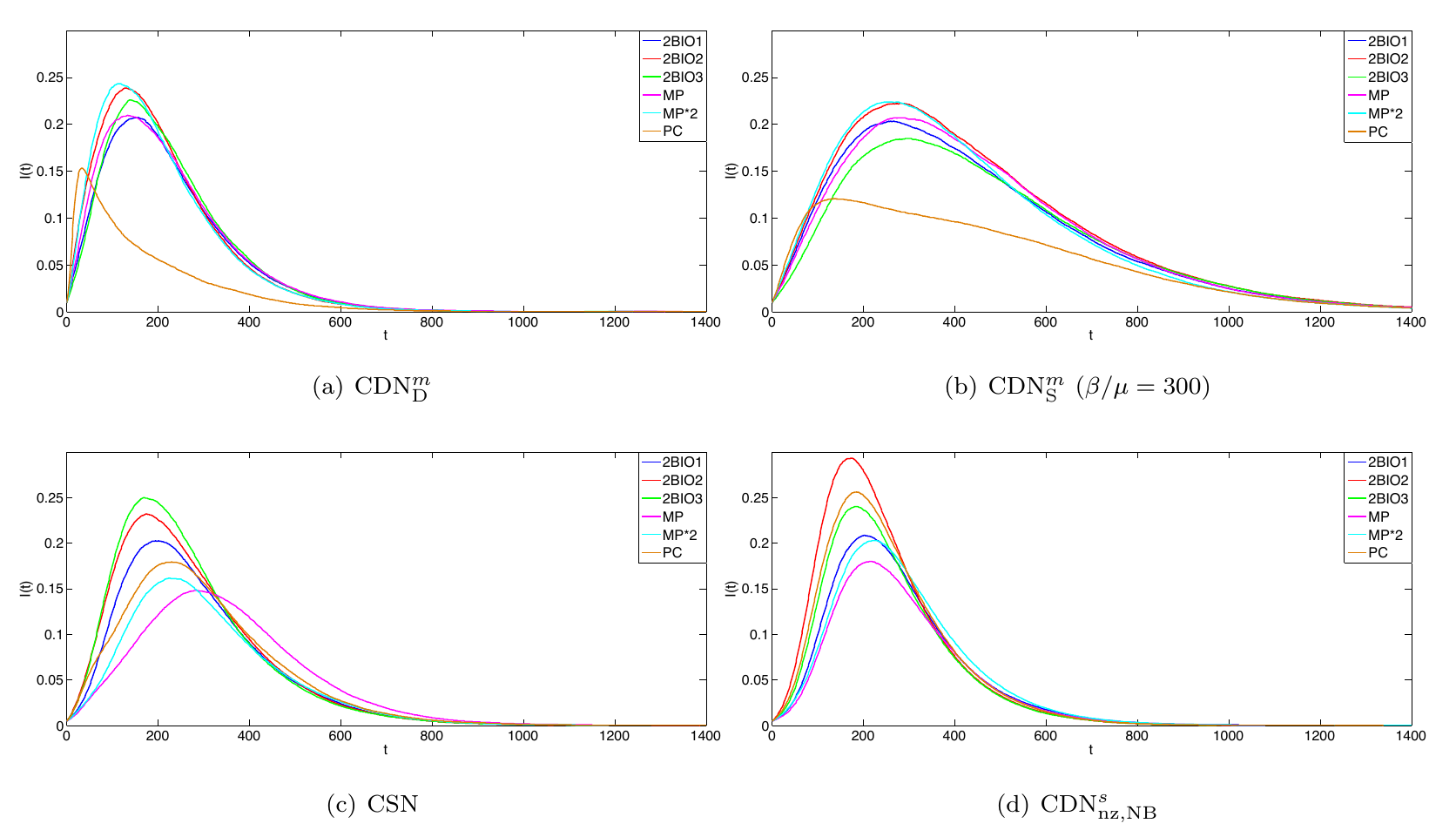}}
\caption{{\bf Temporal evolution of the fraction of infectious individuals.}  
We show the evolution with time of the density of infectious individuals, averaged
over $1000$ realizations and separated depending on the class of the
initial seed, for the contact diaries networks with weights  durations respectively reported by students
    ($\text{CDN}^m_{\text{D}}$) and registered by sensors
    $\text{CDN}^m_{\text{S}}$), the contact sensor network restricted to $6$ classes and
the surrogate network  $\text{CDN}^m_{\text{nz,NB}}$. In order to show 
the effect of the seed for the $\text{CDN}^m_{\text{S}}$, we use a large value of $\beta/\mu=300$, while
for the other cases $\beta/\mu=30$.
\label{I_seed}}
\end{figure}

\begin{figure}[!ht]
\centering
{\includegraphics[width=0.8\textwidth]{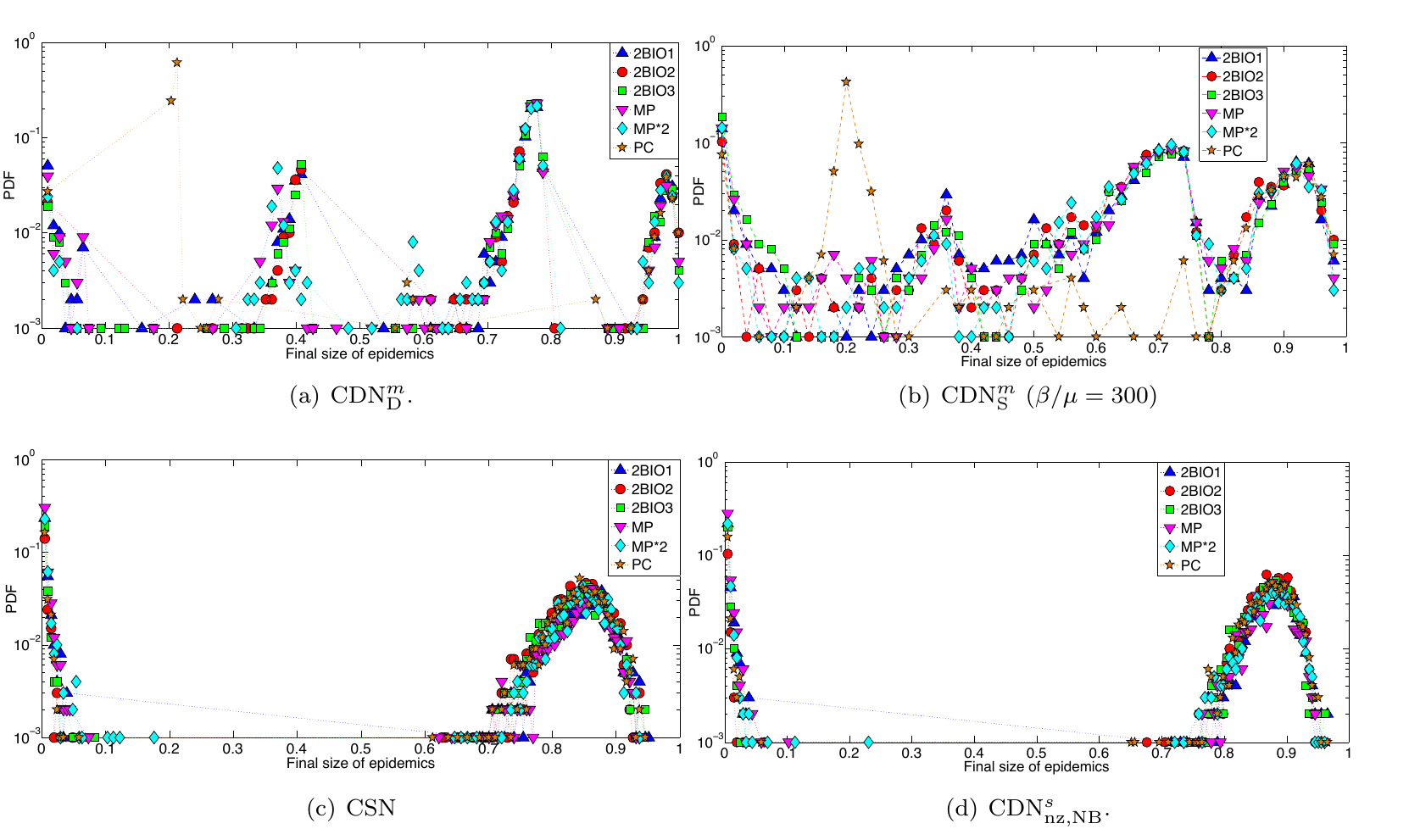}}
\caption{{\bf Impact of the initial seed.}  
Distribution of final sizes of final sizes of epidemics, separated depending on the class of the
initial seed, for the contact diaries networks with weights  durations respectively reported by students
    ($\text{CDN}^m_{\text{D}}$) and registered by sensors
    $\text{CDN}^m_{\text{S}}$), the contact sensor network restricted to $6$ classes and
the surrogate network  $\text{CDN}^m_{\text{nz,NB}}$. In order to show 
the effect of the seed for the $\text{CDN}^m_{\text{S}}$, we use a large value of $\beta/\mu=300$, while
for the other cases $\beta/\mu=30$.
\label{PR_seed}}
\end{figure}

We show in Fig.s \ref{R0_seed}, \ref{I_seed} and \ref{PR_seed} how the
properties of the spread
obtained in the various considered networks depend on the class of the initial seed:
the distribution of the number of secondary infections per index case, the temporal evolution
of the density of infectious individuals, and the distribution of the final size of the epidemics.

Interestingly, the peculiar structure of the contact diaries network leads to a non-trivial dependence
on the initial seed. In particular, if the seed is in PC, which is little connected to the other classes,
the number of secondary infections is larger (Fig. \ref{R0_seed}), leading to an initially faster spread
(\ref{I_seed}) that however decreases also much faster (as with high probability it does not manage
to reach the other classes). 
Some less marked dependency from the initial seed is observed for the CSN, 
with a slower spread if the initial seed is in MP.
The surrogate network also yields a slower spread in this case, even if the effect is not as strong, 
and the peculiar shape of the incidence for an initial seed in PC is not observed.

With respect to the overall impact of the spread, as measured by the distributions of
final sizes of epidemics, the situation is more striking. For the contact diaries network, 
the case of a seed in PC is once again singular, reaching only small sizes with large probability
(as the spread is ``trapped'' in PC and does not easily reach other classes).
For the $\text{CDN}^m_{\text{D}}$ and $\text{CDN}^m_{\text{S}}$ we also
observe similar distributions for seeds in one of the three biology classes, which are indeed
well connected, and a slightly different distribution when the seed is in one of the MP classes.
For CSN and the surrogate network on the other hand, no clear dependency on the initial seed is observed
for the distributions of the final size of epidemics: the differences observed for the 
contact diaries network are thus spurious, and the use of the surrogate networks corrects
this spurious effect.

\clearpage

\section{Spreading simulations using the $\text{CMDN}$}

\begin{figure}[!ht]
\subfigure[$\beta/\mu=6$]
{\includegraphics[width=0.5\textwidth]{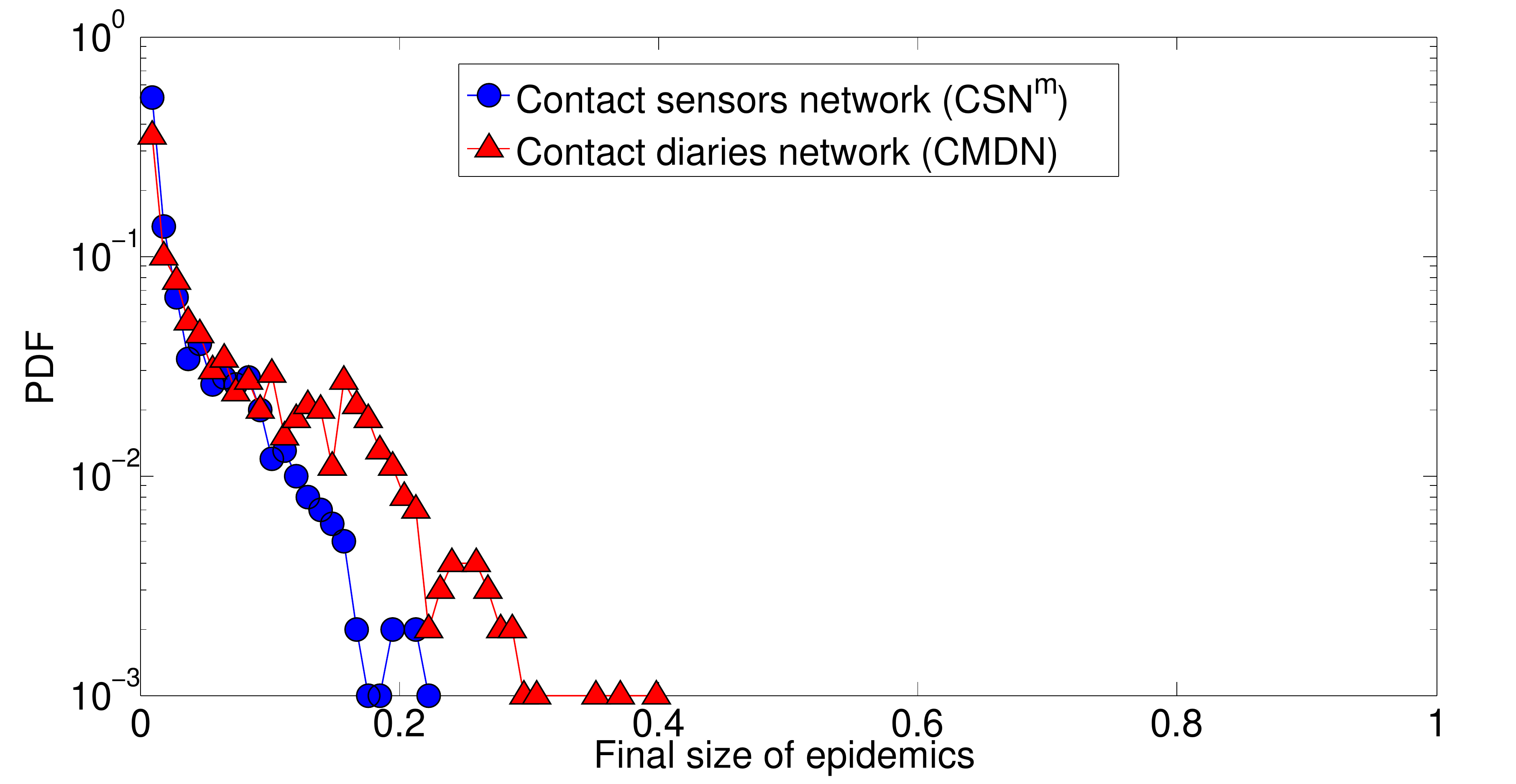}}
\hspace{-5mm}
\subfigure[$\beta/\mu=10$]
{\includegraphics[width=0.5\textwidth]{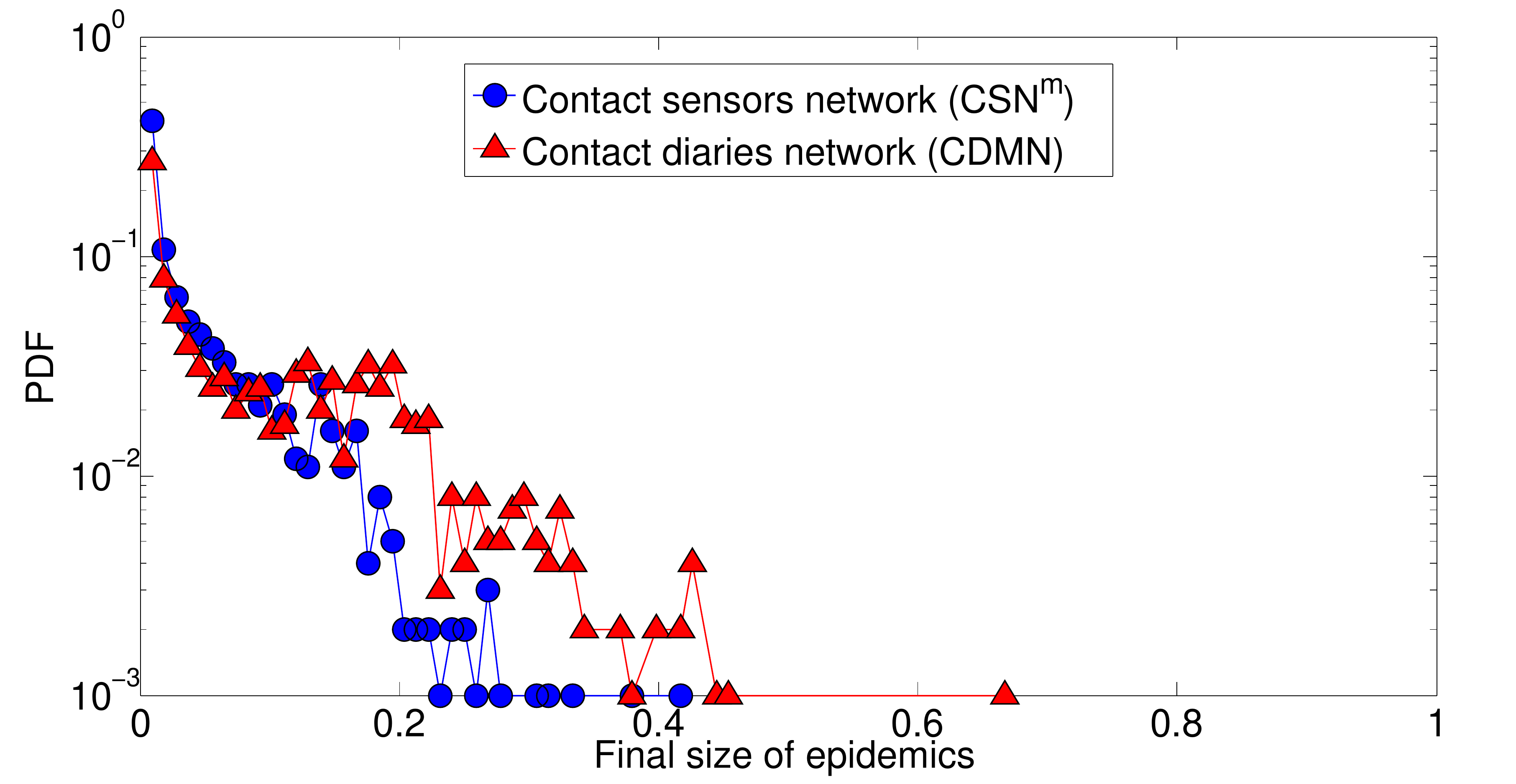}}
\subfigure[$\beta/\mu=14$]
{\includegraphics[width=0.5\textwidth]{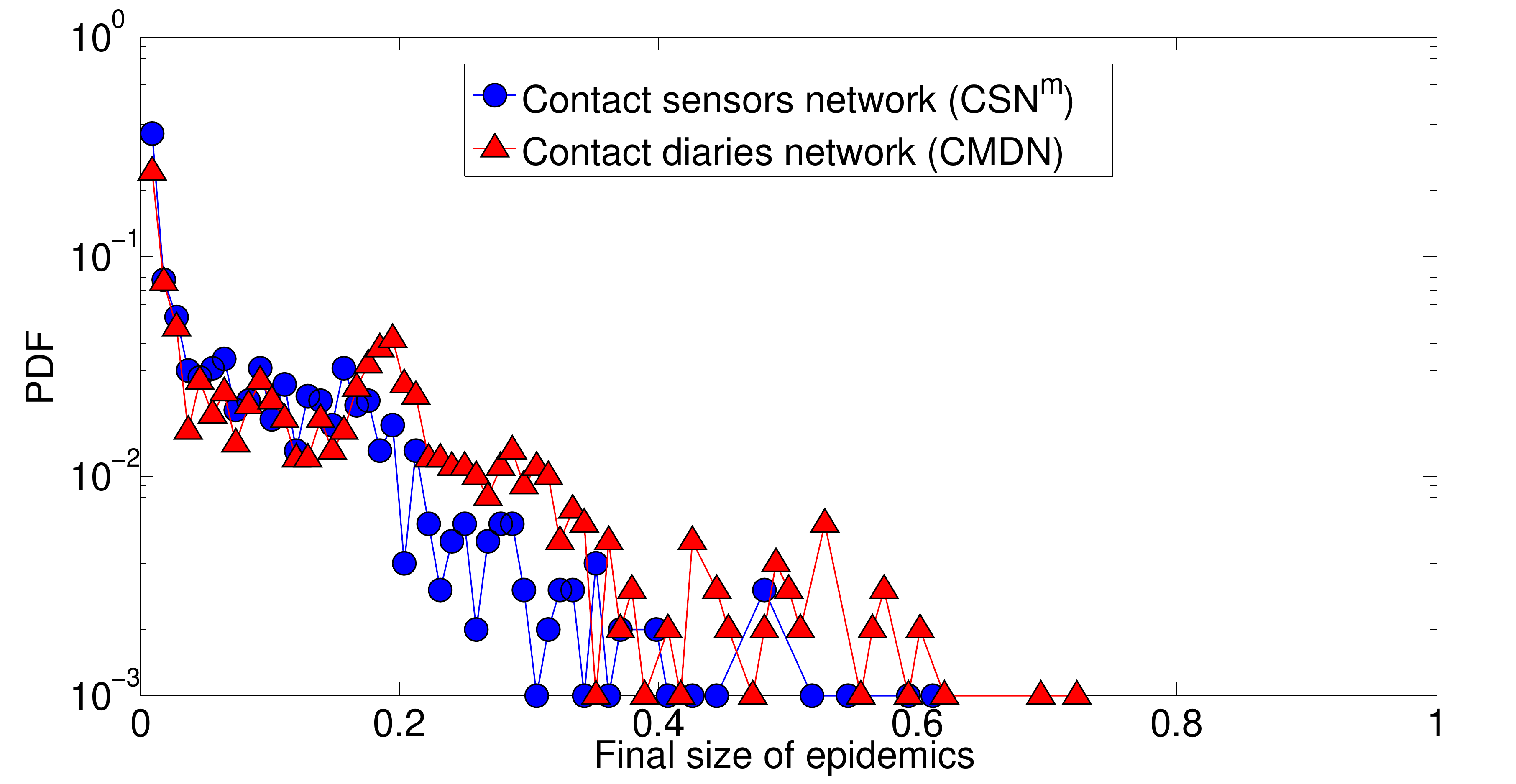}}
\hspace{-5mm}
\subfigure[$\beta/\mu=20$]
{\includegraphics[width=0.5\textwidth]{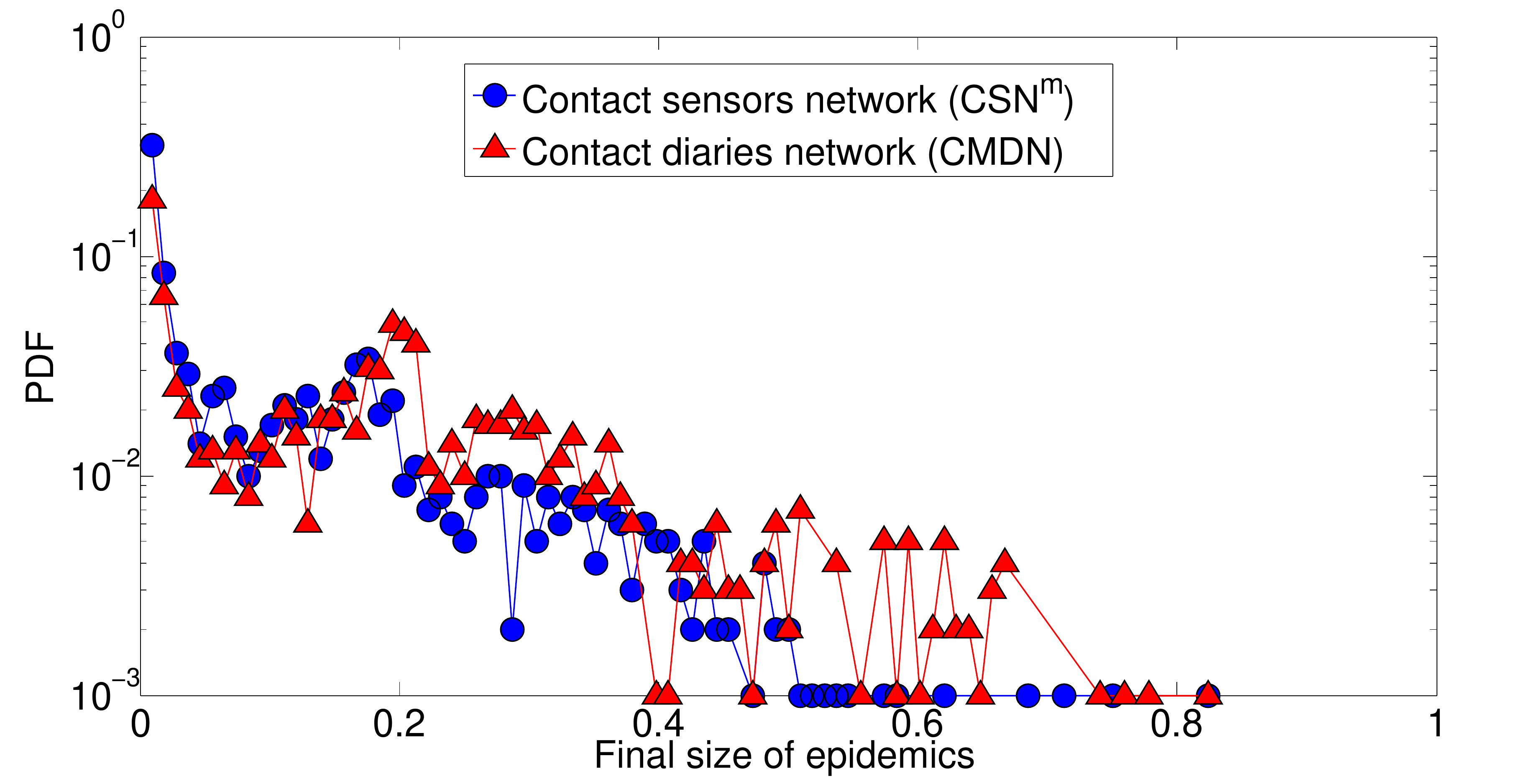}}
\subfigure[$\beta/\mu=30$]
{\includegraphics[width=0.5\textwidth]{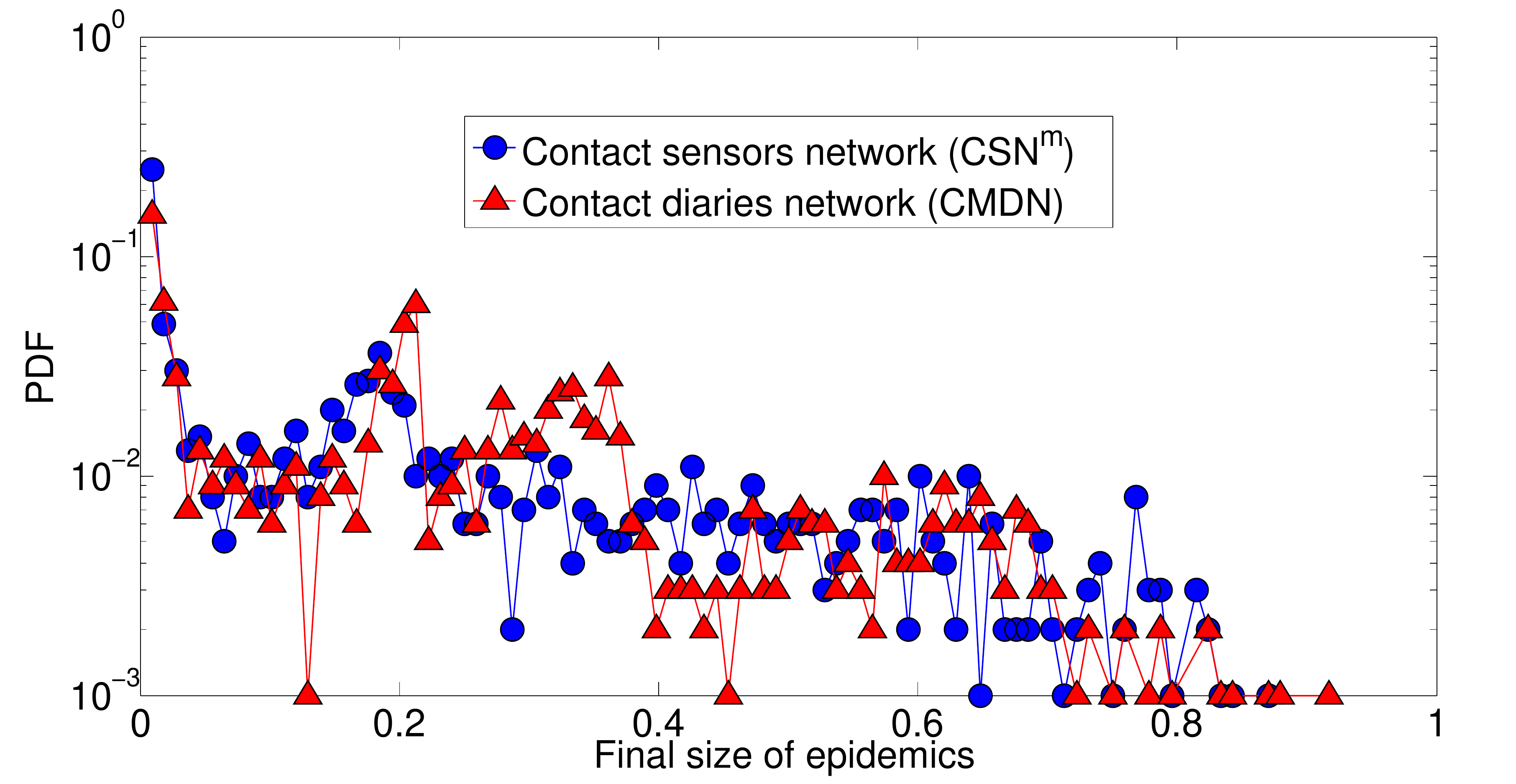}}
\hspace{-2mm}
\subfigure[$\beta/\mu=40$]
{\includegraphics[width=0.5\textwidth]{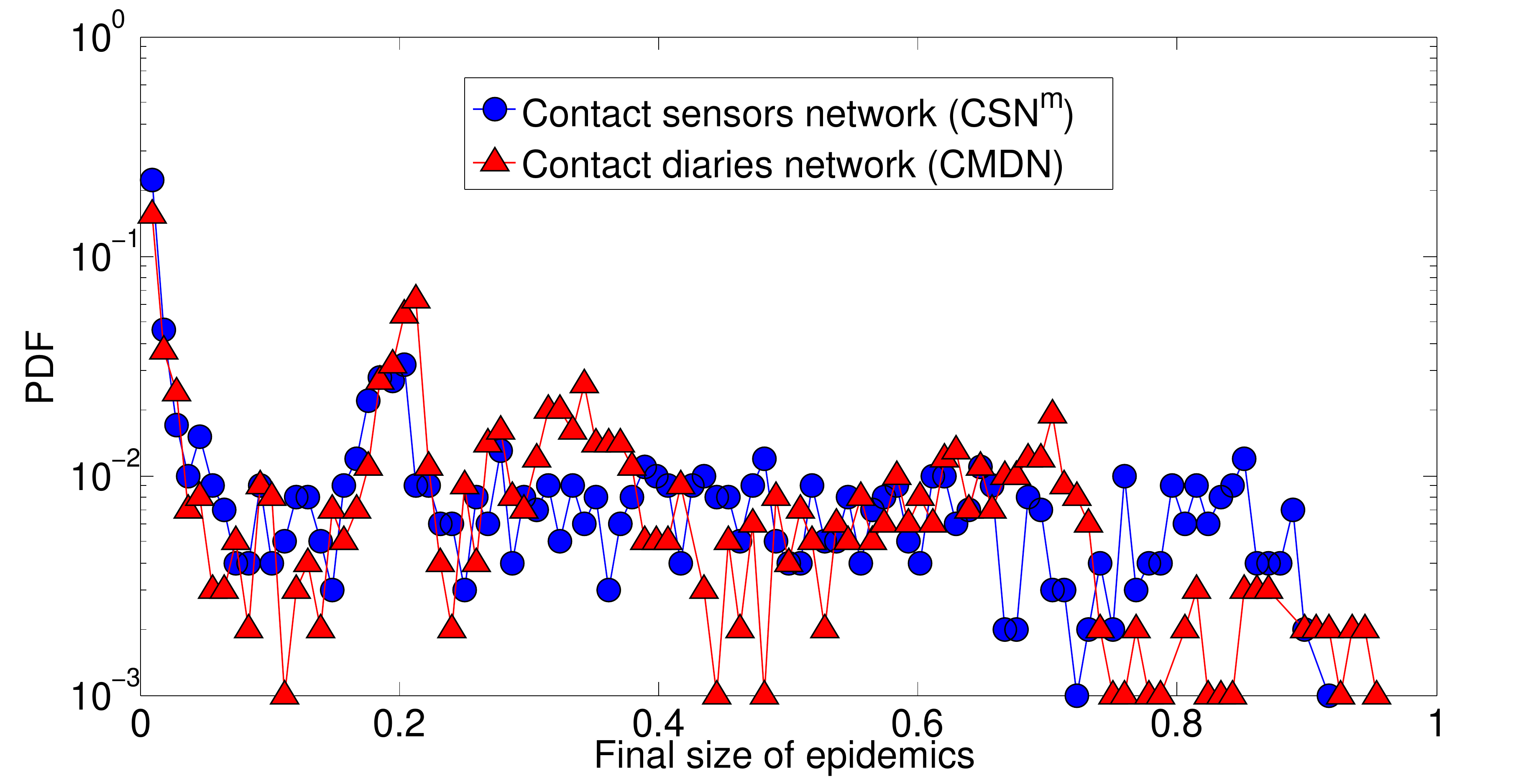}}
\caption{{\bf Distributions of final size of epidemics.}  Outcome of 1000 SIR simulations performed on the contact sensors network ($\text{CSN}^m$) and the 
$\text{CMDN}$ obtained with weights randomly drawn from the negative binomial fit of distributions of contact durations within 
and between classes registered by sensors. 
Each process starts with one random infected seed.
\label{cmd}}
\end{figure}

\begin{figure}[!ht]
\centering
{\includegraphics[width=1\textwidth]{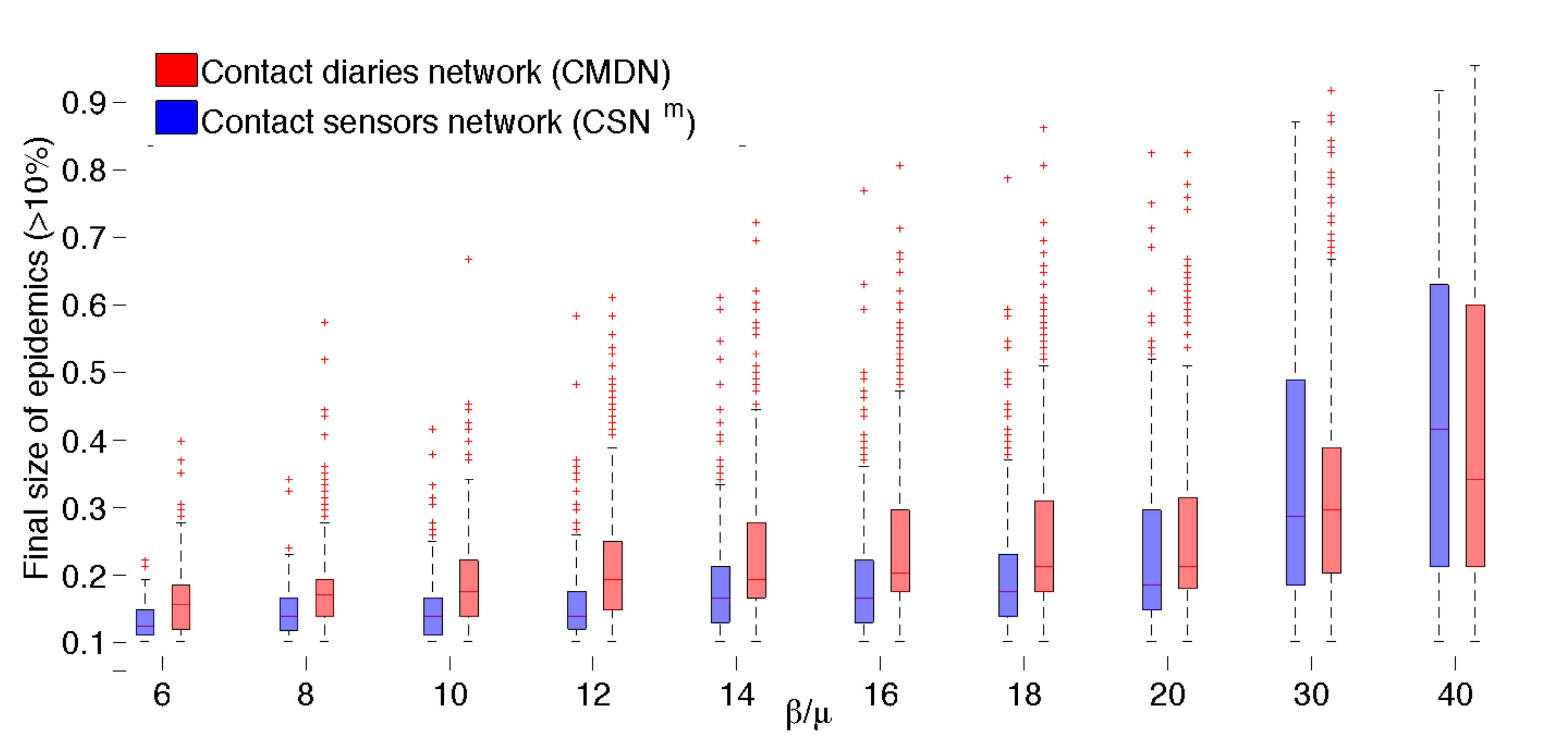}}
\caption{{\bf Box-plot of final size of epidemics.} Comparison of the distributions of final size of epidemics, for sizes larger than $10\%$, for SIR simulations 
performed on the contact sensors network ($\text{CSN}^m$) and the CMDN obtained with weights 
randomly drawn from the negative binomial fit of the distribution of contact durations within and between classes registered by sensors.  
For each box, the central mark stands for the median, its edges represent the $25^{th}$ and $75^{th}$ percentiles.  The whiskers extend to the most extreme data points not considered outliers, while outliers are plotted individually (1000 simulations for each value of $\beta/\mu$).
\label{BoxCmd}}
\end{figure}

In figure \ref{cmd} we show the distribution of final epidemic sizes for SIR simulations performed on the $\text{CSN}^m$ and the $\text{CMDN}$. 
The two networks have the same nodes but different links and weights. Weights in the CMDN 
are randomly drawn from distributions obtained from a negative binomial fit, within and between classes, 
performed on the distributions of durations reported by students. 
We chose several values of parameters $\beta$ and $\mu$ to check the robustness of the outcome shown in Fig. 2 in the paper. 
In Fig. \ref{BoxCmd} we compare the boxplots of the two distributions for epidemic sizes larger than $10\%$.

\newpage

\section{Original, reduced and reshuffled contact sensors networks}

\begin{figure}[!ht]
\centering
{\includegraphics[width=1\textwidth]{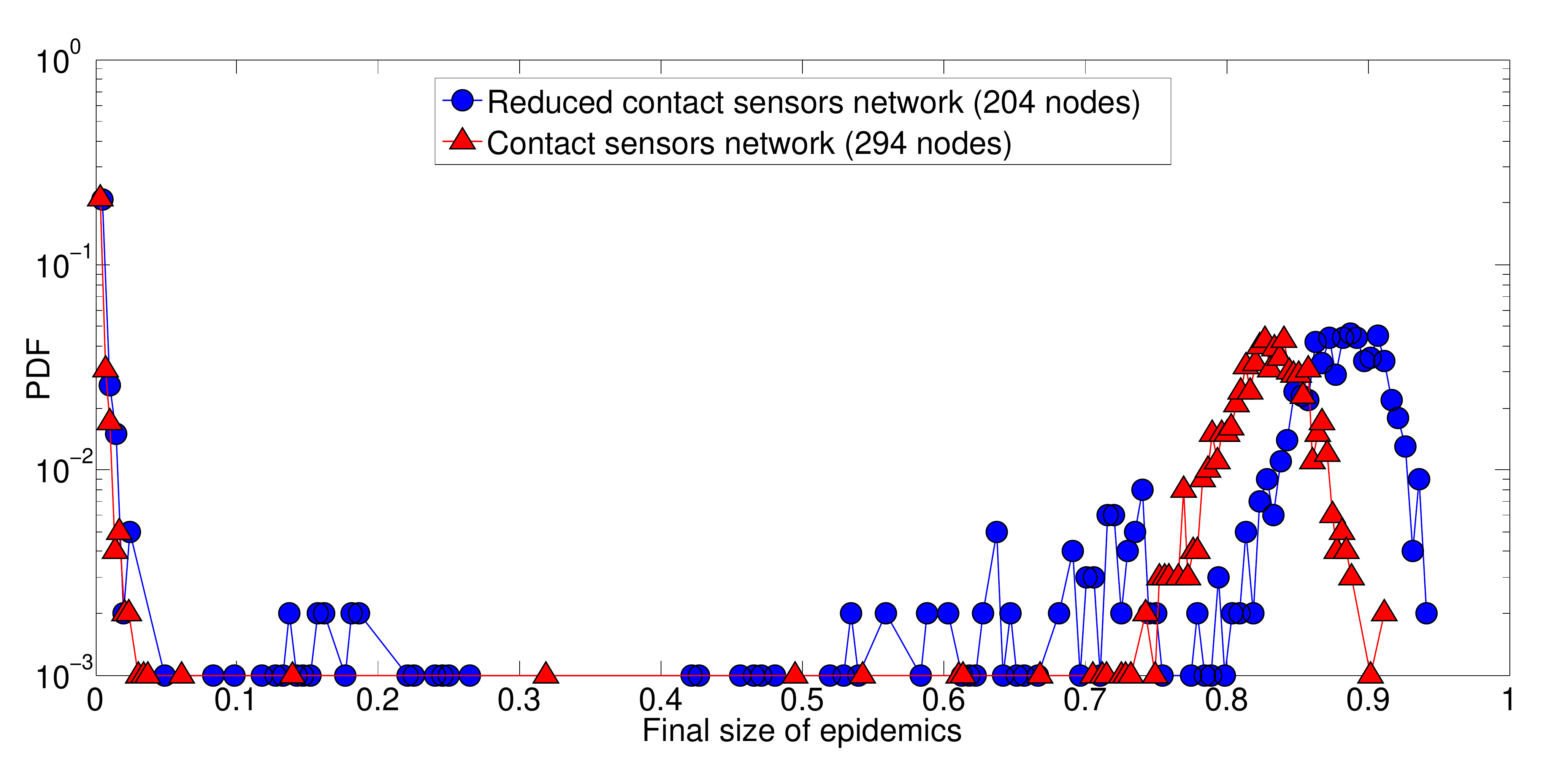}}
\caption{{\bf Distribution of final size of epidemics.} Comparison between the distributions obtained 
from 1000 SIR simulations performed on the original contact sensors network and its reduced form obtained by removing the three classes 
not or under-represented in the contact diaries network.  Each process starts with one random infected seed.  $\beta/\mu=30$.
\label{sens}}
\end{figure}

In Fig. \ref{sens} we explore the change in the prediction of the epidemic risk associated to the contact 
sensors network when some classes are removed from the sample. This choice depends on the fact that 
such classes are absent or under-represented by the contact diaries network, and so for them a 
construction of surrogate links using the available information would be impossible. The original 
network has 295 nodes and 2162 links, its reduced form has 204 nodes and 1600 links.  

The shape of the distributions are quite similar with an overestimation of the epidemic peak for the reduced network with respect to the original contact sensors network. This outcome suggests that the removed classes are not well connected with the rest of the population, 
preventing the diffusion of the epidemics and reducing the epidemic risk.

\begin{figure}[!ht]
\centering
{\includegraphics[width=1\textwidth]{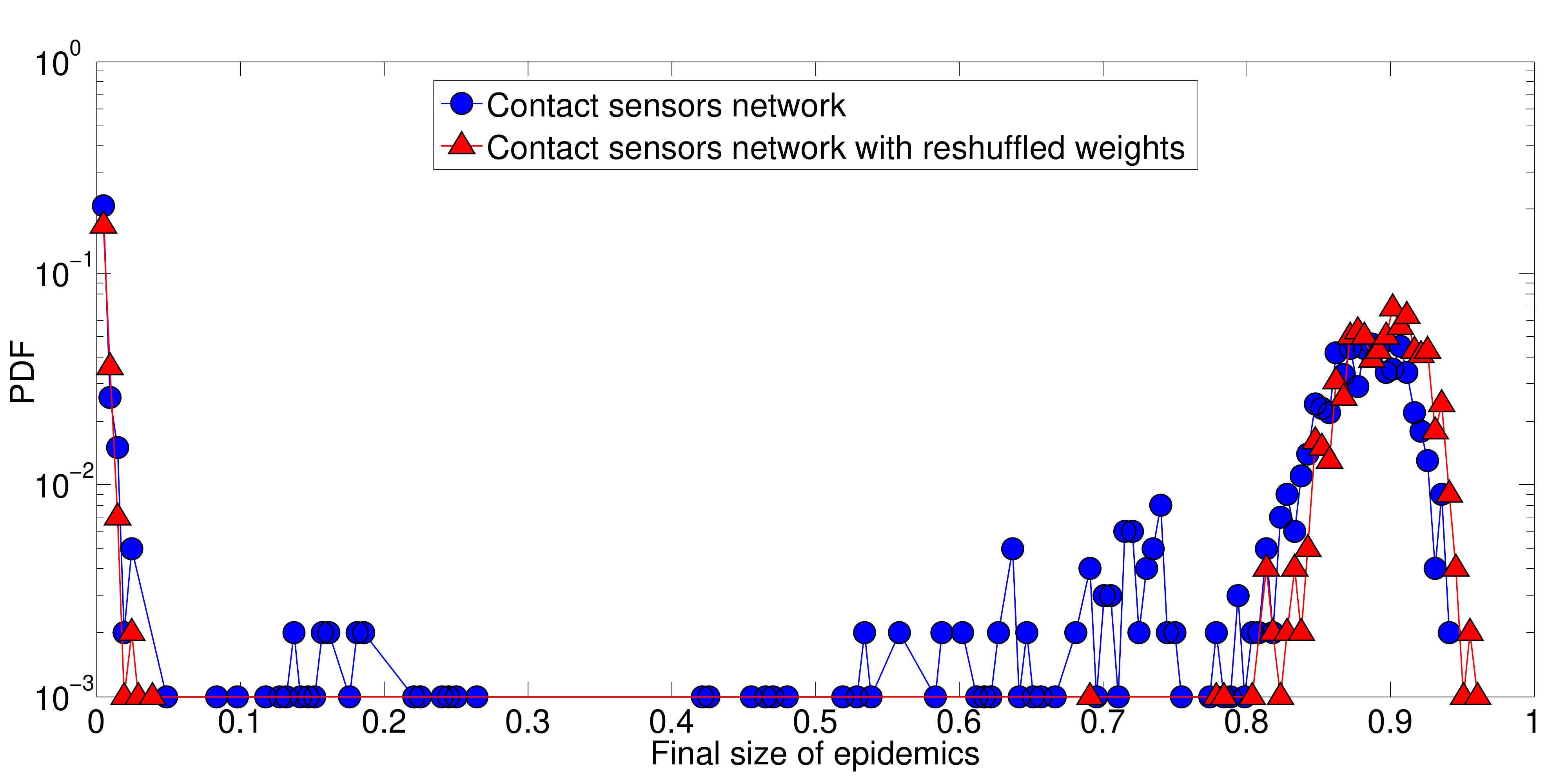}}
\caption{{\bf Distribution of final size of epidemics.} Comparison of the distributions obtained from 1000 SIR simulations  performed on the reduced contact sensors network and on a null model with the same binary structure but reshuffled weights. Each process starts with one random infected seed.  
$\beta/\mu=30$.
\label{oss2}}
\end{figure}

In Fig. \ref{oss2} we show the outcome of  SIR simulations on the reduced contact sensors network and on a 
null-model with the same link structure but reshuffled weights. The comparison between the two distributions reveals a good agreement for the peak of the distribution, while the probability to observe epidemics involving intermediate shares of population is almost zero when weights are reshuffled. 
This suggests that the occurrence of epidemics of intermediate size depends on specific correlations between structure and weights.

\clearpage

\section{Simulations on surrogate networks}

\subsection{Surrogate contact network with homogeneous durations}

\begin{figure}[!ht]
\centering
\subfigure[]
{\includegraphics[width=0.51\textwidth]{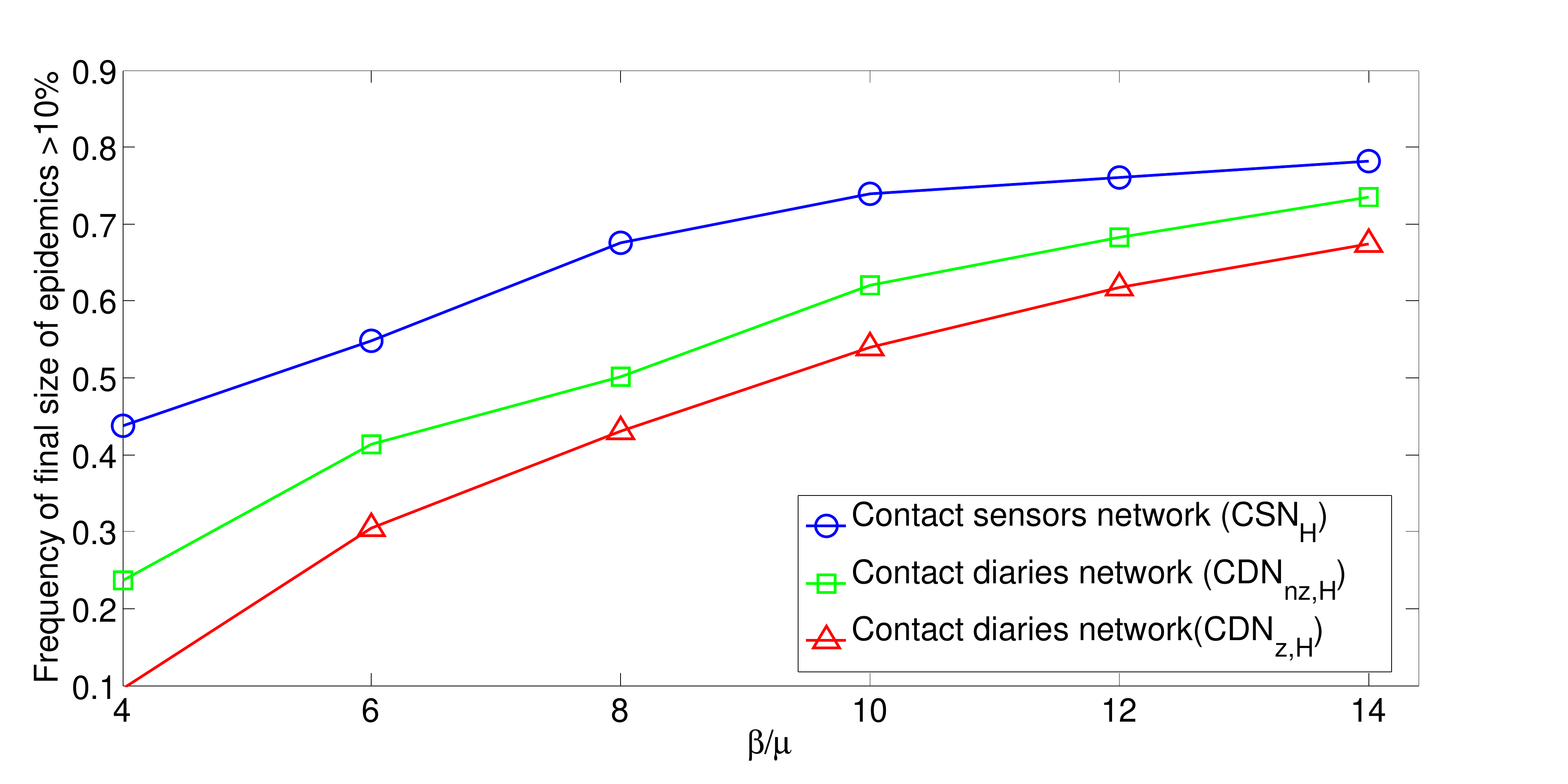}}
\hspace{-4mm}
\subfigure[]
{\includegraphics[width=0.5\textwidth]{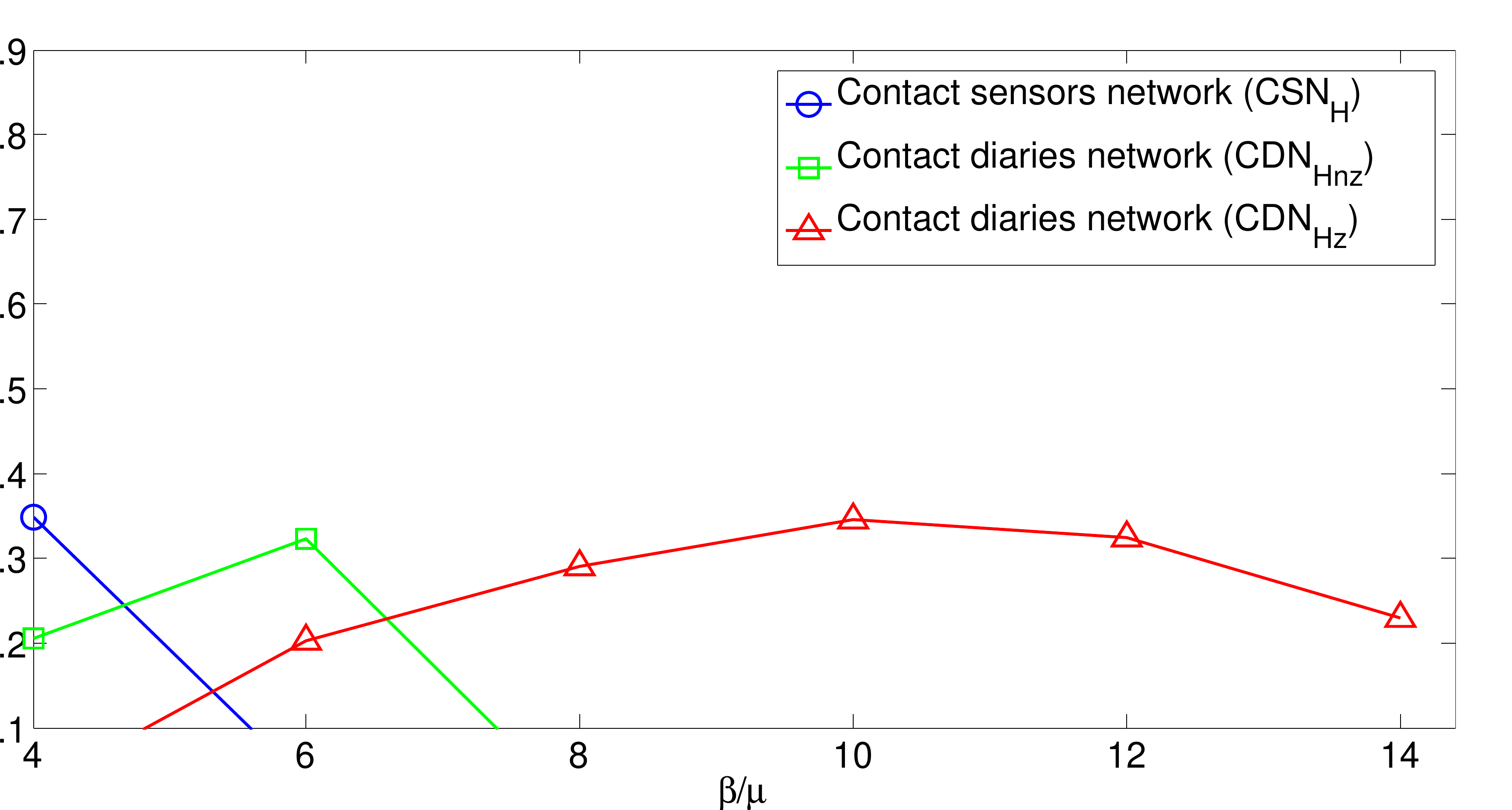}}
\caption{{\bf Fraction of epidemics reaching specific shares of the population.} 
Frequency of processes involving (a) more than $10\%$  and (b) between $20\%$ and $70\%$ 
of the whole population for the contact sensors network ($\text{CSN}_\text{H}$) and the surrogate contact networks under the hypothesis 
of homogeneous cumulative durations ($\text{CDN}^s_\text{z,H}$ and $\text{CDN}^s_\text{nz,H}$) . 
Values computed over 1000 SIR simulations. Each process starts with one random infected seed. 
\label{dist6}}
\end{figure}

Figure \ref{dist6} complements the result presented in Fig. 4 of the main text 
on the comparison between the contact sensors network and the surrogate contact network under the hypothesis of homogeneous cumulative durations. 
Here, we focus on the share of epidemic processes reaching (a) more than $10\%$, (b) between $20\%$ and $70\%$.

\subsection{Surrogate contact diaries network with weights registered by sensors}

\begin{figure}[!ht]
\centering
\subfigure[]
{\includegraphics[width=0.51\textwidth]{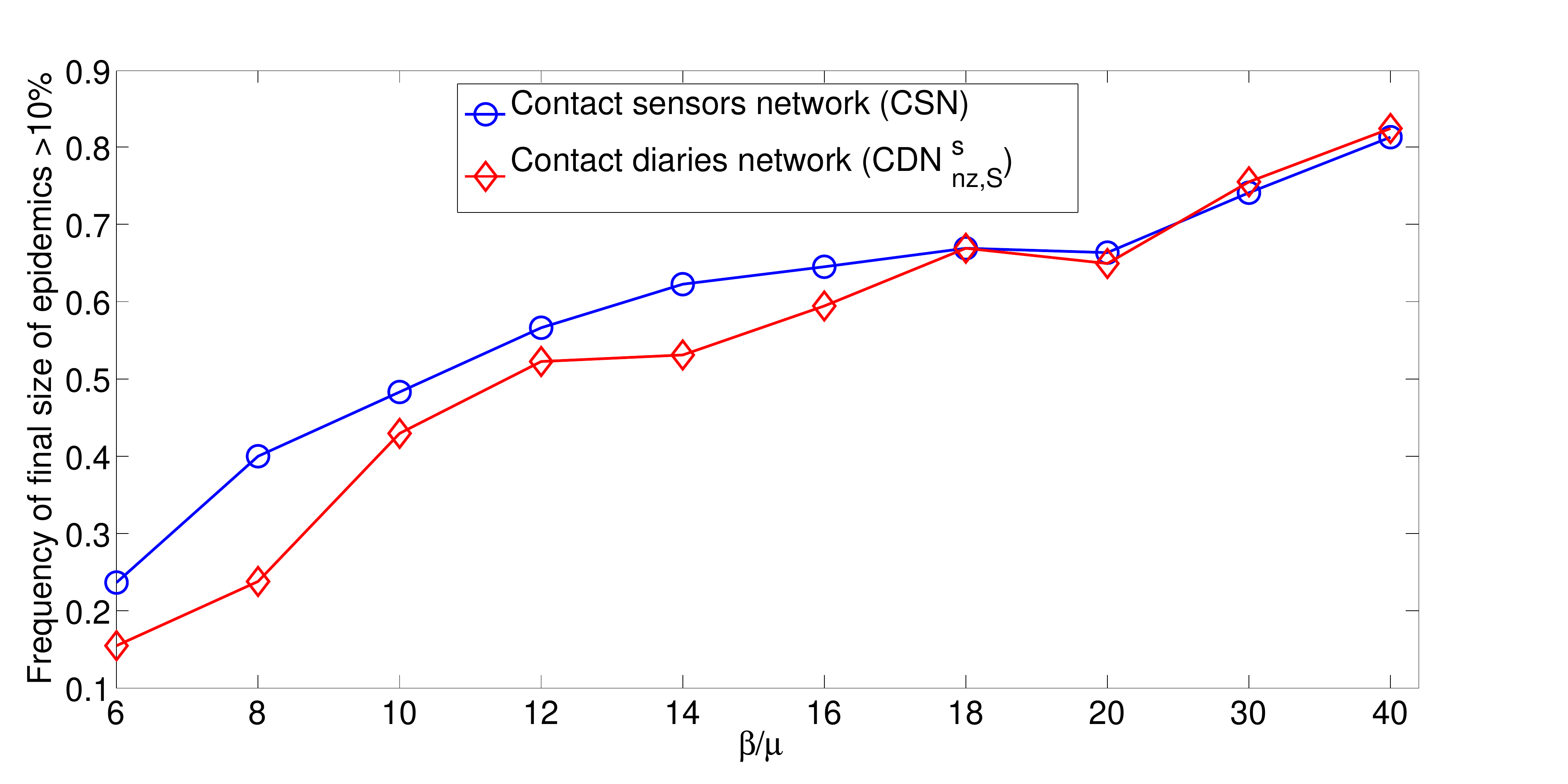}}
\hspace{-4mm}
\subfigure[]
{\includegraphics[width=0.5\textwidth]{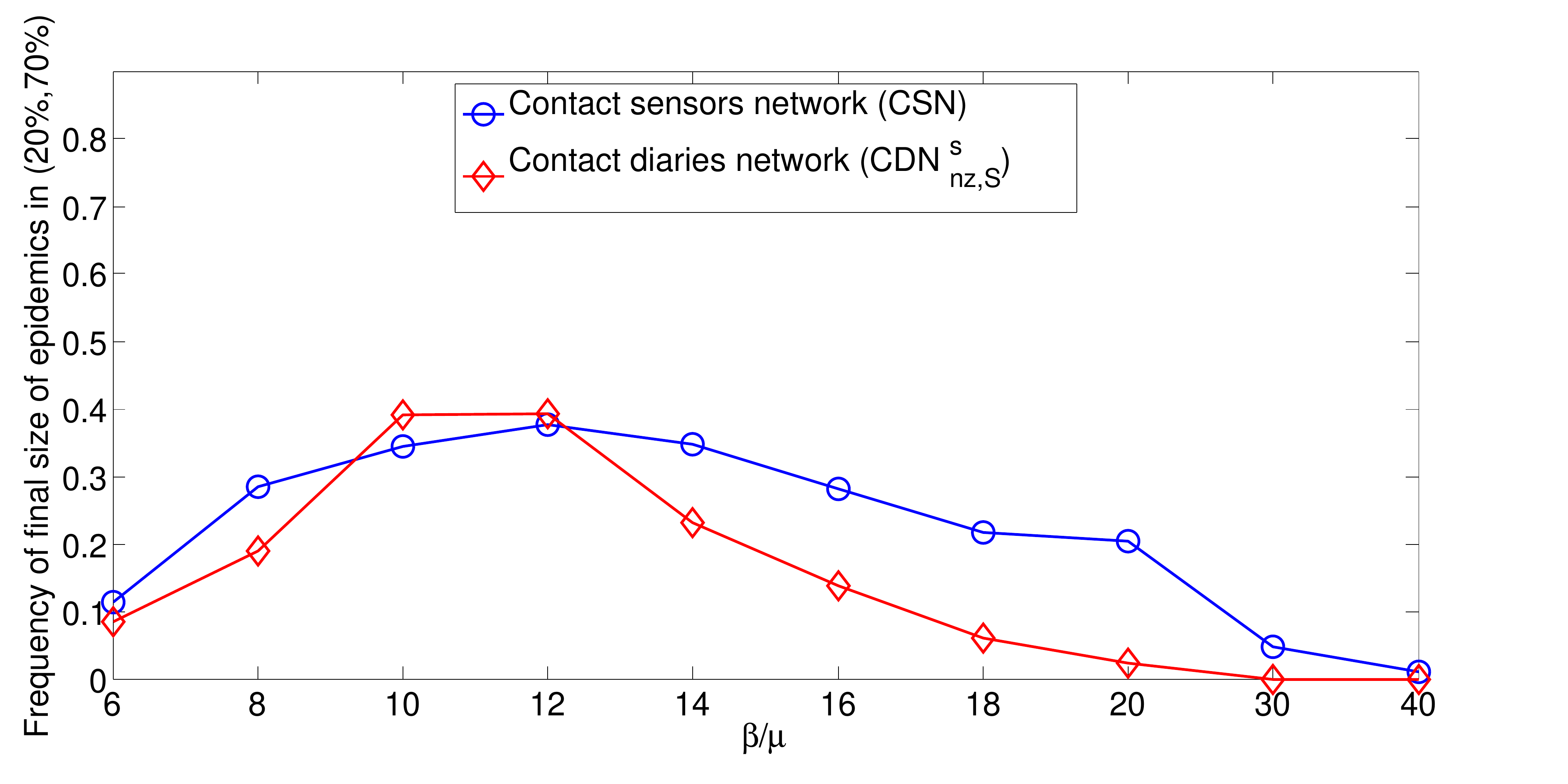}}
\caption{{\bf Fraction of epidemics reaching specific shares of the population.} 
Fraction of simulations involving (a) more than $10\%$  and (b) between $20\%$ and $70\%$ of the whole population for the contact sensors network and the surrogate contact  network ($\text{CDN}^s_\text{nz,S}$). Values computed over 1000 SIR simulations. 
Each process starts with one random infected seed. 
\label{dist7}}
\end{figure}

Figure \ref{dist7} complements the result presented in Fig. 6 of the main text on the comparison between the contact sensors network and the surrogate contact  
network. As in Fig. \ref{dist6}, we focus on the share of epidemic processes reaching (a) more than $10\%$, (b) between $20\%$ and $70\%$.

\subsection{Surrogate contact network with durations taken from publicly available data}

Table \ref{table} reports the parameters (and related confidence intervals) of a negative binomial fit performed on the distribution 
of (non-zero) durations registered by sensors for several datasets. There are several equivalent ways to describe the functional form of a negative binomial distribution, here we use the following definition for a random variable $x$ and parameters $r$ and $p$. In general, when $r$ is an integer:
\begin{equation*}
y=f(x|r,p)=\binom{r+x-1}{x} p^r q^x I_{\{0,1,\dots\} }(x)
\end{equation*}

If $r$ is not integer, the binomial coefficient is replaced by the equivalent expression (using the gamma function):

\begin{equation*}
\dfrac{\Gamma (r+x)}{\Gamma(r)\Gamma(x+1)}
\end{equation*}

\begin{savenotes}
\begin{table}[!ht]
\centering
\medskip
\begin{tabular}{l|c|c}
\toprule
\bf{\Large{Dataset}} & \bf{\Large{Parameters}} \\
\hline
 \multirow{2}{*}{\bf High-school} \protect\cite{Mastrandrea:2015} &   $(0.0137,0.0002)$   \\
                                                     & ([0.001329,0.001404], [0.00018, 0.00021]) \\
\hline
\multirow{2}{*}{\bf American primary school} \protect\cite{Toth:2015} & $(0.0455, 0.0005)$    \\
 & ([0.04470,0.04621],[0.00044,0.00047])\\ 
\hline

\multirow{2}{*}{\bf French primary school} \protect\cite{Stehle:2011} &   $(0.0483, 0.0005)$ \\
& ([0.04710,0.04943],[0.00049,0.00058])\\
\hline

\multirow{2}{*}{\bf Office building}  \protect\cite{Genois2:2015} & $(0.0276, 0.0006)$   \\
 & ([0.02543, 0.02970],[0.00044,0.00066])\\
\hline

\multirow{2}{*}{\bf Conference} \protect\cite{Stehle2:2011} &  $(0.0187, 0.0011)$ \\
& ([0.01828,0.01909],[0.00102,0.00113])\\
\hline
\multirow{2}{*}{\bf Hospital} \protect\cite{Mastrandrea2:2015} &  $(0.2036, 0.0001)$ \\
& ([0.1704,0.2368],[0.00007,0.00014])\\
\hline
\multirow{2}{*}{\bf Combined datasets} & (0.0265,0.0004)\\
 & ([0.02622,0.02678],[0.00040,0.00044])\\
\bottomrule
\end{tabular}
\caption{{\bf Parameters of the negative binomial fits.} 
For each dataset, we fitted the distribution of contact durations by using a maximum likelihood estimates (MLEs) for parameters $r$ and $p$. 
In the first row we report the parameters describing the negative binomial distribution, $r$ and $p$ respectively, 
while in the second row we show the related $95\%$ confidence intervals. }\label{table}
\end{table}
\end{savenotes}

\begin{figure}[!ht]
\centering
\subfigure[]
{\includegraphics[width=0.5\textwidth]{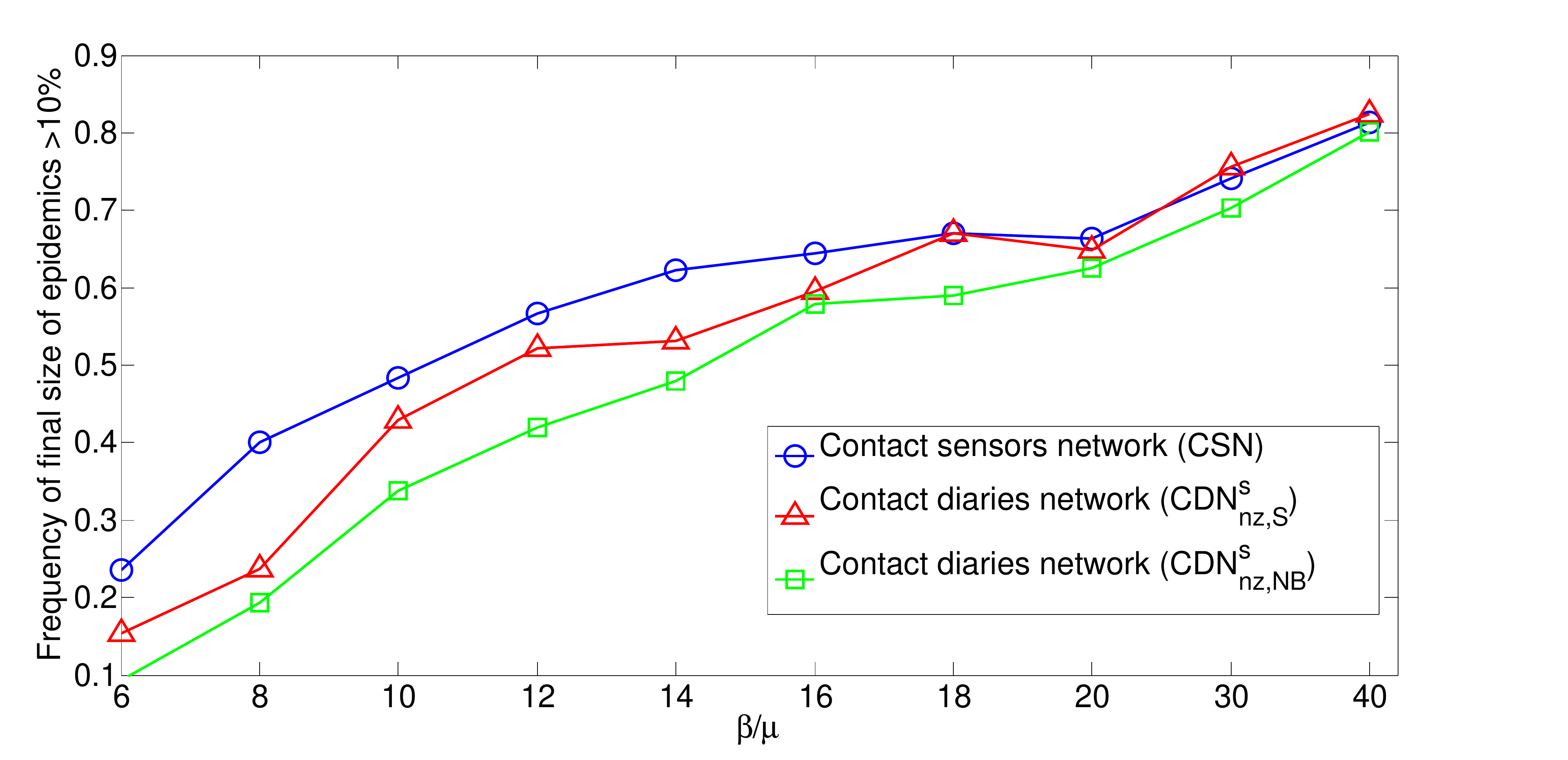}}
\hspace{-5mm}
\subfigure[]
{\includegraphics[width=0.5\textwidth]{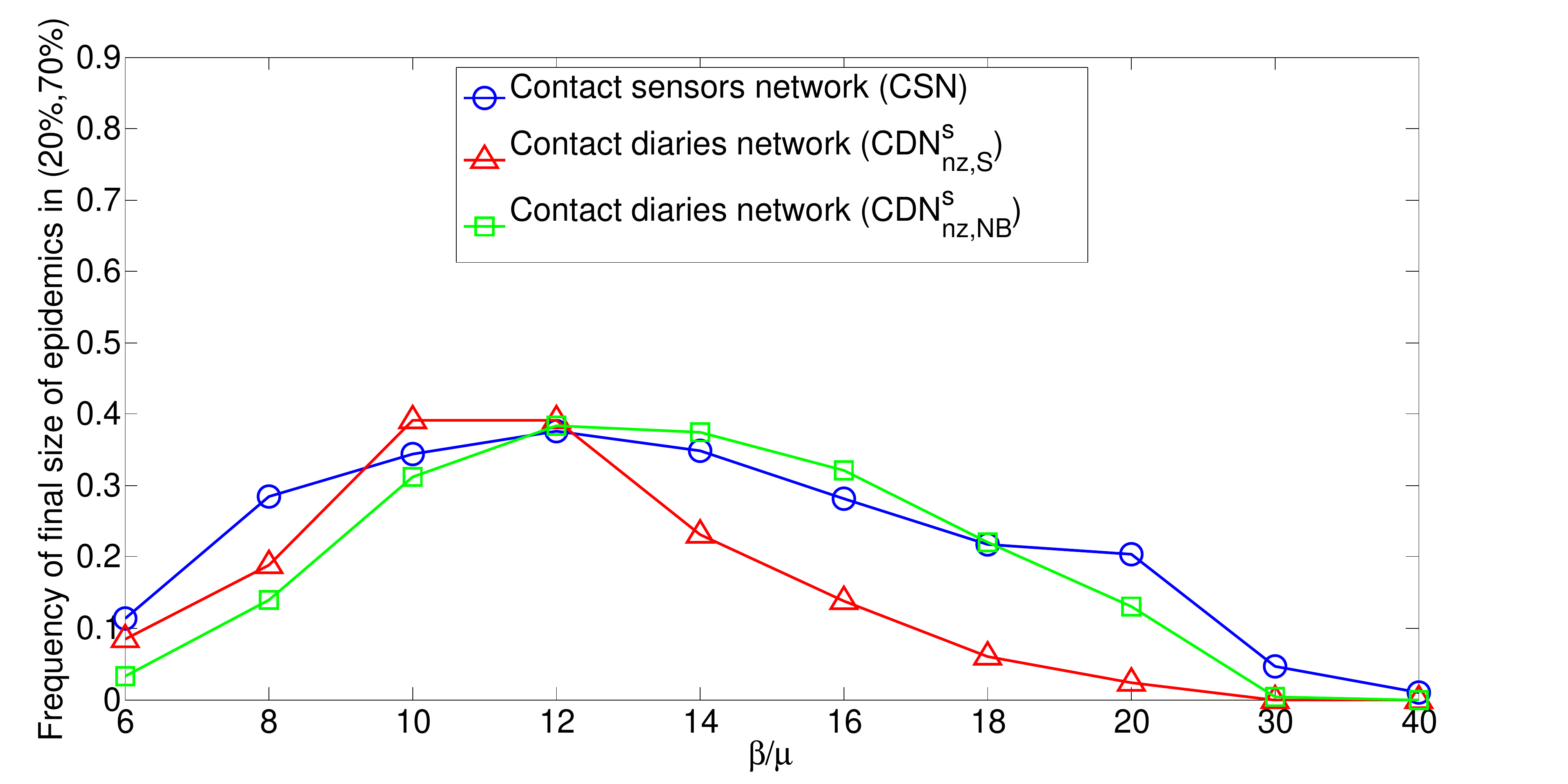}}
\caption{{\bf Fraction of epidemics reaching specific shares of the population.} 
Fraction of epidemics involving (a) more than $10\%$  and (b) between $20\%$ and $70\%$ of the whole population for the contact sensors network and the surrogate contact network ($\text{CDN}^s_\text{nz,NB}$). Values computed over 1000 SIR simulations. Each process starts with one random infected seed.
\label{dist}}
\end{figure}

Fig. \ref{dist} complements the results of Fig. 7 of the main text concerning the distributions of epidemic sizes 
for SIR simulations performed on the contact sensors network and the surrogate contact network 
with weights randomly drawn from the 
negative binomial fit of the publicly available
distributions of contact durations registered by sensors, for several values of $\beta/\mu$.

\newpage

\section{Case of the friendship survey}

We apply here the same procedure to another survey collected in the same school about friendship among students. Since friendship can be considered time-invariant in a short period such as a week, we focus on two contact sensors networks: a daily network as in the main text and a weekly aggregated network. 

Indeed, a high cosine similarity is obtained between the contact matrices of edge densities for the contact and the friendship networks
for both cases (respectively 89\% and 95\%). 
Here we show that such reported relations are poorly informative in terms of the estimation of epidemic risk 
in the considered context, with respect to contact diaries. This can be related to the fact that they capture a different type of social interaction not necessarily 
connected to the number or duration of contacts in the school \cite{Mastrandrea:2015}. This result is in agreement with a recent 
paper \cite{Coviello:2015} showing that friendship networks are not able to predict individual risk. Note however that 
\cite{Coviello:2015} focuses on identifying at-risk individuals while we consider here the global risk of the population, as quantified by the distribution of
epidemic sizes. 

\begin{figure}[!ht]
\centering
{\includegraphics[width=1\textwidth]{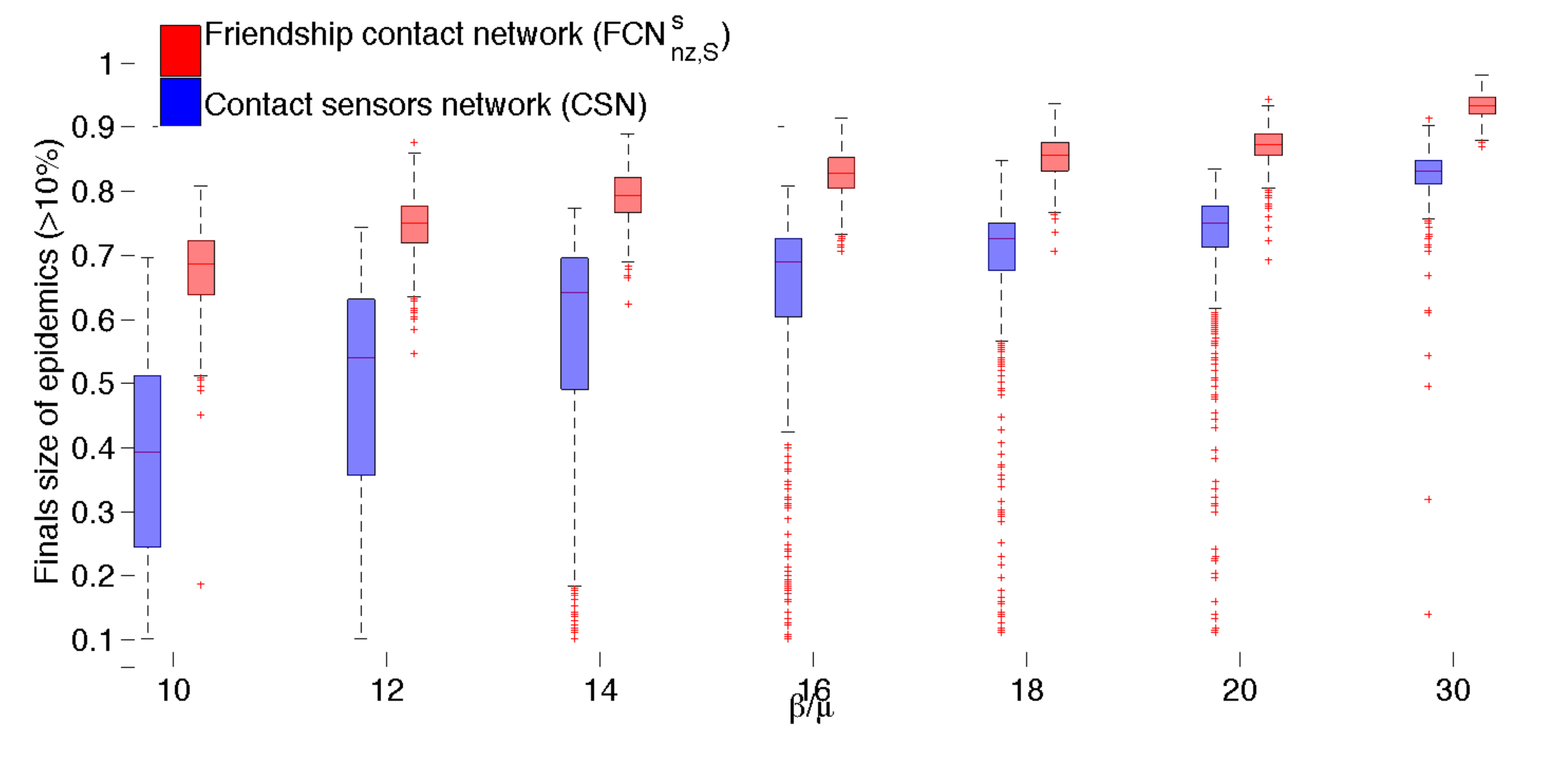}}
\caption{{\bf Box-plots of the distribution of final epidemic sizes.} Comparison of the distributions  obtained from 1000 SIR simulations 
performed on the contact sensors network (CSN) and the surrogate friendship networks without zero-densities in the contact matrix of link densities
and with weights randomly drawn from the distribution of contact durations registered by sensors ($\text{FCN}^s_\text{nz,S}$). For each box, 
the central mark stands for the median, its edges represent the $25^{th}$ and $75^{th}$ percentiles.  The whiskers extend to the most extreme data points not considered outliers, while outliers are shown individually.
\label{fri}}
\end{figure}

\begin{figure}[!ht]
\centering
\subfigure[]
{\includegraphics[width=0.48\textwidth]{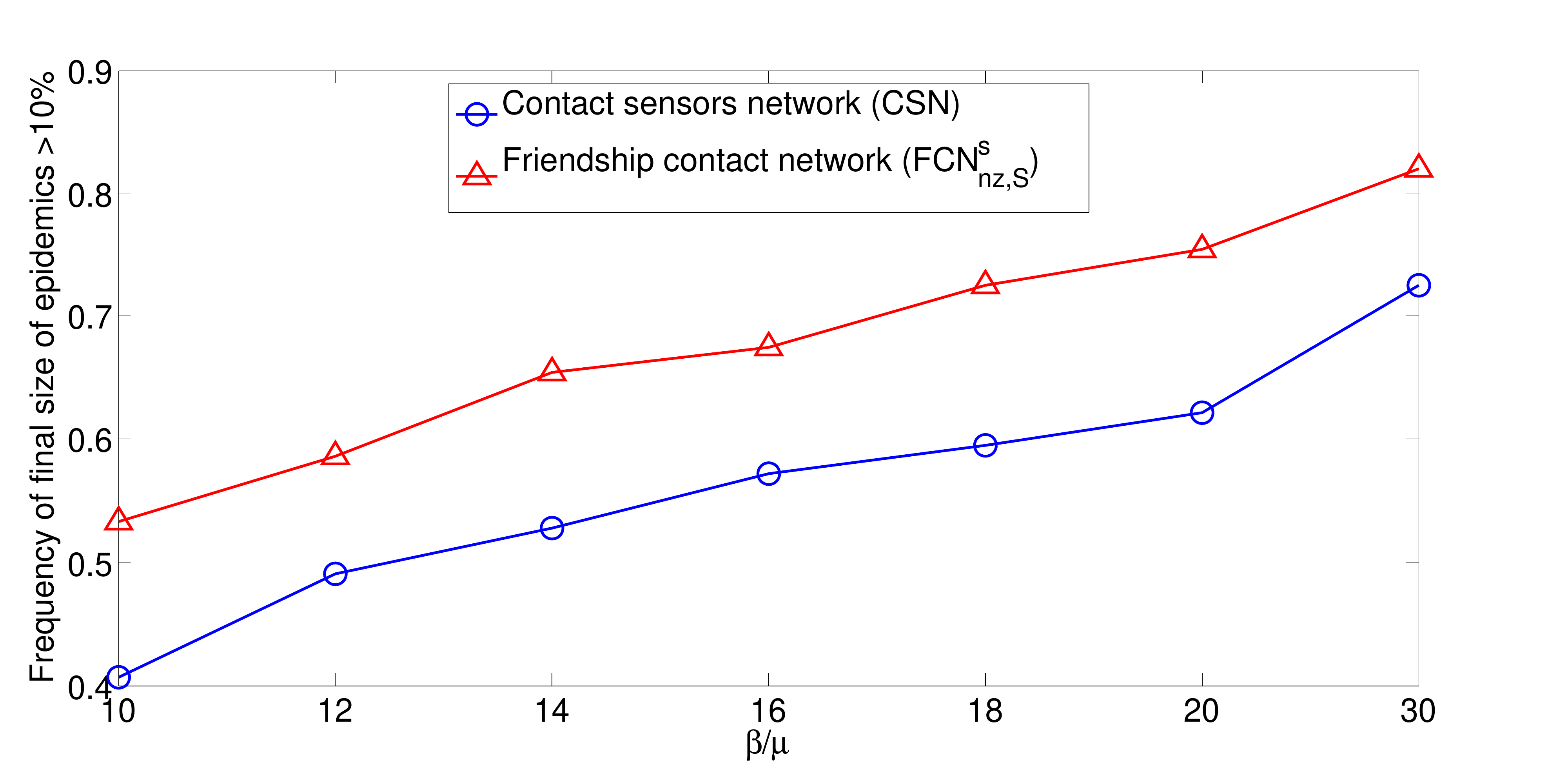}}
\hspace{1mm}
\subfigure[]
{\includegraphics[width=0.48\textwidth]{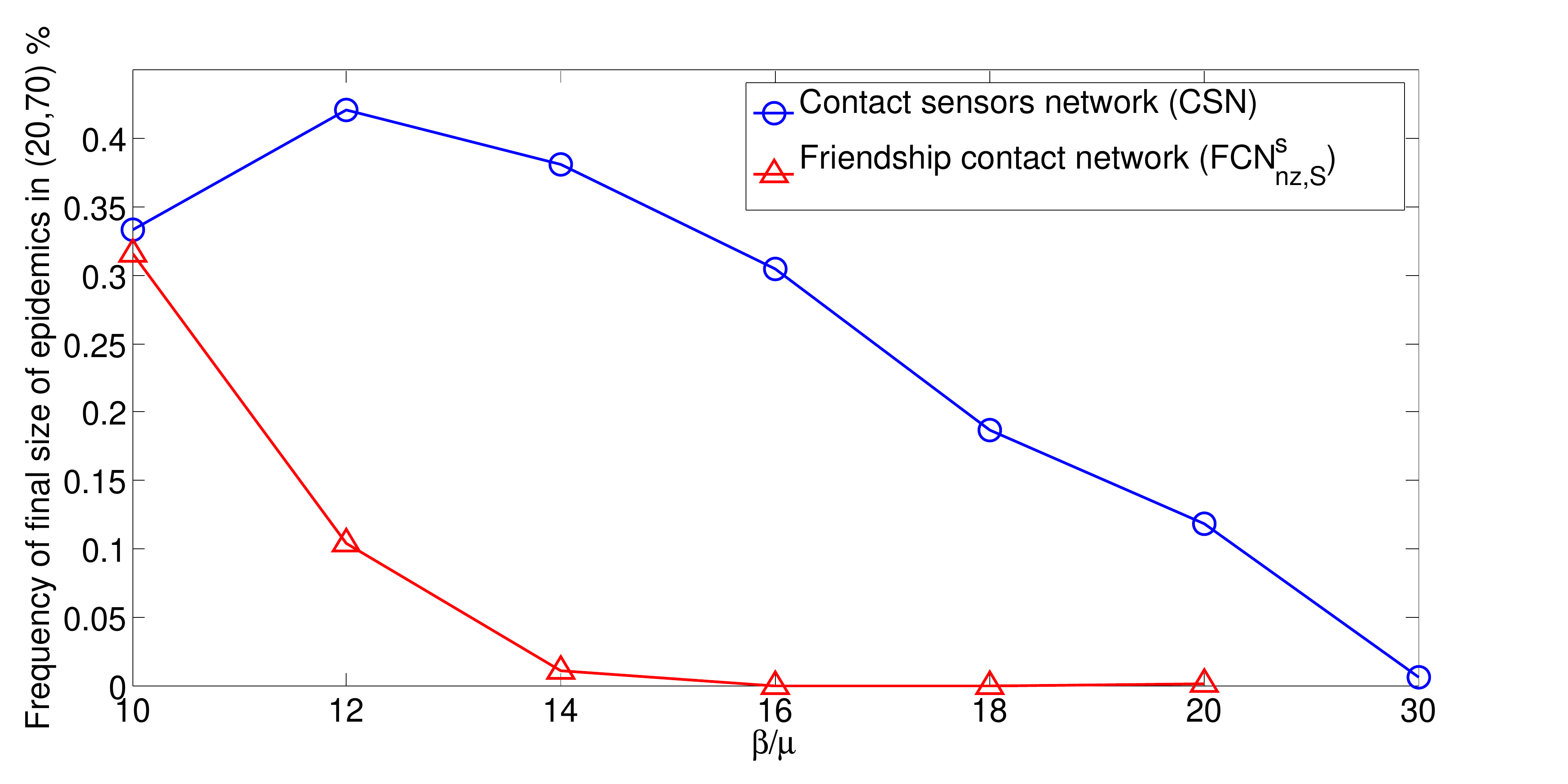}}
\caption{{\bf  Fraction of epidemics reaching specific shares of the population.} 
Fraction of epidemics involving (a) more than $10\%$  and (b) between $20\%$ and $70\%$
of the whole population for the contact sensors network (CSN) and the surrogate friendship network ($\text{FCN}^s_\text{nz,S}$). 
Values computed over 1000 SIR simulations. Each process starts with one random infected seed.
\label{frifre}}
\end{figure}

In Fig. \ref{fri} and \ref{frifre} indeed, we  compare the distributions of final size of epidemics for SIR simulations performed on the contact sensors network and on
surrogate networks obtained through the same procedure as in the main text but taking as starting point the friendship network ($\text{FCN}^s_\text{nz,S}$). 

\begin{figure}[!ht]
\centering
{\includegraphics[width=0.8\textwidth]{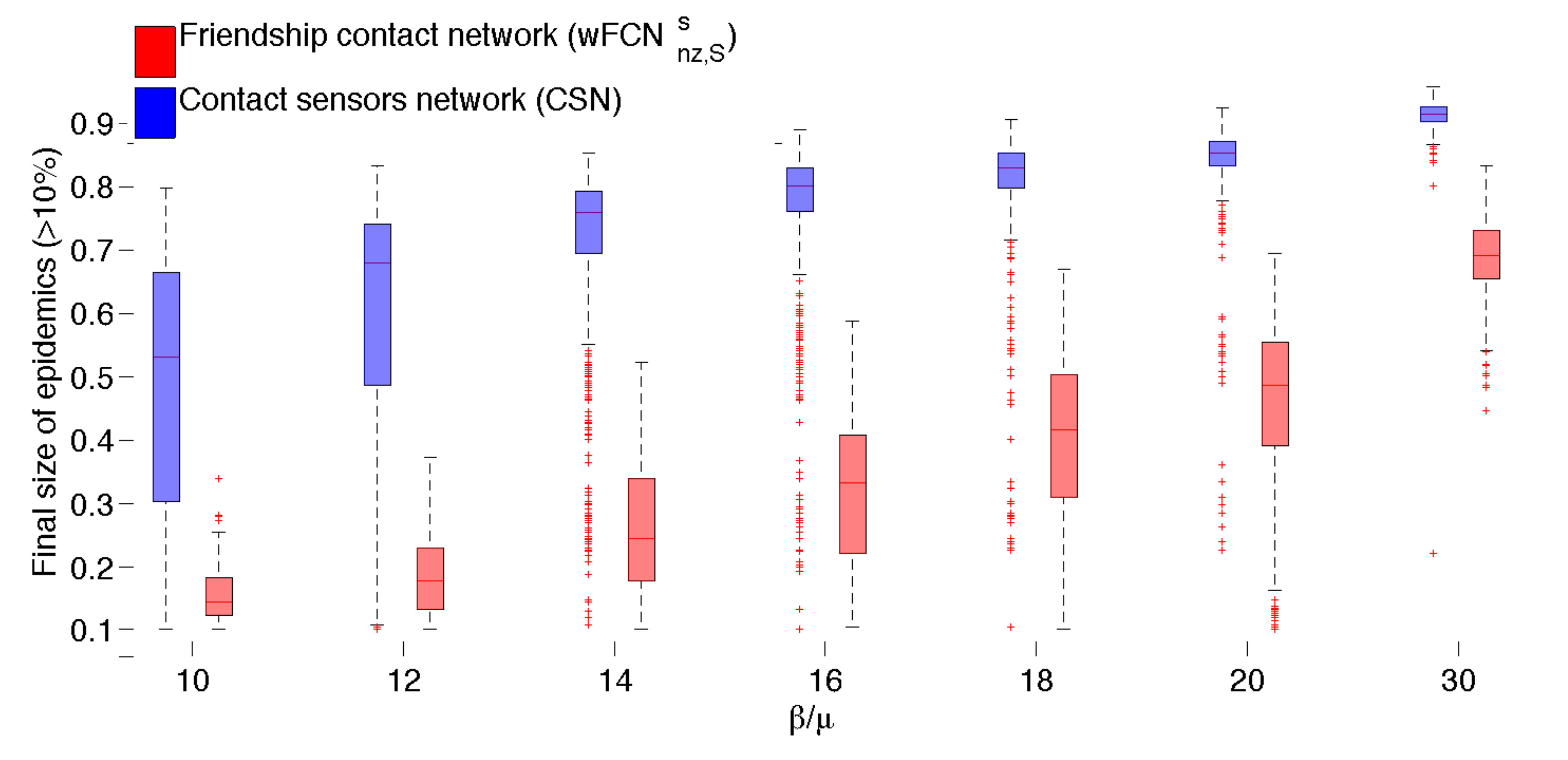}}
\caption{{\bf Box-plots of the distribution of final epidemic sizes.} 
Comparison of the distributions of final sizes of epidemics for 1000 SIR simulations performed on the weekly contact sensors network (wCSN) and 
the weekly surrogate friendship networks without zero-densities and with weights randomly drawn from the distribution of contact durations 
registered by sensors ($\text{wFCN}^s_\text{nz,S}$). In each box, the central mark stands for the median, 
its edges represent the $25^{th}$ and $75^{th}$ percentiles.  The whiskers extend to the most extreme data points not considered outliers, while outliers are plotted individually. 
\label{friw}}
\end{figure}

\begin{figure}[!ht]
\centering
\subfigure[]
{\includegraphics[width=0.48\textwidth]{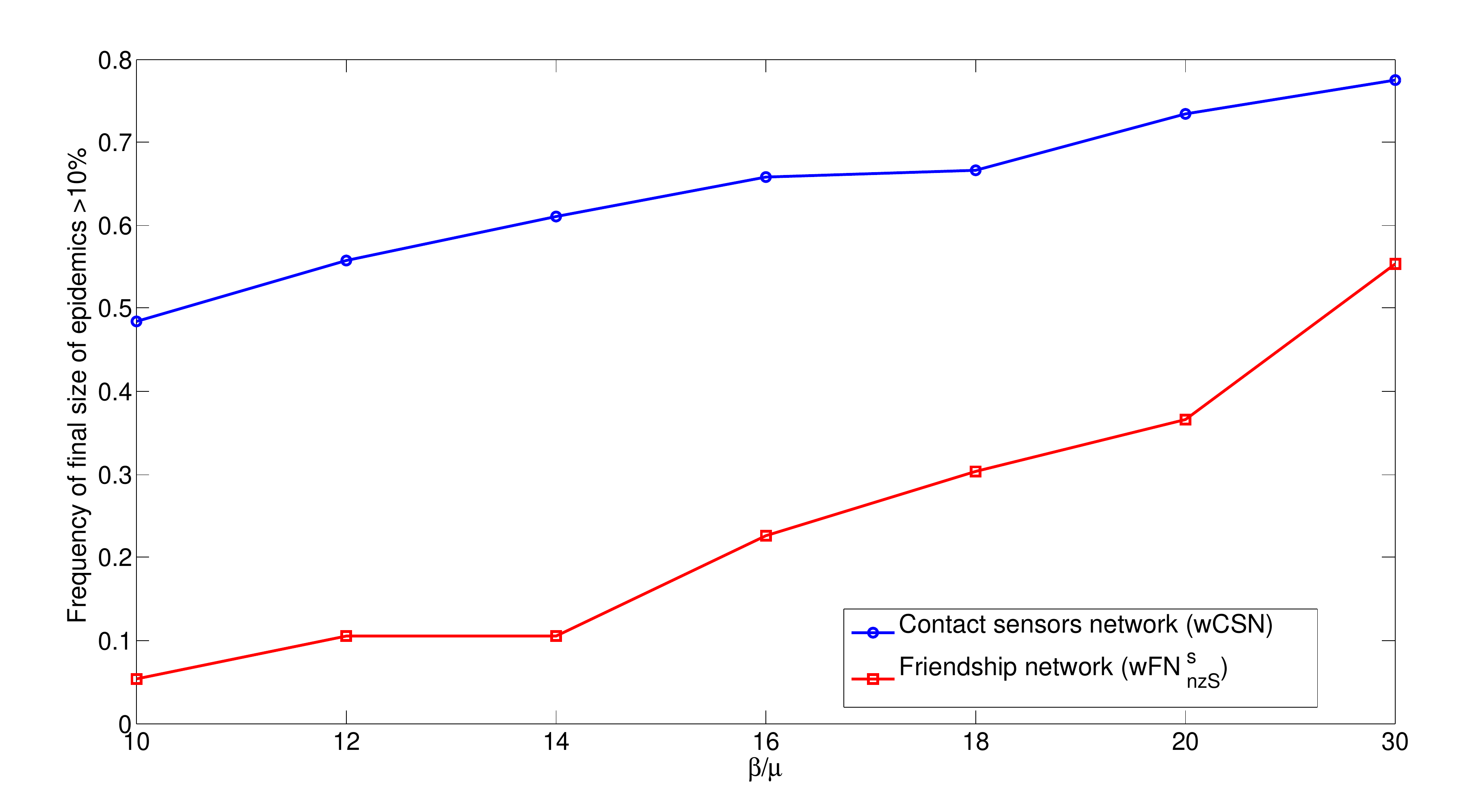}}
\hspace{1mm}
\subfigure[]
{\includegraphics[width=0.48\textwidth]{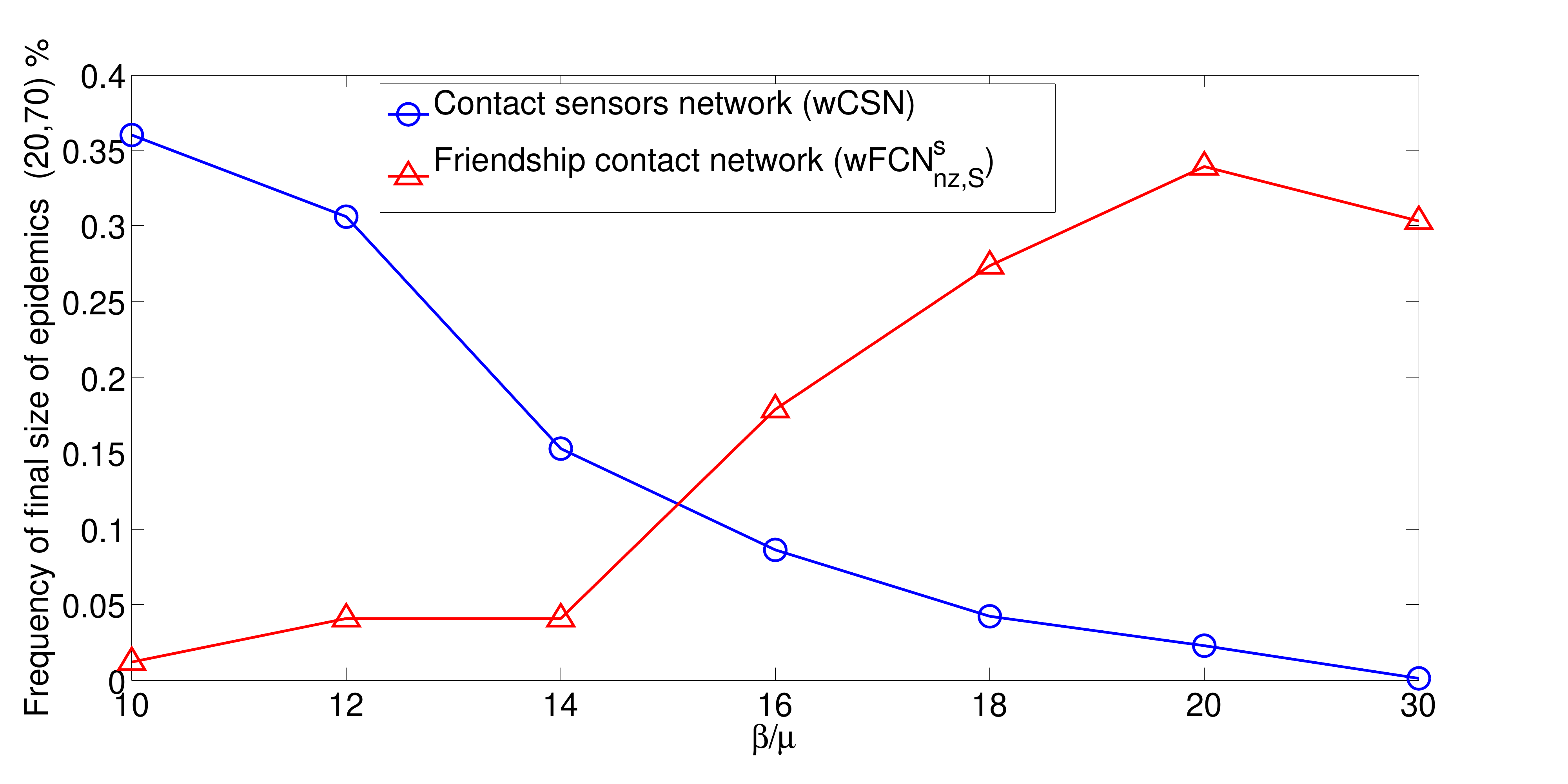}}
\caption{{\bf Fraction of epidemics reaching specific shares of the population.} 
Fraction of epidemics involving (a) more than $10\%$  and (b) between $20\%$ and $70\%$
of the whole population for the  weekly contact sensors network (wCSN) and the weekly surrogate friendship network ($\text{wFCN}^s_\text{nz,S}$). 
Values computed over 1000 SIR simulations. Each process starts with one random infected seed. 
\label{frifrew}}
\end{figure}

The distributions of final size of epidemics for the daily surrogate friendship network shows a clear overestimation of the epidemic risk, 
with a narrow peak concentrated around high shares of population. On the contrary, if the weekly friendship network is used
to build the surrogate network (Figs. \ref{friw} and \ref{frifrew}) we find an important underestimation of the epidemic risk 
and a distribution mostly concentrated around small shares of population. 
Despite the high cosine similarity between the contact matrices of networks, the two social ties (friendship and physical proximity) 
imply a different underlying structure of interactions which does not allow to have a good prediction of the epidemic risk 
associated to the network of contacts between students.

\clearpage

\end{document}